\documentclass[aps,prx,twocolumn,nofootinbib,superscriptaddress,export,eqsecnum]{revtex4-2} 
\usepackage[colorlinks=true,linkcolor=blue,citecolor=blue,urlcolor=blue]{hyperref}

\usepackage[utf8]{inputenc}
\usepackage[german, english]{babel}
\usepackage[dvipsnames]{xcolor}

\usepackage{times}

\usepackage{dcolumn}  
\usepackage{bm}        
\usepackage{bbold}
\usepackage{amsthm}
\usepackage{mathtools}
\usepackage{tikz}
\usetikzlibrary{shapes.geometric, arrows, shapes.multipart}
\usepackage{booktabs, tabularx}

\usepackage{blkarray}
\usepackage{xfrac}

\usepackage[export]{adjustbox}
\usepackage{titlesec}
\usepackage{amsfonts}
\usepackage{amssymb}
\usepackage{braket}

\usepackage{subfigure}
\usepackage{comment}

\usepackage{twemojis}

\newcommand{\mbb}{\mathbb}

\newcommand{\overbar}[1]{\mkern 1.5mu\overline{\mkern-1.5mu#1\mkern-1.5mu}\mkern 1.5mu}

\newcommand{\mD}{\mathcal{D}}
\newcommand{\mA}{\mathcal{A}}
\newcommand{\zotimes}{\otimes_{\mbb{C}[Z(k)]}}

\newcommand{\cX}{\mathcal{X}}
\newcommand{\cZ}{\mathcal{Z}}
\newcommand{\cH}{\mathcal{H}}
\newcommand{\cG}{\mathcal{G}}
\newcommand{\cS}{\mathcal{S}}
\newcommand{\cE}{\mathcal{E}}

\newcommand{\kket}[1]{\ket{#1}\rangle}
\newcommand{\bbra}[1]{\langle\bra{#1}}
\newcommand{\Tr}{\mathrm{Tr}}
\newcommand{\bbraket}[1]{\langle\langle #1 \rangle\rangle}

\usepackage[normalem]{ulem}

\newcommand{\staco}{SymTaco}

\newcommand{\nocontentsline}[3]{}
\let\origcontentsline\addcontentsline
\newcommand\stoptoc{\let\addcontentsline\nocontentsline}
\newcommand\resumetoc{\let\addcontentsline\origcontentsline}

\makeatletter
\def\@fnsymbol#1{\ensuremath{\ifcase#1\or \twemoji{taco}\or \dagger\or \ddagger\or
   \mathsection\or \mathparagraph\or \|\or **\or \dagger\dagger
   \or \ddagger\ddagger \else\@ctrerr\fi}}
\makeatother

\begin{document}

\title{The Symmetry Taco: Equivalences between Gapped, Gapless, and Mixed-State SPTs}

\date{\today}

\author{Marvin Qi}\thanks{Corresponding Author: marvinqi@uchicago.edu}
\affiliation{Leinweber Institute for Theoretical Physics, University of Chicago, Chicago, IL 60637, USA}
\affiliation{Department of Physics and Center for Theory of Quantum Matter, University of Colorado Boulder, Boulder, CO 80309 USA}
\affiliation{Theoretical Sciences Visiting Program, Okinawa Institute of Science and Technology Graduate University, Onna, 904-0495, Japan}

\author{Ramanjit Sohal}
\affiliation{Pritzker School of Molecular Engineering, University of Chicago, Chicago, IL 60637, USA}

\author{Xie Chen}
\affiliation{Institute for Quantum Information and Matter and Department of Physics, California Institute of Technology, Pasadena, CA, 91125, USA}

\author{David T. Stephen}\thanks{Current address: Quantinuum, 303 S. Technology Ct., Broomfield, CO 80021, USA}
\affiliation{Department of Physics and Center for Theory of Quantum Matter, University of Colorado Boulder, Boulder, CO 80309 USA}
\affiliation{Institute for Quantum Information and Matter and Department of Physics, California Institute of Technology, Pasadena, CA, 91125, USA}

\author{Abhinav Prem}
\affiliation{School of Natural Sciences, Institute for Advanced Study, Princeton, NJ 08540, USA}
\affiliation{Physics Program, Bard College, 30 Campus Road, Annandale-on-Hudson, NY 12504, USA}
\affiliation{Theoretical Sciences Visiting Program, Okinawa Institute of Science and Technology Graduate University, Onna, 904-0495, Japan}

\begin{abstract}
Symmetry topological field theory (SymTFT), or topological holography, offers a unifying framework for describing quantum phases of matter and phase transitions between them. While this approach has seen remarkable success in describing gapped and gapless pure-state phases in $1+1$d, its applicability to open quantum systems remains entirely unexplored. In this work, we propose a natural extension of the SymTFT framework to mixed-state phases by introducing the \textit{symmetry taco}: a bilayer topological order in $2+1$d whose folded geometry naturally encapsulates both strong and weak symmetries of the $1+1$d theory. We use this perspective to identify a series of correspondences, including a one-to-one map between intrinsically gapless SPTs (igSPTs) and certain gapped SPTs, and a mapping between igSPTs and intrinsically average SPTs (iASPTs) arising in $1+1$d mixed states. More broadly, our framework yields a classification of short-range correlated $G$-symmetric Choi states in $1+1$d, provides a route for systematically generating mixed-state SPTs via local decoherence of igSPTs, and allows us to identify a new mixed-state ``anomaly". Besides folding in mixed-state phases into the SymTFT paradigm, the symmetry taco opens new avenues for exploring dualities, anomalies, and non-equilibrium criticality in mixed-state quantum matter.
\end{abstract}

\maketitle

\tableofcontents


\begin{figure*}[t]
    \centering
    \includegraphics[width=\textwidth]{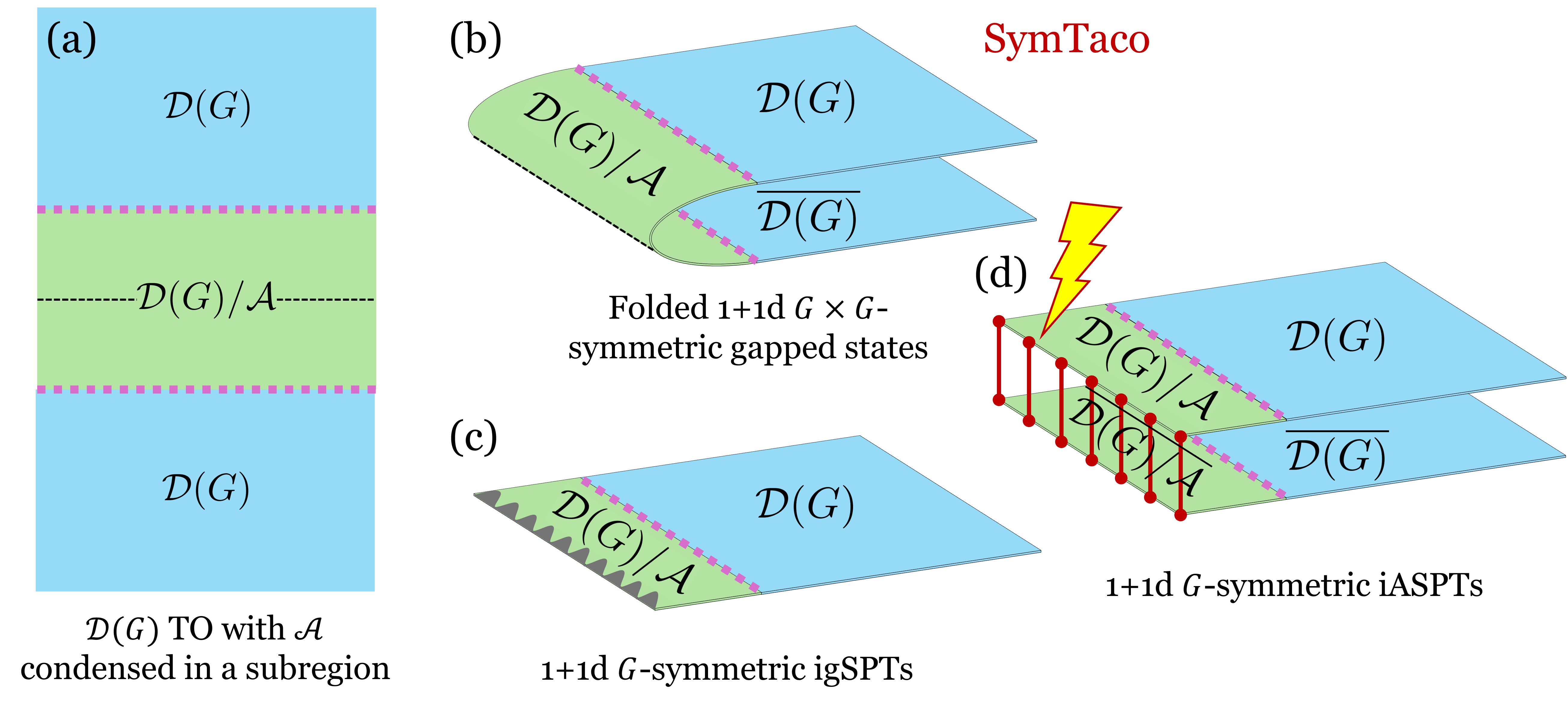} 
    \caption{The Symmetry Taco (sym\twemoji{taco}) framework: (a) starting with a $2+1$d topological order described by the quantum double $\mD(G)$, condense a subset of anyons--specified by a condensable subalgebra $\mA$--in a subregion to obtain a new topological order $\mD(G)/\mA$. (b) Folding this theory along the mirror axis (dashed black line) produces the SymTaco. This setup describes gapped boundaries from the folded $\mD(G) \times \overline{\mD(G)} \cong \mD(G\times G)$ topological order to vacuum. Such \textit{folded} boundaries characterize a subset of $1+1$d gapped phases with $G\times G$ symmetry. (c) Breaking the \staco\, in half along the dashed black line produces a gapless boundary from the $\mD(G)$ topological order to vacuum. Such boundaries characterize $G$-symmetric intrinsically gapless SPT phases. (d) Constraints imposed by folding along the dashed black line precisely correspond to the positivity and Hermiticity constraints on Choi states of density matrices. In the Choi space, folding corresponds to maximal decoherence which couples the two layers, represented here via the bold red lines. The \staco\, hence provides the SymTFT for $1+1$d $G$-symmetric mixed-state SPTs. In all cases, the symmetry boundary (from $\mD(G)$ to vacuum) is given by the canonical charge condensed boundary. See the summary of results~\ref{sec:summary} for details.}
    \label{fig:main}
\end{figure*}

\section{Introduction}
\label{sec:intro}

As Anderson famously observed, ``It is only slightly
overstating the case to say that physics
is the study of symmetry"~\cite{anderson1972}. This is especially true in the context of many-body physics, where symmetry plays a central role in organizing and classifying phases of matter. Within the Landau paradigm, phases are distinguished by distinct patterns of spontaneous symmetry breaking of a given symmetry. In contrast, symmetry protected topological (SPT) phases are quantum phases that fall outside this paradigm: they do not spontaneously break any symmetry, yet remain distinct from each other and from trivial, short-range entangled states provided a global symmetry is enforced~\cite{chen2012spt,chen2013SPT}. Recently, generalized notions of symmetry have led to fresh perspectives on familiar systems: for instance, topological orders (TOs)--long-range entangled phases which support fractionalized anyon excitations--do not require the imposition of any microscopic (UV) symmetries but can nevertheless be cast as phases of matter arising from the spontaneous breaking of emergent higher-form symmetries, which act on extended, rather than point-like, objects. Separately, non-local duality transformations such as the paradigmatic Kramers-Wannier duality~\cite{kramerswannier} have been reinterpreted as ``non-invertible" symmetries, described not by symmetry groups but rather by fusion categories~\cite{shao2023review,sakurareview}. Such generalized notions of symmetry build on the one-to-one correspondence between a symmetry group $G$ and codimension-1 topological defects, which fuse according to group multiplication rules. From this perspective, every topological defect in a theory--which can have higher codimension and can obey non-invertible fusion rules--corresponds to a generalized global symmetry~\cite{barkeshli2014sym,ggs2015,cordova2022,mcgreevy2023review}. Such defects are expected to be described via fusion higher-categories~\cite{etingof2009,kapustin2010,kong2014bfc,douglas2018f2c,gaiotto2019,Johnson_Freyd_2022,yu2024f2c}. 

While such generalized notions of symmetry and the corresponding phases of matter they characterize initially appear disparate, they can in fact be unified within a single framework known as the \textit{symmetry topological field theory} (SymTFT) (and its closely related cousin, \textit{topological holography})~\cite{freed2012,kong2015bdry, kong2017,kong2018emc,freed2018,kong2020holo,kong2020class,chatterjee2022,ji2020,freed2022,ji2023,chatterjee2023shadow,apruzzi2023,lin2023,benini2023,kaidi2023,kaidi2023anomaly,bhardwaj2023ii,moradi2023,Wen2023,Huang2023,Chatterjee2023,bhardwaj2025club,bhardwaj2024landau,bhardwaj2024lattice,warman2024,tiwari2024fermion,wen2024fermion,huang2025, Kong_2020, Kong_2021}. Within this framework, the different phases of a system with a given symmetry structure are described in terms of a topologically ordered ``bulk" in one higher dimension. The SymTFT perspective has most thoroughly been studied in the context of $1+1$d systems (see Refs.~\cite{ji2024top,wen2024string,wen2025top,antinucci2025sym,putrov2025braid,bhardwaj2025gapless} for higher-dimensional generalizations). The central concept of this framework is that the symmetry data of a $1+1$d system with global symmetry $G$ can be encoded, via the bulk-boundary correspondence, within a $2+1$d slab with TO given by the quantum double $\mD(G)$, with appropriately chosen boundary conditions\footnote{Any theory in $d+1$ spacetime dimensions with symmetry $\mathcal{C}$ (a fusion $d$-category in general) can be recovered via its SymTFT, which is the interval compactification of a $d+2$ TO given by the Drinfeld center $\mathcal{Z}_1(\mathcal{C})$ with the following boundary conditions: on one end (called the ``symmetry" boundary), topological boundary conditions corresponding to the canonical gapped boundary of $\mathcal{Z}_1(\mathcal{C})$ are enforced, while the other boundary (called the ``physical" boundary) is allowed to have arbitrary non-topological boundary conditions.}.

Different phases of the $G$-symmetric $1+1$d theory can then be identified with different choices of boundary conditions (i.e., anyon condensates) on the physical boundary of the $\mD(G)$ TO. Crucially, properties of the symmetry operators of the $1+1$d theory--such as their anomalies--can be precisely identified with properties of the anyon operators in the SymTFT. For instance, the SymTFT of a $1+1$d $\mathbb{Z}_2$ symmetric system is given by the $2+1$d $\mathbb{Z}_2$ Toric Code, the TO of which is given by $\mD(\mathbb{Z}_2) = \mathbb{Z}_2 \times \mathbb{Z}_2$. The different phases of the former--the symmetry breaking and symmetry preserving gapped phases, as well as symmetric gapless phases--can be identified with the different boundary anyon condensates of the latter. Moreover, the Kramers-Wannier duality of the $1+1$d theory can be seen as arising from the electromagnetic duality of the Toric Code. 

An appealing aspect of this framework is that it encodes all topological or kinematic properties of a $G$-symmetric theory in the corresponding higher-dimensional SymTFT while disregarding non-universal dynamical data, providing a route to making non-perturbative statements about phases and phase transitions. Consequently, this framework not only reproduces the full classification of gapped $1+1$d $G$-symmetric phases~\cite{kong2020class,kong2020holo,chatterjee2022,kong2022one,bhardwaj2024landau} but also neatly incorporates duality transformations as changes of boundary conditions, which correspond to certain anyon permutation symmetries of the bulk TO~\cite{freed2012,freed2018}. More surprisingly, the SymTFT has also proven useful in classifying \textit{gapless} phases of matter~\cite{Wen2023,Huang2023,bhardwaj2025club,Chatterjee2023,warman2024,tiwari2024fermion,wen2024fermion,huang2025}, especially intrinsically gapless SPTs (igSPTs)~\cite{verresen2021,Thorngren2021}. Despite their gapless nature, igSPTs support topological edge modes, which are determined by an emergent anomaly that arises at low energies. As we will review, igSPTs correspond to partially gapped boundaries of the bulk TO. While the SymTFT does not capture dynamical properties of igSPTs, it does describe their emergent anomaly and hence also their topological features. Given that the SymTFT can describe such a diverse array of gapless and gapped phases of matter, it is natural to ask whether we can also use this framework to establish correspondences between their classifications. This will be one of the primary themes of the present work.

Concurrent with the development of the SymTFT framework, there has recently been rapid progress in characterizing phases of matter in \textit{open} quantum systems, i.e., quantum systems which are coupled to some external environment. Developing a systematic framework for phases in open systems is both of inherent theoretical interest and of immediate practical importance, especially since incoherent processes are unavoidable in current quantum simulators, where a plethora of nontrivially entangled many-body states have already been realized~\cite{satzinger2021,semeghini2021,iqbal2023,xu2023,andersen2023,fossfeig2023,goel2024,fib1}. In the context of open systems, two distinct notions of symmetry arise: a symmetry is \textit{strong} if the system does not exchange any symmetry charges with the environment, and \textit{weak} if the total system-plus-bath remains symmetric, even though the system itself does not conserve the symmetry charge.

While one might expect environmental decoherence to destroy all quantum correlations, it turns out that certain pure-state phases of matter, such as SPTs and TOs, can persist under sufficiently weak noise (up to finite timescales). Remarkably, appropriately engineered environmental couplings can in fact generate \emph{intrinsically} mixed-state phases of matter, with no analogues in isolated quantum systems. As we will review, these include strong-to-weak spontaneous symmetry breaking (SWSSB) states \cite{lee2023,sala2024,lessa2025ssb}, intrinsically average SPTs (iASPTs) \cite{deGroot2022,lee2023aspt,zhang2022strange,Ma2023,Ma2025}, and intrinsically mixed-state topological orders (imTOs) \cite{Sohal2025,Ellison2025,hsin2025lre}. The emergence of a panoply of novel mixed-state phases of matter raises the compelling question of whether the SymTFT framework can be extended to describe the rich structure of mixed-state phases by incorporating notions of strong and weak symmetry. We set ourselves to this task in the present work.

An important goal of this work is to provide the first step towards extending the SymTFT framework to the $1+1$d mixed-state setting\footnote{See also Refs.~\cite{sun2025holo,heckman2025holo} which discuss distinct holographic perspectives for mixed-states.}; along the way, this will lead us to argue for correspondences between both gapless and gapped pure-state phases with intrinsically mixed-state phases. Our central results are illustrated in the tetraptych of Fig.~\ref{fig:main}. We first establish a one-to-one correspondence between igSPTs and certain bilayer gapped phases. We then move to the open system setting, arguing that the bilayer gapped SPTs arising from igSPTs have a natural interpretation as Choi states (i.e., vectorizations) of mixed-state SPTs. Following recent work on classifying mixed-state phases (see e.g., Refs.~\cite{ma2024,zhang2025}), we restrict our analysis to mixed states whose corresponding Choi states are gapped in the usual (pure state) sense. We conclude that the appropriate SymTFT for such mixed states is given by the \textit{symmetry taco} (\staco), i.e., a bilayer TO whose two layers describe the ``ket" and ``bra" spaces, which are not fully independent (hence a taco and not a sandwich). This allows us to establish a correspondence between mixed-state iASPTs and pure-state igSPTs, as well as more generally provide a partial classification of all short-range correlated $1+1$d mixed states. Our perspective motivates a systematic approach for generating certain gapped SPTs via deformations of igSPT Hamiltonians and of obtaining iASPTs by decoherence of igSPTs. Finally, we also show how the symmetry taco can be further exploited to explore novel forms of gauging unique to the mixed-state setting. Before proceeding, we first summarize our main results below.


\subsection{Summary of Main Results}
\label{sec:summary}

The principal character in this work is a folded $2+1$d bulk topological order, whose anyon theory is specified by representations of the quantum double $\mD(G)$ (for $G$ a finite group) and whose boundary conditions we will specify later. We will show that this setup, which we dub the \textit{symmetry taco} (sym\twemoji{taco}), naturally allows us to utilize the theory of anyon condensation in the $2+1$d TO to understand and relate gapped, gapless, and mixed-state SPTs in $1+1$d. Note that while our focus is $\mD(G)$ TOs here, the \staco\, can be defined more generally for a bulk TO whose anyon theory is described by a unitary modular tensor category $\mathcal{C}$ (see Sec.~\ref{sec:gapped}).

To understand the utility of the \staco, consider the quantum double $\mD(G)$ describing the $2+1$d bulk TO. Given this anyon theory and a condensable subalgebra of anyons $\mA$, \textit{anyon condensation}\footnote{In this paper, we will only be discussing the algebraic procedure of anyon condensation and not the energetic process of tuning a Hamiltonian across a phase transition separating distinct topological orders, which is oftentimes referred to by the same moniker.} refers to a formal map between two gapped phases--the original uncondensed TO $\mD(G)$ and the condensed TO $\mD(G)/\mA$. This map provides a systematic approach for identifying the anyons in the former that become condensed (identified with the vacuum superselection sector), are confined, or split into distinct superselection sectors in the latter (see Ref.~\cite{burnellrev} for a discussion). The classification of condensable subalgebras $\mA$ of $\mD(G)$ is well understood~\cite{Davydov2009} and the condensed TO $\mD(G)/\mA$ is generally described by a twisted quantum double $\mD_\omega(G/K)$ specified by a normal subgroup $K\lhd G$ and $[\alpha] \in H^3(G/K,U(1))$ a 3-cocycle class~\cite{Davydov2016}. Besides the bulk picture, anyon condensation can also be understood holographically, i.e., in terms of states defined on a $1+1$d boundary; we now outline three distinct physical pictures that follow from this holographic dictionary.

Suppose we perform anyon condensation specified by the condensable subalgebra $\mA$, but only in some connected subregion of the $\mD(G)$ TO--this leads to the subregion containing $\mD(G)/\mA$ TO, which is separated from the original TO by a pair of gapped domain walls (each specified by a module category~\cite{ostrik2001}), as shown in Fig.~\ref{fig:main}. We can then pick these domain walls in a canonical manner such that anyons from $\mD(G)/\mA$ can pass through a domain wall into the $\mD(G)$ TO without leaving any nontrivial excitations that are confined on the domain wall~\cite{Chatterjee2023}. Now, we fold this $2+1$d system with a subregion containing the condensed anyon theory into a taco, as illustrated in Fig.~\ref{fig:main}. In accordance with the folding trick, the bulk TO is now doubled, i.e., it is given by $\mD(G')$ (with $G' = G\times G$)\footnote{This follows from the fact that $\mathcal{Z}(G\times G) \cong \mathcal{Z}(G) \boxtimes \mathcal{Z}(G)$~\cite{Davydov2009,Beigi2011}.}. We will refer to this bulk TO folded along the condensed subregion as the \staco.

The first way to interpret the condensed subregion via this \staco\, is as follows: after folding, we obtain a gapped boundary between the folded $\mD(G')$ TO and vacuum, with the particular choice of gapped boundary determined by the condensable algebra $\mA$. Thus, distinct allowed choices of gapped boundaries between the folded TO and vacuum correspond to distinct condensable subalgebras $\mA$. On the other hand, gapped boundaries between $\mD(G')$ TO (not necessarily obtained via folding) and vacuum are classified via $1+1$d gapped phases with symmetry $G'$, i.e., in terms of a subgroup $H \subset G'$, which characterizes symmetry breaking, and a cohomology class $[\omega] \in H^2(H, U(1))$, which characterizes SPT order ~\cite{Beigi2011}. Imposing the constraint that the bulk TO is obtained via folding, only a subset of all possible gapped boundaries between $\mD(G')$ and vacuum can be obtained, corresponding to a subset of all $G'$-symmetric $1+1$d gapped phases. We characterize this subset of phases precisely in Sec.~\ref{sec:gapped}, and show how it relates to the classification of condensable subalgebras obtained in Ref.~\cite{Davydov2009}. We remark that this correspondence between condensable subalgebras $\mA$ of $\mD(G)$ and $1+1$d $G'$ symmetric gapped phases was previously obtained in Ref.~\cite{Duivenvoorden2017} using tensor network techniques. 

A second perspective relating anyon condensation to $G$-symmetric $1+1$d phases appears when we now break the \staco\, in half, i.e., cut the system with the condensed subregion in half, as shown in Fig.~\ref{fig:main}. 
Now, we obtain a boundary between $\mD(G)$ TO and vacuum, but this boundary is generically gapless because the algebra that we condensed is non-Lagrangian. Physically, this gaplessness can be enforced if the bulk $1$-form symmetry generated by the uncondensed anyons acts in an anomalous manner on the boundary, forbidding a trivial symmetric gapped boundary theory. 
In other words, the boundary theory must either break this symmetry spontaneously or it must be gapless. As discussed in Refs.~\cite{Wen2023,Huang2023}, the gapless boundary states can be interpreted as gapless SPT phases and hence, the \staco\, also encodes the fact that the classification of $1+1$d gapless SPT phases protected by $G$ symmetry is in one-to-one correspondence with the condensable subalgebras $\mA$ of $\mD(G)$ and, following the preceding paragraph, with a subclass of $1+1$d gapped phases with $G'$ symmetry. Indeed, we show how to directly translate between the algebraic data specifying a $1+1$d gapless phase and that of the corresponding gapped phase, and discuss the physical meaning of this correspondence. 

Finally, we will put these results together to show that the folded boundary of the \staco\, also describes mixed-state SPT phases, providing a third manner in which anyon condensation in the bulk theory encodes the data of $G$-symmetric $1+1$d phases and extending the SymTFT paradigm to the mixed-state setting. We obtain this novel correspondence by making a precise identification between the constraints characterizing the folded boundary and the constraints characterizing Choi states within a doubled Hilbert space. Alternatively, we can also view the folded bulk TO as a mixed-state topological order, providing a SymTFT picture for the boundary mixed state. 

Thus, the holographic correspondence as encoded in the \staco\, allows us to interpret anyon condensation in $\mD(G)$ TOs in terms of $1+1$d (i) gapped $G\times G$-symmetric phases, (ii) gapless $G$-symmetric phases, or (iii) $G$-symmetric mixed-state phases. This leads to a nontrivial correspondence between these three kinds of $1+1$d phases; for instance, it is known that $G$-symmetric gapless SPTs and mixed-state SPTs can exhibit symmetry fractionalization patterns that are impossible in $G$-symmetric gapped SPT phases~\cite{Thorngren2021}. Nevertheless, our correspondence provides a way of relating these phases: given a gapless/mixed-state SPT with symmetry $G$, we can determine the corresponding condensable subalgebra $\mA$ in the bulk \staco, which then allows us to determine the corresponding gapped $G\times G$-symmetric phase. In this way, we can establish equivalences between the tetraptych shown in Fig.~\ref{fig:main}, which summarizes our main results.

The balance of this paper is organized as follows: we review the SymTFT construction for gapped $1+1$d phases in Sec.~\ref{sec:genreview}, followed by a review of partial anyon condensation and its relation to gapless $1+1$d SPT phases in Sec.~\ref{sec:gapless}. In both of these review Sections, we introduce the requisite formalism and discuss explicit examples which will inform our discussion in the following Sections. We introduce the \staco\, in Sec.~\ref{sec:gapped} and use it to establish our first non-trivial correspondence between $G$-symmetric $1+1$d igSPTs and a subset of gapped $G \times G$-symmetric $1+1$d phases, which can be understood as \textit{folded} boundaries of a $2+1$d $\mD(G \times G)$ TO. Here, we show how to relate the symmetry fractionalization data characterizing these seemingly distinct phases, and use this to identify the Hamiltonian deformations that take a stack of igSPTs to a folded SPT. We develop the SymTaco perspective for $1+1$d $G$-symmetric mixed states in Section~\ref{sec:mixed}, where we show how the strong (exact) and weak (average) symmetries inherent to mixed states are encoded in the SymTFT for the Choi state. Crucially, we identify the positivity and Hermiticity constraints on density matrices as being precisely the constraints characterizing folded boundaries of the SymTaco. This in turn allows us to relate igSPTs and iASPTs, and to provide a classification for the latter (under certain physically reasonable assumptions that we clarify). Finally, in Sec.~\ref{sec:apps} we discuss some applications of the SymTaco to mixed states, which include the identification of a new mixed-state ``anomaly," a description of mixed-state SPTs (for Abelian $G$) as symmetric Pauli subsystem codes, and a method for generating intrinsically average SPTs from igSPTs. We conclude in Sec.~\ref{sec:cncls} with a discussion of open questions and future directions. Appendix~\ref{app:notes} contains a brief discussion of the various notations used throughout the text.


\section{Review: Anyon Condensation and the SymTFT}
\label{sec:genreview}

We begin by reviewing the basic features and some elementary examples of the SymTFT construction. In this paper, we will primarily focus on $1+1$d systems with global symmetries given by a finite group $G$ and will hence limit our discussion to this setting. The central conceit of SymTFT is that, for a $1+1$d theory with $G$ symmetry, the kinematic data for the symmetry can be encoded in a $2+1$d TO which is specified by the quantum double $\mD(G)$ of $G$ (i.e., a discrete $G$ gauge theory) and is defined on a strip geometry. It is in this context that the TO is referred to as the SymTFT. The symmetry data of the $1+1$d theory is encoded in the SymTFT as follows: the symmetry operators, local charged operators, and their commutation relations, are related to the ``magnetic" anyons,  ``electric" anyons, and braiding statistics of the $G$ gauge theory, respectively. Since the $1+1$d system can be thought of as living on the boundary of the $2+1$d gauge theory (in a manner that we make more precise shortly), we will often refer to the gauge theory as the ``bulk" and the one-dimensional system as the ``boundary."  In the following, we elaborate on the details of the SymTFT construction by focusing on the classification of gapped phases, providing explicit examples for Abelian symmetries, and reviewing how the gauging of symmetries is incorporated into this framework.


\subsection{The SymTFT Construction}
\label{sec:symtft}

As outlined above, the central object in the SymTFT for $1+1$d, $G$-symmetric systems is the quantum double $\mD(G)$. The topologically distinct superselection sectors (anyons) of this $2+1$d TO (or $G$ gauge theory) are classified by the irreducible representations of $\mD(G)$, which is the quantum double of the group algebra $\mathbb{C}[G]$~\cite{drinfeld}. An anyon $a = ([g],\pi)$ is determined by a conjugacy class $[g]$ and an irreducible representation (irrep) of the centralizer of $g$, which can be chosen to be any representative of the conjugacy class. Anyons of the form $a = ([e], \pi)$ are called (electric) charges, while anyons of the form $a = ([g], 1)$ are called (magnetic) fluxes. All other composite anyons are referred to as ``dyons". 

\begin{figure}[t]
    \centering
    \includegraphics[width=\linewidth]{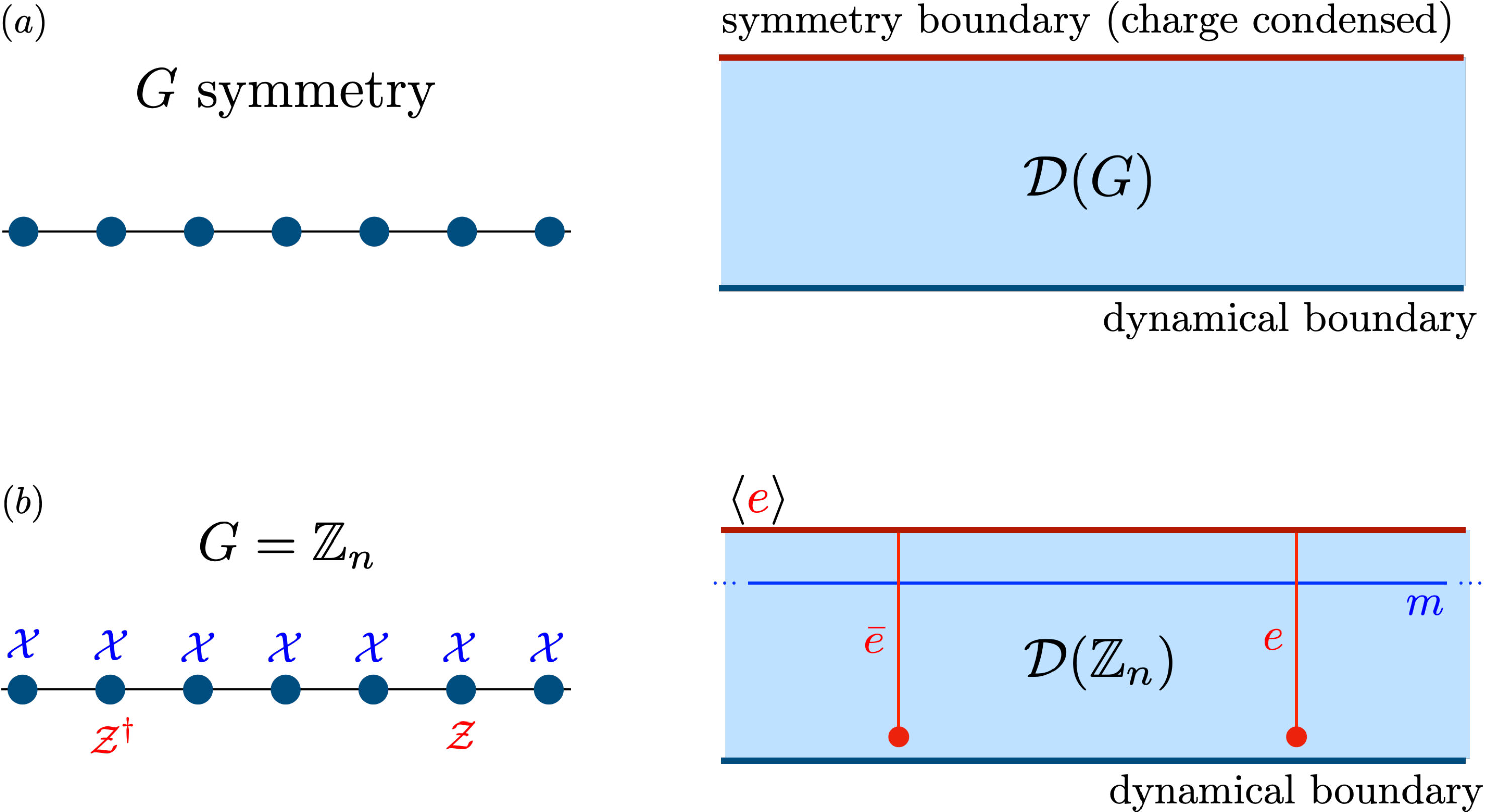}
    \caption{(a) Thin-slab construction of SymTFT for general $1+1$d phases. (b) Illustration of bulk-boundary correspondence for symmetry operator $\prod \mathcal{X}_i$ and local order parameter $\mathcal{Z}_i^k$ for $\mathbb{Z}_n$ symmetry (shown here for $k=1$). }
    \label{fig:slab}
\end{figure}

The mathematical correspondence between gapped boundaries of $2+1$d TO and gapped $1+1$d phases can be pictorially summarized by the ``sandwich" picture, where we place the bulk $\mD(G)$ TO on a strip geometry with finite width (see Fig.~\ref{fig:slab}(a)). Adopting the lexicon of Ref.~\cite{Wen2023}, we denote the top boundary as the ``symmetry boundary" and the bottom boundary as the ``dynamical boundary", which we also refer to as the reference and physical boundaries, respectively. Gapped boundary conditions for $2+1$d TOs are generally described via Lagrangian subalgebras, which physically correspond to maximal collections of anyons that can simultaneously condense at a boundary\footnote{There are two types of anyon condensations: if the subalgebra is Lagrangian, then all anyons (except the vacuum) are confined in the resulting theory. In contrast, if the subalgebra is non-Lagrangian, then the resulting theory retains nontrivial deconfined anyons and is only partially confining.}. In the SymTFT perspective, as applicable to $1+1$d $G$-symmetric phases, the reference boundary is canonically taken as the charge condensed gapped boundary, on which all anyons of the form $([e], \pi)$ are condensed. This corresponds to the Lagrangian algebra $\mathcal L = \text{Rep}(G)$ for a $\mD(G)$ TO.

This reference boundary encodes the kinematic data of the symmetry--the symmetry operators and the local charged operators. The symmetry operators correspond to the line operators for the confined anyons, which form the fusion category $\text{Vec}(G)$. Acting with $g \in G$ on the $1+1$d system amounts to inserting a $g$ line operator oriented parallel to the reference boundary. Since the corresponding anyons are not condensed on the reference boundary (and hence cannot be absorbed by it), such line operators act nontrivially on the dynamical boundary.
In contrast, the local charged operators correspond to the anyons condensed on the reference boundary and are given by line operators which begin at the reference boundary and traverse the finite width of the slab. Such an operator can be thought of as one that tunnels an anyon  $([e],\pi)$ from the reference boundary to the dynamical boundary. Since the collection of these line operators is given by the condensed anyons at the reference boundary, the local charged operators are specified by the category $\text{Rep}(G)$, where the action of the symmetry on a local charged operator is given by the action of an element $g \in G$ on the irrep $\pi$.

As suggested by the name, the dynamical boundary encodes the manner in which the symmetry is dynamically realized on the $1+1$d theory--namely, whether it is preserved or is spontaneously broken. In the remainder of this Section, we will focus on the case where the data of the dynamical boundary corresponds to gapped $G$-symmetric $1+1$d phases. In this case, the dynamical boundary is fully characterized by the Lagrangian subalgebra of condensed anyons: for $\mD(G)$ TO, these are classified by the pair $(H,[\omega])$, where $H$ is some (not necessarily normal) subgroup $H \in G$ and $[\omega] \in H^2(H,U(1))$ is a 2-cocycle class which describes the SPT class of the residual unbroken subgroup $H$~\cite{Davydov2009}. This data precisely coincides with the classification of $1+1$d $G$-symmetric gapped phases. Thus, for each choice of gapped dynamical boundary, the sandwich/slab construction provides a map for identifying the universal topological data characterizing the phase--the ground state degeneracy and the (string) order parameters. We now illustrate how this works in practice via simple, yet instructive, examples where $G$ is taken to be Abelian.


\subsection{Explicit Construction for Abelian $G$ Symmetries}
\label{sec:gapped_abelian_spt}

For $1+1$d gapped phases with Abelian $G$ symmetry, the anyons in the corresponding bulk TO form an Abelian group under fusion, significantly simplifying the analysis (see Appendix~\ref{app:nonabelian} for details on the non-Abelian case). In particular, all anyons $a$ in this case have quantum dimension $d_a = 1$ and can be obtained as fusion products of pure charges and pure fluxes: specifying an anyon $a = (g,\pi)$ simply amounts to specifying a charge $\pi$ and a flux $g$. Since both the group of fluxes and the group of charges are isomorphic to $G$, the group formed by the anyons under fusion is $G\times G$. The topological spin $\theta(a)$ of an anyon $a = (g, \pi)$ is given by 
\begin{equation}
\label{eq:spin}
    \theta((g, \pi)) = \pi(g) \, ,
\end{equation}
and the counterclockwise braiding between two anyons $a = (g, \pi)$ and $b=(h, \chi)$ is given by
\begin{equation}
\label{eq:braiding}
    \mathcal{B}_\theta(a,b) = \pi(h) \chi(g) \, .
\end{equation}
Physically, the phase $\pi(g)$ is the phase obtained via a counterclockwise braiding of the charge $\pi$ around the flux $g$.

For an Abelian anyon theory described by $\mD(G) \cong G\times G$, a Lagrangian subalgebra is entirely specified by a Lagrangian subgroup $\mathcal{L}$, which is a subgroup of anyons which satisfy the following properties:
\begin{itemize}
    \item Each anyon $\ell \in \mathcal L$ is a boson, i.e., it has topological spin $\theta(\ell) = 1$. 
    \item All anyons in $\mathcal{L}$ have trivial mutual braiding, i.e., $\forall \, \ell, \tilde{\ell} \in \mathcal{L}\,, \mathcal{B}_\theta(\ell,\tilde{\ell}) = 1$.
    \item Any anyon not in $\mathcal{L}$ braids nontrivially with some anyon in $\mathcal{L}$, i.e., for any $a \notin \mathcal{L}$, $\exists \, \ell \in \mathcal{L}: \mathcal{B}_\theta(a,\ell) \neq 1$. 
\end{itemize}
The first two conditions ensure that all anyons in $\mathcal L$ can simultaneously condense, while the last condition guarantees that the resulting boundary is fully gapped. For Abelian $G$, the last condition is equivalent to the condition that $|\mathcal{L}| = |G|$. 

For Abelian anyon theories, the Lagrangian subgroup $\mathcal L$ of condensed anyons completely determines the data of the gapped boundary\footnote{Note that when $G$ is non-Abelian, the full data of a gapped boundary $(H, [\omega])$ cannot be completely recovered by the set of condensed anyons~\cite{davydov2014,kobayashi2025soft}, but this set nevertheless provides valuable intuition regarding the gapped boundary. In Appendix~\ref{app:nonabelian}, we compute which anyons from a bulk non-Abelian TO $\mD(G)$ are condensed at the boundary labeled by $(H, [\omega])$.}. In particular, the data $H \subset G$ and the $2$-cocycle class $[\omega]$ can be obtained from the Lagrangian subgroup by viewing $\mathcal L \subset G \times \hat{G}$ as a subgroup of anyons, where the first and second factors correspond to the groups of fluxes and charges, respectively. Then, the subgroup $H \subset G$ is the image of the projection of $\mathcal L$ onto the first factor. Physically, $H$ corresponds to the subgroup of fluxes that appear in the Lagrangian subgroup of condensed dyons. 

To obtain the cocycle class, recall first that for an arbitrary $2$-cocycle $\gamma: H \times H \to U(1)$ we can define the function
\begin{equation} 
\label{eq:Omegadefinition}
    \Omega(g,h) = \frac{\gamma(h,g)}{\gamma(g,h)} \, .
\end{equation}
This function $\Omega$ has the following properties:
\begin{equation} \label{eq:Omegaproperties} 
\begin{aligned} 
    \Omega(g,h) &= \Omega(h,g)^{*} \, ,\\
    \Omega(gh,k) &= \Omega(g,k) \Omega(h,k) \, ,\\
    \Omega(g, hk) &= \Omega(g,h) \Omega(g,k) \, .
\end{aligned}
\end{equation}
The first condition (antisymmetry) follows directly from the definition of $\Omega$, while the second and third properties (bilinearity) follow from $H$ being Abelian and $\gamma$ being a $2$-cocycle. The second and third conditions imply that, for every $h \in H$, the function $\Omega_h: h' \mapsto \Omega(h,h')$ is an irreducible representation of $H$, and the map $h \mapsto \Omega_h$ is a group homomorphism. For Abelian groups, the function $\Omega(g,h)$ satisfying the above properties completely determines the class $[\gamma] \in H^2(H, U(1))$. 

The Lagrangian subgroup $\mathcal L \subset G \times \hat{G}$ specifies a function $\Omega$ with the above properties as follows: first, we obtain $H \subset G$ by projecting $\mathcal{L}$ onto the first factor and taking its image. Then, for each $h \in H$, we choose a charge $\pi_h$ such that $(h, \pi_h) \in \mathcal L$. For $g, h \in H$, we define $\Omega$ to be
\begin{equation}
    \Omega(g, h) = \Omega_g(h) = \pi_g(h) \, .
\end{equation}
To see that the function $\Omega$ is well-defined, i.e., independent of the choice of $\pi_h$ for all $h \in H$, suppose that $(h, \pi_h)$ and $(h, \chi_h)$ are two anyons in $\mathcal L$. Since $\mathcal L$ forms a group, the anyon $(e, \pi_h \chi^*_h)$ also belongs to $\mathcal L$ and (by definition) braids trivially with any anyon $(g, \pi_g) \in \mathcal L$. This implies that, for any $g,h \in H$, we have $\pi_h(g) = \chi_h(g)$. Therefore, different choices of the charge lead to the same function $\Omega$. 

Antisymmetry of $\Omega$ follows from the fact that the anyons $(g, \pi_g)$ and $(h, \pi_h)$ in $\mathcal L$ braid trivially. Indeed, by Eq.~\eqref{eq:braiding}, trivial braiding implies that
\begin{equation}
    B_\theta((g, \pi_g), (h, \pi_h)) = \pi_g(h) \pi_h(g) = 1\, ,
\end{equation}
such that $\Omega(g,h) = \Omega(h,g)^*$. The second condition of Eq.~\eqref{eq:Omegaproperties} follows from the fact that $\mathcal L$ is an Abelian group. Let $(g, \pi_g)$ and $(h, \pi_h)$ be anyons in $\mathcal L$. It follows that $(gh, \pi_g \pi_h)$ is in $\mathcal L$. Since any choice of $\pi_{gh}$ gives the same $\Omega_{gh}$ as shown above, we find that $\pi_{gh} = \pi_g \pi_h$, proving the second property. The third property of Eq.~\eqref{eq:Omegaproperties} immediately follows from the fact that $\Omega_h$ is a representation. 

Conversely, given $H \subset G$ and $\Omega$, we can also determine the corresponding Lagrangian subgroup of anyons. The Lagrangian subgroup $\mathcal L \subset G \times \hat{G}$ consists of anyons $(h, \tilde{\Omega}(h, \cdot))$ for $h \in H$ such that $\tilde{\Omega}(h, \cdot) = \Omega(h, \cdot)$ when the second argument is restricted to $H \subset G$\footnote{There are $|G|/|H|$ possible choices of $\tilde{\Omega}(h, \cdot)$ that restrict to $\Omega(h, \cdot)$ when the second argument is restricted to $H$, and all of the corresponding anyons are condensed.}. Importantly, this group of anyons contains the pure charges $(e, \Omega(e, \cdot))$ that are neutral under $H$ and are responsible for spontaneous symmetry breaking. Antisymmetry of $\Omega$ implies that the anyons are all bosons and have trivial mutual brading, while bilinearity of $\Omega$ ensures that the anyons $(h, \tilde{\Omega}_h)$ form a group. Finally, the Lagrangian condition follows from the fact that for each $h$, there are $|G|/|H|$ charges $\tilde{\Omega}(h, \cdot)$ that restrict to $\Omega(h, \cdot)$, leading to $|G|$ condensed anyons in total.

\subsection*{Examples}
We are now in a position to illustrate the SymTFT correspondence for Abelian $G$ symmetries through examples. For each example, we will describe the data of the gapped dynamical boundary of the SymTFT in terms of both the subgroup $H$ and the $2$-cocycle class $[\omega]$, as well as via the corresponding Lagrangian subgroup of condensed anyons. From the data of the condensed anyons, we will describe how to construct a stabilizer Hamiltonian which realizes the corresponding $G$-symmetric $1+1$d gapped phase (in the non-Abelian case, these will be commuting projector Hamiltonians, which is guaranteed by the condensable algebra being Lagrangian~\cite{cong2016top}). Since these examples and the construction of the stabilizer Hamiltonians will inform our subsequent analysis of mixed-state SPTs, we would encourage even SymTFT experts to peruse the remainder of this Section.

\paragraph*{Example 1: $G = \mbb{Z}_n$.}

In this case, the bulk TO of the SymTFT is given by $\mD(\mbb{Z}_n) = \mbb{Z}_n \times \mbb{Z}_n$, which is generated by the charge $e$ and the flux $m$. Both $e$ and $m$ are self-bosons and satisfy $e^n = m^n = 1$. The topological spin and braiding statistics are given by
\begin{equation}
    \theta(e^a m^b) = e^{2 \pi i \frac{ab}{n}}\, , \,\, \mathcal{B}_\theta(e^a m^b,e^c m^d) = \frac{\theta(e^{a+c}m^{b+d})}{\theta(e^am^b)\theta(e^c m^d)} \, ,
\end{equation}
respectively.

In accordance with the classification of Lagrangian subalgebras, the classification is given by a subgroup of preserved symmetries and a $2$-cocycle. In this case, subgroups of $G$ are given by $H = \mbb{Z}_k$ for any $k$ that divides $n$, while all $2$-cocycles for $H$ are trivial. This data can be equivalently written in terms of the subgroup of condensed anyons, which are given by $(h, \pi)$ where $h \in H$ and $\pi$ is a representation of $G$ that restricts to the trivial representation on $H$. The group of condensed anyons is generated by 
\begin{equation} \label{eq:L_k}
    \mathcal{L}_k = \langle e^{k}, m^{n/k} \rangle.
\end{equation}
where $m^{n/k}$ corresponds to the generator of $H$ and $e^k$ generates the charges that are neutral under $H$. 

Given the bulk SymTFT, one can, in principle, construct an effective lattice Hamiltonian for the boundary which is in correspondence with the set of condensed anyons on the dynamical boundary (see e.g., Refs.~\cite{schmitz2019,chatterjee2022,chatterjee2023shadow,schuster2023holographic,liang2024operatoralgebra,bhardwaj2024lattice}.) We will not reconstruct these arguments here but, rather, will demonstrate how the anyon data of the SymTFT can be used to construct such a Hamiltonian. In the present case, since the bulk TO is given by $\mD(\mathbb{Z}_n)$, we choose the boundary degrees of freedom to be $n$-qudits, with generalized Pauli operators $\mathcal{X}_j,\mathcal{Z}_j$ acting on each site. These operators satisfy the $\mbb{Z}_n$ Pauli algebra, $\mathcal{X}_j^n = \mathcal{Z}_j^n = 1$ and $\mathcal{Z}_j \mathcal{X}_j = e^{2\pi i/n} \mathcal{X}_j \mathcal{Z}_j$. As noted in Section~\ref{sec:symtft} and depicted in Fig.~\ref{fig:slab}(b), the $\mathbb{Z}_n$ symmetry is generated by the magnetic (or $m$) Wilson line--on the edge, we can choose it to be generated by $\mathcal{X} \equiv \prod_i \mathcal{X}_i$. A boundary on which $\langle \mathcal{X}_i \rangle = +1$ would then be understood as one on which $m$ is condensed. Similarly, the $\mathbb{Z}_n$ order parameter is given by the $e$ line stretching from the reference boundary, where it is condensed, to the dynamical boundary. Since $e$ braids with a phase of $e^{2\pi i / n}$ with $m$, it is natural to associate the action of the $e$ line terminating on the dynamical boundary with the operator $\mathcal{Z}_i$, since the commutation relation $ \mathcal{Z}_i \mathcal{X} = e^{2\pi i / n} \mathcal{X} \mathcal{Z}_i$ matches the bulk anyon braiding. A boundary corresponding to condensation of $e$ will thus exhibit long-range order in $\mathcal{Z}_i$; for instance, a state with $\langle \mathcal{Z}_i^\dagger \mathcal{Z}_j \rangle = 1$ would correspond to such an $e$-condensed boundary. This latter operator may be understood as an $e$ line emanating from a point on the dynamical boundary and terminating at an adjacent point on the same boundary, indicating that an $e$-$\overline{e}$ anyon pair can emerge from the bulk and condense on the boundary. This motivates the correspondence
\begin{equation}
\label{eq:anyontostabilizer}
    e \leftrightarrow \mathcal{Z}^\dagger_i \mathcal{Z}_{i+1}, \;\; m \leftrightarrow \mathcal{X}_i \, ,
\end{equation}
between the bulk anyons and the operators which would condense them on the dynamical boundary. Let us emphasize that, at this level, Eq.~\eqref{eq:anyontostabilizer} should be viewed as a heuristic correspondence which can be made precise following the discussion in Refs.~\cite{chatterjee2022,chatterjee2023shadow,schuster2023holographic,liang2024operatoralgebra}. For the purposes of this paper, we find that this intuitive correspondence will suffice. 

Exploiting this correspondence, we can write down a stabilizer Hamiltonian whose ground state subspace realizes a $1+1$d gapped phase with the symmetry data encoded in the bulk SymTFT: 
\begin{equation}
    H_{1d} = - \sum_i \left[ (\mathcal{Z}_i^\dagger \mathcal{Z}_{i+1})^{k} + (\mathcal{X}_i)^{n/k} + \text{ h.c.} \right]
\end{equation}
The nonzero expectation value (evaluated in the ground space) which characterizes the phase is, 
\begin{equation}
    \langle (\mathcal{Z}_i^\dagger \mathcal{Z}_j)^k \rangle = 1 \, .
\end{equation}
The local order parameter $\mathcal{Z}_i^k$ exhibits long-range order, indicating that the symmetry $G=\mathbb{Z}_n$ is indeed spontaneously broken down to $H=\mathbb{Z}_k$, as expected.

\paragraph*{Example 2: $G = \mbb{Z}_2^A \times \mbb{Z}_2^B$.}

Next, let us consider $G = \mbb{Z}_2^A \times \mbb{Z}_2^B$. The bulk $\mD(G)$ TO admits a total of six gapped boundaries to vacuum, of which five correspond to picking subgroups of $G$ but only admit a trivial $2$-cocyle (these gapped boundaries thus all correspond to $1+1$d phases where the $G$ symmetry is spontaneously broken). The remaining gapped boundary, which is obtained by picking the subgroup $H = \mbb{Z}_2 \times \mbb{Z}_2$ (thus there is no spontaneous symmetry breaking) along with a nontrivial cocycle, realizes the nontrivial $\mbb{Z}_2 \times \mbb{Z}_2$ SPT phase. The corresponding Lagrangian subgroup in this case is generated by 
\begin{equation}
\mathcal L_{cluster} = \langle e_A m_B, m_A e_B \rangle.
\end{equation}

As in the preceding example, we may use the SymTFT data to motivate a Hamiltonian realizing the SPT phase. We consider two chains of qubits, labeled as $A$ and $B$, with Pauli algebras $Z_i^\alpha X_i^\alpha = - X_i^\alpha Z_i^\alpha$. The heuristic correspondence of Eq.~\eqref{eq:anyontostabilizer} (now applied separately to the $A$ and $B$ qubits) motivates the Hamiltonian
\begin{equation}
    H_{1d} = - \sum_i \left[Z_i^A Z_{i+1}^A X_i^B + Z_i^B Z_{i+1}^B X_{i+1}^A \right] \, ,
\end{equation}
as realizing the $\mathbb{Z}_2 \times \mathbb{Z}_2$ $1+1$d SPT. This Hamiltonian is indeed the familiar cluster-state Hamiltonian~\cite{raussendorfMBQC, son2012cluster}. From Eq.~\eqref{eq:anyontostabilizer}, we see that the two terms in this Hamiltonian can be interpreted as corresponding to the condensation of $e_A m_B$ and $e_B m_A$, respectively, on the dynamical boundary of the bulk SymTFT. Note that, in the second term we chose $X_{i+1}^A$ rather than $X_{i}^A$ in order to ensure that the stabilizers commute; this choice is not dictated by the heuristic correspondence of Eq.~\eqref{eq:anyontostabilizer}. In the ground state, the stabilizers all have eigenvalue $+1$. The string order parameters characterizing this SPT phase are
\begin{equation}
\begin{aligned}
    \langle Z_{A,i} X_{B,i} \ldots  X_{B, j-1} Z_{A,j} \rangle &= 1 \, ,\\
    \langle Z_{B,i} X_{A,i+1}  \ldots X_{A, j} Z_{B,j} \rangle &= 1 \, ,
\end{aligned}
\end{equation}
where the $\ldots$ denote products of $X_A$ ($X_B$) from sites $i$ to $j-1$ (sites $i+1$ to $j$) in the first (second) line. 

\paragraph*{Example 3: $G = \mbb{Z}_4^A \times \mbb{Z}_4^B$.}

Finally, in anticipation of our analysis for mixed-state SPTs, we consider $G = \mbb{Z}_4^A \times \mbb{Z}_4^B$ and focus on a particular $1+1$d phase exhibiting both symmetry breaking and SPT order under the residual symmetry. We choose $H = \mbb{Z}_4^{AB} \times \mbb{Z}_2^{A}$ to be the subgroup generated by the diagonal $\mbb{Z}_4$ action and $\mbb{Z}_2^A \subset \mbb{Z}_4^A$ acting on the first factor\footnote{This symmetry structure will reappear in our analysis of mixed-state examples, with the ``$AB$" and ``$A$" symmetries corresponding, respectively, to \emph{weak} and \emph{strong} symmetries. See Sec.~\ref{sec:mixed}.}. We choose the Lagrangian subgroup 
\begin{equation}
\label{eq:L_SSBSPT}
    \mathcal L_{SSB+SPT} = \langle e^2_A e^2_B, e_A^2 m_A^2, e_A e^3_B m_A m_B \rangle \, .
\end{equation}
Note that this subgroup contains the anyons $e_B^2 m_B^2$ as well as $m_A^2 m_B^2$. Condensation of the diagonal charge $e^2_A e^2_B$ breaks the symmetry down to $H$, which matches the subgroup of fluxes generated by the remaining anyons. Using this SymTFT data and motivated by the heuristic correspondence provided in Eq.~\eqref{eq:anyontostabilizer}, we can write down a stabilizer Hamiltonian on a two-chain system of qudits for this phase. We find, 
\begin{equation}
\begin{aligned}
    H_{1d} = -\sum_i \Big( &\cZ_{A,i-1}^2 \cX_{A,i-1}^2 \cZ_{A,i-1}^2 + \cZ_{B,i-1}^2 
    \cX_{B,i-1}^2 \cZ_{B,i}^2 \\
    + &\cZ^2_{A,i-1} \cZ^2_{A,i} \cZ^2_{B,i-1}  \cZ^2_{B,i} + \cX^2_{A,i} \cX^2_{B,i} \\
    + &\cZ^\dagger_{A,i-1} \cZ_{A,i} \cX_{A,i} \cZ_{B,i-1} \cZ^\dagger_{B,i} \cX_{B, i} + \text{ h.c.}  \Big)
\end{aligned}.
\end{equation}
As in the $\mbb{Z}_2 \times \mbb{Z}_2$ case, when obtaining this Hamiltonian via the heuristic correspondence in Eq.~\eqref{eq:anyontostabilizer}, some care must be taken to ensure that all terms commute. The (string) order parameters characterizing this phase are 
\begin{equation}
\label{eq:exham}
\begin{aligned}
    \langle \mathcal{Z}_{A,i}^2 \mathcal{Z}_{B,i}^2 \mathcal{Z}_{A,j}^2 \mathcal{Z}_{B,j}^2 \rangle &= 1 \\
    \langle \mathcal{Z}_{A,i}^2 \mathcal{X}_{A,i}^2 \ldots \mathcal{X}_{A,j-1}^2 \mathcal{Z}_{A,j}^2 \rangle &= 1 \\
    \langle \mathcal{Z}^\dagger_{A,i} \mathcal{Z}_{B,i} \mathcal{X}_{A,i+1} \mathcal{X}_{B,i+1} \ldots \mathcal{X}_{A,j} \mathcal{X}_{B,j} \mathcal{Z}_{A,j} \mathcal{Z}^\dagger_{B,j} \rangle &= 1
\end{aligned}.
\end{equation}
The first, local order parameter indicates spontaneous breaking of the symmetry $G$ down to $H$, while the latter two non-local string order parameters encode nontrivial SPT order. The symmetry fractionalization can be seen from the fact that the endpoints of the string order parameters carry charge under the residual symmetry. We can modify the Hamiltonian to explicitly break the symmetry from $\mbb{Z}_4^A \times \mbb{Z}_4^B$ to $\mbb{Z}_4^{AB} \times \mbb{Z}_2^A$ by replacing the first term in the second line of Eq.~\eqref{eq:exham} with $\cZ^2_{a,i} \cZ^2_{b,i}$. Adding this perturbation does not modify the behavior of the string order parameters (they still remain nonzero as $|i-j| \to \infty$, and the SPT order survives). 


\subsection{SymTFT Description of Gauging}
\label{sec:gauging}

The preceding discussion focused on how the SymTFT construction can be used to identify distinct gapped phases of a $1+1$d $G$-symmetric system. In fact, these methods have further utility in that the SymTFT framework can be employed to relate the universal aspects of $1+1$d theories which may be invariant under different symmetry groups but are nonetheless related via \textit{gauging}. Consequently, dualities which arise from gauging--such as the Kramers-Wannier duality~\cite{kramerswannier}--can also be incorporated within this perspective. 

Stated more explicitly, dualities of a $1+1$d gapped $G$-symmetric phase are realized by changing the gapped boundary condition at the \textit{reference} boundary~\cite{kaidi2023,Huang2023}. Consider a $G$-symmetric $1+1$d gapped phase and let $H$ be a subgroup of $G$ that we wish to gauge, with $\mathcal{L}_H$ the Lagrangian subgroup that is condensed on the dynamical boundary in the SymTFT to realize SSB to $H$. In the slab picture, gauging $H$ is realized by changing the reference boundary condition to the $\mathcal{L}_H$ condensed boundary. If $\mathcal{L}'_H$ is a different Lagrangian subgroup that realizes SSB to $H$ (i.e., these differ by an $H$-SPT), then changing the reference boundary to $\mathcal{L}'_H$ in the SymTFT implements twisted gauging in the $1+1$d theory.

This prescription allows us to straightforwardly identify the kinematic data of the new symmetry. As an example, let us take $G = \mbb{Z}_4$ and suppose that $H = \mbb{Z}_2$ is the subgroup we wish to gauge. By Eq.~\eqref{eq:L_k}, the reference boundary is specified by the choice $\mathcal L = \langle e^2, m^2 \rangle$. The symmetry operators are the confined line operators at the reference boundary, generated by $e$ and $m$. Since $e^2$ and $m^2$ are condensed at the reference boundary, $e$ and $m$ generate a $\mbb{Z}_2 \times \mbb{Z}_2$ symmetry. Moreover, the nontrivial braiding between $e$ and $m$ corresponds to a mixed 't Hooft anomaly between the two $\mbb{Z}_2$ symmetries at the boundary. Note that this braiding cannot be removed by simply attaching line operators corresponding to condensed anyons at the reference boundary to the line operators for $e$ and $m$~\cite{freed2014,kaidi2023anomaly}. This mixed 't Hooft anomaly enforces the constraint that there does not exist a short-range entangled, gapped ground state that is invariant under the global $\mbb{Z}_2 \times \mbb{Z}_2$ symmetry. Indeed, any pattern of anyon condensation on the dynamical boundary must share at least one nontrivial anyon with the reference boundary, i.e., there does not exist any magnetic Lagrangian subgroup on the dynamical boundary~\cite{zhang2024_categorical}. This implies that any gapped dynamical boundary \textit{must} partially break the symmetry; this is the SymTFT restatement of the familiar fact that there is no trivial symmetry-preserving gapped state for the anomalous $\mbb{Z}_2 \times \mbb{Z}_2$ symmetry in $1+1$d. 

Finally, note that a special case of gauging occurs when the original charge-condensed reference boundary and the new reference boundary are related by an anyon-permuting symmetry~\cite{Huang2023}. In this case, the original and gauged theories have the same symmetry, which leaves open the possibility that the two corresponding theories are self-dual; indeed, this possibility is realized when the dynamical boundary is itself invariant under the anyon permutation symmetry.


\section{Review: Partial Anyon Condensation and Gapless SPTs} \label{sec:gapless}

While one naturally expects the SymTFT to accurately encapsulate all $G$-symmetric gapped phases (given the bulk-boundary correspondence), it has recently become clear that the utility even extends to certain families of \textit{gapless} phases of matter~\cite{Wen2023,Huang2023,bhardwaj2025club,Chatterjee2023,warman2024,tiwari2024fermion,wen2024fermion,huang2025}. Indeed, when anyon condensation at the dynamical boundary is not fully confining (equivalently, the condensable algebra is not maximal), the dynamical boundary describes the universal kinematic properties of a gapless $1+1$ $G$-symmetric phase. In this Section, we review this connection between partial anyon condensation and gapless $1+1$d phases, focusing in particular on the description of intrinsically gapless SPTs (igSPTs) proposed in Ref.~\cite{Wen2023}.

In order to set the context for igSPTs and partial anyon condensation, let us
consider a $1+1$d lattice spin system, which has a tensor product structure generated by finite-dimensional local Hilbert spaces at each site, and an on-site symmetry $G$. As shown recently, this is equivalent to the statement that $G$ is non-anomalous (so that $H^3(G,U(1)) = \mbb{Z}_1$)~\cite{wilbursahand}. At low energies, it may so happen that the full symmetry group does not act faithfully; specifically, there may be a normal subgroup $A\lhd G$ which only acts nontrivially on some gapped degrees of freedom, such that the effective symmetry group of the low-energy theory is given by the quotient group $G_{low}=G/A$. This structure is described by the short exact sequence, 
\begin{equation}
\label{eq:groupext}
    1 \to A \to G \to G_{low} \to 1 \, .
\end{equation}
The key idea is that while $G$ is on-site and therefore non-anomalous, the action of $G_{low}$ on the low-energy degrees of freedom \textit{can} be anomalous, with an anomaly captured by a cohomology class $[\omega] \in H^{3}(G_{low}, U(1))$. 

Since $G$ is an on-site symmetry, this \emph{emergent anomaly} of $G_{low}$ is trivialized by the symmetry extension to $G$~\cite{Thorngren2021}. In this way, the low energy physics of the $1+1$d $G$-symmetric lattice spin system can mimic the behavior of the boundary of a $2+1$d $G_{low}$-SPT, the anomaly of which enforces either an SSB or a gapless, symmetric phase. By extending $G_{low}$ to $G$ by including gapped degrees of freedom transforming nontrivially under $A$, it is hence possible to realize a protected gapless phase whose robustness is guaranteed by the emergent $G_{low}$ anomaly. Such anomalous gapless phases are referred to as \textit{intrinsically gapless SPTs (igSPTs)}. Despite being gapless, igSPTs are characterized by nontrivial string order parameters and support topologically protected edge modes~\cite{Thorngren2021}, justifying the appellation ``topological". The ``intrinsically gapless" nature of igSPTs refers to the fact that there do not exist purely gapped SPT analogues of these states in $1+1$d. For instance, as we review below, there exists a $1+1$d igSPT protected by $\mathbb{Z}_4$ symmetry which, on its own, is \emph{not} sufficient to protect any gapped SPTs in $1+1$d (since $H^2(\mbb{Z}_4,U(1)) = \mbb{Z}_1$).

Although \textit{a priori} one may not expect the topological data defining a quantum double $\mD(G)$ to be sufficient to fully constrain the dynamics of a $G$-symmetric gapless theory, it turns out that certain boundaries of the former do encode the universal, kinematic data of the emergent anomaly discussed above. However, this requires relaxing the gapped boundary condition enforced on the dynamical boundary in Section~\ref{sec:genreview} and considering gapless boundaries of $\mD(G)$, which permit (partial) characterization in terms of the algebraic data governing anyon condensation. Lifting the gapped boundary condition and allowing gapless edges of the bulk TO is achieved by relaxing the constraint that the corresponding subalgebras of condensed bulk anyons $\mA$ are Lagrangian. As illustrated in Fig.~\ref{fig:main}(a), condensing a condensable but non-Lagrangian subalgebra of anyons $\mA$ within a subregion of the bulk TO leads to a new TO $\mD(G)/\mA$ within the subregion, with gapped domain walls separating the two distinct anyon theories. On the other hand, condensing a non-Lagrangian $\mA$ at a boundary leads to a gapless boundary theory (see Fig.~\ref{fig:main}(c)). As we will review below, the reduced $\mD(G)/\mA$ TO encodes the effective low-energy symmetry group $G_{low}$ of the corresponding $1+1$d gapless theory.

The condensable subalgebras of quantum double models were classified in Ref.~\cite{Davydov2009} and their correspondence to gapless SPTs \textit{via} the SymTFT construction was established in Ref.~\cite{Wen2023}. We review the classification of condensable subalgebras below, after which we explain the paradigmatic example of a $\mbb{Z}_4$ igSPT through the lens of the SymTFT.


\subsection{Classification of Condensable Subalgebras and Partial Anyon Condensation}
\label{sec:class}

In Sec.~\ref{sec:genreview}, we reviewed the correspondence between gapped boundaries of $\mD(G)$ topological orders, Lagrangian subalgebras, and $1+1$d $G$-symmetric gapped phases. Here, we review the classification of \textit{partial} anyon condensation in $2+1$d anyon theories described by the quantum double $\mD(G)$; such partial condensations mathematically correspond to condensable (but not Lagrangian) subalgebras $\mA$ of $\mD(G)$, i.e., sets of bosonic anyons which all braid trivially with each other. Relaxing the Lagrangian condition amounts to lifting the condition that every anyon $b \notin \mathcal{A}$ braid nontrivially with at least one anyon $a \in \mA$. In this scenario, anyons which lie outside $\mathcal{A}$ but braid trivially with anyons in $\mathcal{A}$ will remain deconfined anyons after $\mathcal{A}$ is condensed. 

As discussed in Ref.~\cite{Davydov2009}, the condensable subalgebras $\mathcal{A}$ of $\mD(G)$ are classified by a subgroup $H\subset G$, a normal subgroup $A\lhd H$, a 2-cocycle $\eta\in Z^2(A,U(1))$, and a function $\epsilon:A\times H\rightarrow U(1)$, which satisfy the conditions
\begin{subequations}
\label{eq:condensationdata}
\begin{align} 
    \frac{\epsilon(a,h)\epsilon(b,h)}{\epsilon(ab,h)} &= \frac{\eta(a,b)}{\eta(hah^{-1},hbh^{-1})} \, ,\\
    \epsilon(a,h_1h_2) &=\epsilon(h_2ah_2^{-1},h_1)\epsilon(a,h_2) \, ,\\
    \epsilon(a,b)&=\frac{\eta(b,a)}{\eta(bab^{-1},b)}\, , \quad \forall b\in A \, .
\end{align}
\end{subequations}
The holographic dictionary described in Ref.~\cite{Wen2023} showed how this data can be understood as describing the symmetry fractionalization of $A$ and $H$ on the boundary. In the case when $H=G$, the condensable algebra $\mathcal A$ contains no pure charges, and the resulting boundary does not break any symmetry; since we are interested in $G$-symmetric $1+1$d phases, we restrict to this case for the remainder of this Section. Note that the condensable algebra is Lagrangian if and only if $A = H$. In this case, $\epsilon$ is fully determined by $\eta$, and the classification reduces to the case discussed in Sec.~\ref{sec:genreview}. 

In general for a non-Abelian group $G$, the subalgebra $\mA$ can also be decomposed into the set of condensed anyons as follows: 
\begin{equation}
    \mathcal A = \bigoplus_a n_a a \, ,
\end{equation}
where $a$ labels the condensed anyons and the integer $n_a \in \mbb{Z}_{>0}$ labels the corresponding fusion multiplicity. However, this decomposition does not fully characterize the condensable algebra~\cite{davydov2014,kobayashi2025soft}. Nevertheless, this decomposition into condensed anyons is useful in developing an intuitive picture for the condensation. Since the precise relation between the data $A \subset H$, $\epsilon$, and $\eta$ with the set of condensed anyons is somewhat involved for general $G$, we defer a detailed discussion to Appendix~\ref{app:nonabelian} and discuss the simpler case when $G$ is Abelian here.

For Abelian groups $G$, the conditions in Eqs.~\eqref{eq:condensationdata} simplify:
\begin{subequations}
\label{eq:abeliandata}
\begin{align}
    \epsilon(a,g)\epsilon(b,g) &= \epsilon(ab,g) \, , \label{eq:abeliandataa} \\
    \epsilon(a,g_1g_2) &=\epsilon(a,g_1)\epsilon(a,g_2) \, , \label{eq:abeliandatab} \\
    \epsilon(a,b)&=\frac{\eta(b,a)}{\eta(a,b)} \, , \quad \forall b\in A \, . \label{eq:abeliandatac}
\end{align}
\end{subequations}
In this case, the data encoded in $\epsilon$ and $\eta$ can be repackaged into a (not necessarily Lagrangian) subgroup of condensed anyons. First, note that the condition Eq.~\eqref{eq:abeliandatab} shows that $\epsilon(a, \cdot): H \to U(1)$ is a $1$d irrep of $H$. This gives the charge of the condensed dyon with flux $a$: for every flux $a \in A$, the dyon labeled by $(a, \epsilon(a, \cdot) )$ is condensed. The condition Eq.~\eqref{eq:abeliandataa} ensures that the condensed anyons form a group: the charge bound to the flux $a b$ is equal to the product of the charges associated to $a$ and $b$. To understand the final condition Eq.~\eqref{eq:abeliandatac}, note that since $\eta$ is a 2-cocycle on $A$, the ratio $\Omega(a,b) = \eta(b,a)/\eta(a,b)$ satisfies the same properties in Eq.~\eqref{eq:Omegaproperties} as the function $\Omega$ defined in Eq.~\eqref{eq:Omegadefinition}. As before, $\Omega(a, \cdot)$ can be interpreted as the $A$-charge associated to the flux $a$. Thus, the final condition is a consistency condition that states that the restriction of the $H$-charge $\epsilon(a, \cdot)$ agrees with the $A$-charge $\Omega(a, \cdot)$ when $H$ is restricted to $A$. Finally, we note that the third condition Eq.~\eqref{eq:abeliandatac} implies that
\begin{align}
    \epsilon(a,a) = \epsilon(a,b) \epsilon(b,a)^* = 1, \quad a, b \in A \label{eq:antisymm}
\end{align}
which is equivalent to the statement that the condensed anyons $(a, \epsilon(a, \cdot))$ and $(b, \epsilon(b, \cdot))$ are bosons and braid trivially with one another. 


\subsection{SymTFT for igSPTs}
\label{sec:symigspt}

The key insight of Ref.~\cite{Wen2023} was that the data classifying condensable subalgebras of a quantum double $\mD(G)$ is in one-to-one correspondence with the data characterizing the symmetry fractionalization patterns of $G$-symmetric igSPTs in $1+1$d. 
Here, we briefly sketch out this correspondence and refer the reader to Ref.~\cite{Wen2023} for a more detailed exposition.

Consider a $G$-symmetric $1+1$d spin system. Let us assume that, at low energies, a normal subgroup $A \lhd G$ acts trivially, such that the effective low-energy symmetry group is given by the quotient $G_{low} = G/A$. Now, on a system with periodic boundary conditions, we expect that the symmetry operator $U_a$, with $a\in A$, acts trivially. However, since $A$ only acts nontrivially on the gapped degrees of freedom, the symmetry may \emph{fractionalize} on a system with open boundary conditions, such that $U_a = U_a^L \otimes U_a^R$, where $U_a^{L/R}$ has spatial support only on the left/right edge of the system. Since the symmetry operators $U_a$ form a linear representation of $A$, the restricted symmetry operators $U_a^{\alpha = L/R}$ need only form a projective representation:
\begin{align}
    U_a^\alpha U_b^\alpha = \eta(a,b) U_{ab}^\alpha , \qquad a,b \in A \, ,
\end{align}
where $\eta \in \cZ(A,U(1))$ is a 2-cocycle. Physically, the presence of a nontrivial $[\eta]$ implies that the gapped degrees of freedom form an $A$-SPT. 

In contrast, since $U_g$ for general $g \in G$ will also act nontrivially on the gapless degrees of freedom, it cannot fractionalize. Nevertheless, there may still be a nontrivial interplay between the gapped and gapless sectors; indeed, since $g a g^{-1} \in A$ for $g \in G$ and $a\in A$, the operator $U_g U_a U_g^{-1} = U_{g a g^{-1}}$ will generally fractionalize into $U_{g a g^{-1}} = U_{g a g^{-1}}^L \otimes U_{g a g^{-1}}^R$. These fractionalized operators need only satisfy
\begin{align}
    U_g U^\alpha_{a} U_g^{-1} = \epsilon(a,g) U^\alpha_{g a g^{-1}} \, ,
\end{align}
where $\epsilon(a , g) \in U(1)$ is a phase that may be interpreted as measuring the fractionalized charge under $G$ that $U^\alpha_a$ carries. As shown in Ref.~\cite{Wen2023}, the 3-cocycle $[\alpha] \in H^3(A,U(1))$ characterizing the emergent anomaly of the igSPT can be extracted from the pair of functions $(\eta,\epsilon)$.

Now, the group structures of $G$ and $A$ impose constraints on $\eta$ and $\epsilon$. Remarkably, these are precisely the same constraints satisfied by the pair $(\eta, \epsilon)$ characterizing condensable subalgebras of the quantum double $\mD(G)$, which are given in Eq.~\eqref{eq:condensationdata}. This establishes a one-to-one correspondence between igSPTs and partially condensed boundaries of quantum doubles, such that the latter provide SymTFTs for the former. Using this SymTFT description, one can in fact extract the boundary modes and string operators characterizing igSPTs (see Refs.~\cite{Wen2023,Huang2023} for details). We now briefly discuss an example of this correspondence, which will also feature prominently in our treatment of mixed-state SPTs.


\subsection{Example: $\mbb{Z}_4$ igSPT}
\label{sec:z4eg}

To illustrate the main idea behind the SymTFT perspective on igSPTs, let us consider in detail the example of the $1+1$d $\mbb{Z}_4$ gapless SPT first introduced in Ref.~\cite{Thorngren2021}. The $\mbb{Z}_4$ gapless SPT is characterized by the short exact sequence 
\begin{equation}
    1 \to \mbb{Z}_2 \to \mbb{Z}_4 \to \mbb{Z}_2 \to 1 \, ,
\end{equation}
where the low-energy symmetry $G_{low} = \mbb{Z}_2$ exhibits an emergent anomaly characterized by the nontrivial 3-cocycle class $\alpha \in H^3(\mbb{Z}_2, U(1)) \cong \mbb{Z}_2$, which is trivialized by the extension to the full symmetry group $G = \mbb{Z}_4$~\cite{Thorngren2021}. The low energy behavior of the $\mbb{Z}_4$ gapless SPT is identical to that of the symmetry-preserving boundary of the $2+1$d $\mbb{Z}_2$ Levin-Gu SPT~\cite{levingu}, which carries the same anomaly $\alpha$. The physical consequence of this anomaly is that it prevents a $1+1$d theory from realizing a short-range entangled state that is symmetry preserving, thereby enforcing either gaplessness or spontaneous symmetry breaking.

\begin{figure}
    \centering
    \includegraphics[width=\linewidth]{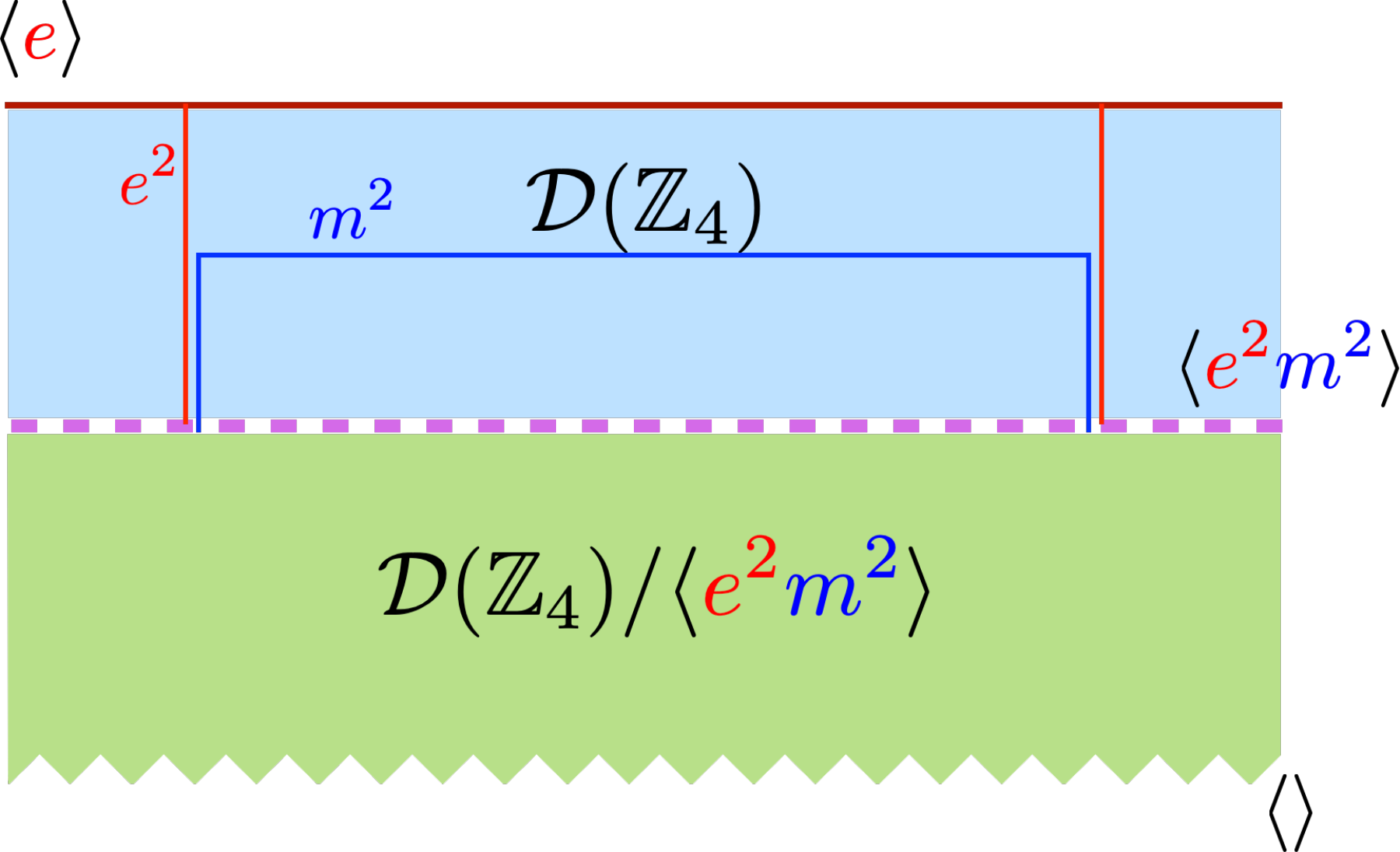}
    \caption{Thin-slab construction of the SymTFT corresponding to the $\mathbb{Z}_4$ igSPT (dotted pink line), which is obtained by condensing the non-Lagrangian subalgebra generated by $\langle e^2 m^2 \rangle$ in the $\mD(\mbb{Z}_4)$ TO. The string order parameter shown here has long range order, and encodes Eq.~\eqref{eq:igspt_stringorder}.}
    \label{fig:gapless}
\end{figure}

That the data encapsulating the universal, low-energy behavior of the $\mbb{Z}_4$ gapless SPT and the boundary of the Levin-Gu SPT are identical suggests that the holographic dual for the former should be the gauged version of the Levin-Gu SPT, which is given by the doubled semion model TO $\mD(\mbb{Z}^\alpha_2)$. On the other hand, since the global symmetry for the gapless SPT is $\mbb{Z}_4$, one might expect that the corresponding bulk anyon theory is given by the $\mD(\mbb{Z}_4)$ TO. As it turns out, the resolution to this ostensible puzzle is that both are true since the doubled semion TO can be obtained via the $\mD(\mbb{Z}_4)$ anyon theory by condensing the $e^2 m^2$ anyon in the latter. The physical picture is provided by Fig.~\ref{fig:gapless}: the blue region denotes the $\mD(\mbb{Z}_4)$ TO while the green region denotes the doubled semion TO, obtained via condensation of $e^2 m^2$ in the former. Thus, the condensable algebra generated by $\langle e^2 m^2\rangle$ in $\mD(\mbb{Z}_4)$ precisely encodes the local low-energy properties of a $1+1$d $\mbb{Z}_4$ symmetric gapless theory (red dotted line in Fig.~\ref{fig:gapless}), which correspond to a $\mbb{Z}_2$ symmetry along with the Levin-Gu anomaly $\alpha$. As discussed above, this is precisely the data that characterizes the $\mbb{Z}_4$ igSPT.

The condensable algebra $\mA$ can alternatively be characterized in terms of the data defined in Sec.~\ref{sec:class}: here, $H = G$ and $A = \mbb{Z}_2 \subset \mbb{Z}_4$, with
\begin{align}
\label{eq:z4_eta_epsilon}
\begin{split}
    \eta(a,b) &= 1 \, ,\\
    \epsilon(a,g) &= i^{ag} \, ,
\end{split}
\end{align}
for $a \in \{0,2\} \cong \mbb{Z}_2 \subset \mbb{Z}_4$ and $g \in \{0,1,2,3\} \cong \mbb{Z}_4$. When $a=0$, $\epsilon(a, \cdot)$ is the trivial representation of $\mbb{Z}_4$; when $a=2$, $\epsilon(a, \cdot)$ is the charge-2 representation of $\mbb{Z}_4$, reflecting the condensation of $e^2 m^2$. 

Let us see how the anyon condensation perspective is realized within a concrete lattice model describing the $\mbb{Z}_4$ igSPT. Consider a one-dimensional lattice with two qubits (labeled by $A$ and $B$) living on each site of the lattice. The on-site $\mathbb{Z}_4$ symmetry action is defined via the unitary operator
\begin{equation}
    U \equiv \prod_i X_i^A \sqrt{X_i^B} \, ,
\end{equation}
where the operator $\sqrt{X_i^B} \equiv \exp(i \frac{\pi}{4} (1-X_i^B))$ squares to $X_i^B$. A local Hamiltonian with this global $\mbb{Z}_4$ symmetry is given by,  
\begin{equation}
    H_{gSPT} = H_{\Delta} + H_{low} \, ,
\end{equation}
where the first term
\begin{equation}
    H_{\Delta} = -\Delta \sum_i Z_{i-1}^A X_{i-1}^B Z_i^A \, ,
\end{equation}
imposes an energetic constraint via $\Delta \gg 1$, which is a large positive constant that defines the low-energy subspace of this model. The second term $H_{low}$ consists of sums of terms which commute with both the symmetry and the constraint imposed by $H_{\Delta}$. One natural choice is given by
\begin{equation}
    H_{low} = -g \sum_i Z^B_{i-1} X^A_i Z^B_i + Y^B_{i-1} X^A_i Y^B_i \, ,
\end{equation}
where the two terms individually commute with the constraint imposed by $H_\Delta$ and are also related to each other via the symmetry operator $U$. As discussed in Ref.~\cite{Wen2023}, this choice realizes a gapless spectrum. Both directly and also from the SymTFT perspective, one can straightforwardly obtain the nontrivial string order parameter and the localized edge modes characterizing the low-energy behavior of this model.

To understand the relation between this Hamiltonian and the partial anyon condensation in the corresponding bulk theory, it is helpful to group the two qubits per site into a single $\mbb{Z}_4$ qudit. Denoting the $\mbb{Z}_4$ Pauli operators by $\mathcal{X}$ and $\mathcal{Z}$, the mapping between the two qubits and the $\mathbb{Z}_4$ qudit is as follows:
\begin{equation}
\begin{aligned}
    X_i^A \sqrt{X_{i}^B} &= \mathcal{X}_i \, , \\
    Z_i^A Z_{i}^B CX_i^{AB} &= \mathcal{Z}_i \, .
\end{aligned}
\end{equation}
The symmetry operator $U$ is then simply giving by the product $U = \prod_i \mathcal{X}_i$, and the energetic constraint $H_{\Delta}$ takes the form 
\begin{equation}
\label{eq:HDelta}
    H_\Delta = -\Delta \sum_i \mathcal{Z}_{i-1}^2 \mathcal{X}_{i-1}^2 \mathcal{Z}_i^2.
\end{equation}
Under the correspondence outlined in Eq.~\eqref{eq:anyontostabilizer}, this is precisely the stabilizer Hamiltonian corresponding to the boundary condensation of $e^2 m^2$ in the $\mD(\mbb{Z}_4)$ TO. The terms in $H_{low}$ do not have a simple or illuminating representation in terms of $\mbb{Z}_4$ degrees of freedom, so we omit them here. However, we note that from Eq.~\eqref{eq:HDelta} one can deduce that the igSPT state satisfies 
\begin{equation}
\label{eq:igspt_stringorder}
    \langle \mathcal{Z}_{i}^2 \mathcal{X}_{i}^2 \ldots \mathcal{X}_{j-1}^2 \mathcal{Z}_{j}^2 \rangle = 1 \, ,
\end{equation}
indicating long-range order in a string order parameter (see Fig.~\ref{fig:gapless}). 
In the following, we will relate this $\mbb{Z}_4$ igSPT to its gapped and mixed-state counterparts. 


\section{From Gapless SPTs to Gapped SPTs via the SymTaco}
\label{sec:gapped}

In this Section, we will leverage the SymTFT framework reviewed in the prior two Sections to establish our first result: namely, a correspondence between $1+1$d $G$-symmetric gapless SPTs and certain $G\times G$-symmetric gapped phases. This correspondence follows from the \staco\, construction, whereby the SymTFTs of the aforementioned families of gapless and gapped phases can be directly related via the \textit{folding} and \textit{cutting} procedures displayed in Fig.~\ref{fig:main}. Here, the folded bulk $\mD(G)$ TO, which is folded along a subregion which hosts $\mD(G)/\mA$ TO, defines the symmetry taco alluded to in Section~\ref{sec:intro}. While our primary objective in introducing the \staco\, is for extending the symTFT perspective to mixed-state phases (which we develop in the following Section~\ref{sec:mixed}), it also sheds new light on the emergence of gapped $G\times G$-symmetric phases in certain tensor network descriptions of $\mD(G)$ TOs~\cite{Duivenvoorden2017} and provides a systematic approach for relating gapless and gapped SPTs. 

Let us first construct the \staco\, for general anyon theories: we begin with a $2+1$d TO whose anyon theory is given by a unitary modular tensor category $\mathcal{C}$ and condense anyons belonging to a condensable subalgebra $\mA$ in some subregion. This reduces the TO in the subregion to $\mathcal{C}/\mA$, which is separated from the $\mathcal{C}$ TO by a pair of gapped domain walls, which we assume are identical, i.e., they are described by the same condensable subalgebras. This choice endows the system with reflection symmetry about an appropriately chosen mirror line passing through the condensed subregion, which maps the two regions with $\mathcal{C}$ TO onto one another. We obtain the \staco\, by folding the system along this mirror line, yielding a bilayer with TO given by $\mathcal{C} \boxtimes \overline{\mathcal{{C}}}$\footnote{Here, $\overline{\mathcal{C}}$ denotes the unitary modular tensor category with the reversed braiding of $\mathcal{C}$, i.e., it is the time-reversed conjugate of $\mathcal{C}$.} which has a gapped boundary to vacuum along the fold. More precisely, starting from the doubled bulk TO and moving towards the fold, there is a gapped boundary from $\mathcal{C} \boxtimes \overline{\mathcal{{C}}}$ to $\left(\mathcal{C} \boxtimes \overline{\mathcal{{C}}} \right)/\left(\mA \boxtimes \bar{\mA}\right)$, followed by a gapped boundary from the latter TO to vacuum. It is this particular setup that we refer to as the \staco.

The key point, which we will expound on shortly in this Section, is that we can only obtain a subset of all possible gapped boundaries between the doubled TO $\mathcal{C} \boxtimes \overline{\mathcal{{C}}}$ and vacuum when the former is obtained via the folding procedure described above. We will refer to this restricted set of gapped boundaries as \textit{folded boundaries}. As reviewed in Sec.~\ref{sec:genreview}, in the case when $\mathcal{C} = \mD(G)$ these gapped boundaries correspond to gapped phases of $1+1$d $G\times G$-symmetric systems; hence, we expect the condensable subalgebras $\mA$ to correspond to a subset of all $G\times G$-symmetric gapped phases in $1+1$d. On the other hand, we can also imagine breaking the \staco\, in half (or cutting the unfolded theory along the mirror line), in which case the resulting edge would support a gapless $G$-SPT whose symmetry data is equivalently encoded in the condensable subalgebra $\mA$, as discussed in Sec.~\ref{sec:gapless}. Through the condensable subalgebra $\mA$, we can thus obtain a one-to-one correspondence between $G$-igSPTs and a subset of gapped $G\times G$-symmetric $1+1$d phases. In the following subsections, we work out the explicit constraints on the condensable subalgebras that characterize folded boundaries of general bilayer topological orders, and then apply these constraints to the case of $\mD(G)\boxtimes \overline{\mD(G)}\cong \mD(G\times G)$. This will allow us to work out the subgroup $H \subset G \times G$ and 2-cocycle $\omega \in H^2(H,U(1))$ characterizing such boundaries and to show that they do in fact correspond to the data defining a $G$-igSPT. 


\subsection{Characterization of Folded Boundaries}
\label{sec:foldedbdy}

Here, we discuss folded boundaries of general $2+1$d TOs given by $\mathcal{C} \boxtimes \overline{\mathcal{{C}}}$ and specialize to the case of interest $\mathcal{C} = \mD(G)$ in the following subsection. For a general gapped boundary to vacuum of the $\mathcal{C} \boxtimes \overline{\mathcal{{C}}}$ TO, we will show that whether or not it corresponds to a folded boundary of the \staco\, is determined by the positivity of a certain matrix, which encodes the data of the condensed anyons in the condensable subalgebra $\mA$. As it turns out, this positivity constraint is closely related to the positivity constraint on physical density matrices, which we leverage in Section~\ref{sec:mixed} to obtain the \staco\, construction for mixed-state SPTs in $1+1$d.

The data for a general $2+1$d TO is specified by a unitary modular tensor category $\mD$, which contains the (finite) set of topologically distinct anyons (or superselection sectors) of the theory, $\{a\}$. We denote the vacuum (or trivial) superselection sector as $1$ and denote the inverse of an anyon $a \in \mD$ as $\bar{a}$, where $\bar{a}$ is the unique anyon in $\mD$ such that its fusion with $a$ contains the vacuum as a fusion product: $a \times \bar{a} = 1 + \dots$.

\paragraph*{Abelian TOs:} Consider first Abelian TOs, for which we will characterize folded boundaries based on the \textit{positivity} of a particular indicator function. Given a gapped boundary of a TO $\mD$, we define the indicator function 
\begin{equation}
\label{eq:indicator-function}
M: \mD \to \{0,1\} \text{ s.t. } 
\begin{cases}
M_a = 1, \quad a \text{ condensed on boundary}, \\
M_a = 0, \quad \text{otherwise} 
\end{cases}
\end{equation}
for each anyon $a \in \mD$. For an Abelian anyon theory $\mD$, the indicator function $M$ fully specifies the gapped boundary and satisfies the following constraints:
\begin{equation}
\label{eq:m_constraints_1}
\begin{aligned} 
    M_a=1 &\implies M_{\bar{a}}=1 \, ,\\
    M_a=M_b=1 &\implies M_{a \times b} = 1 \, .
\end{aligned}
\end{equation}
The first constraint enforces the condition that if an anyon $a$ condenses on the boundary, so does its inverse $\bar{a}$. The second condition enforces the condition that if two anyons $a, b$ condense on a given boundary, so does their fusion product. Taken together, these two conditions encode that the Abelian anyons condensed at a gapped boundary form a Lagrangian subgroup which is, of course, a group. Note that the second constraint need not be satisfied for the gapped boundary of a non-Abelian topological order; in particular, $a$ and $b$ generically will not have a unique fusion product and, even if they do, their fusion products need not condense. Hence, a matrix $M$ satisfying Eq.~\eqref{eq:m_constraints_1} is not capable of describing the gapped boundaries of non-Abelian topological orders. Nevertheless, there exists a different matrix which encodes the condensed anyon content of the gapped boundary of a non-Abelian order--we return to this more general construction following our discussion of the Abelian case.

Let us turn to the object of interest: a bilayer Abelian TO of the form $\mathcal{D} = \mathcal{C} \boxtimes \overline{\mathcal{C}}$. In this case, the anyons of $\mathcal{D}$ can be written as 
$a_+ b_-$ for $a,b \in \mathcal C$, with the understanding that anyons with the $-$ label have braiding opposite to their counterparts in $\mathcal C$. For notational convenience, in the following we denote the indicator function $M_{a,b} \equiv M_{a_+ b_-}$. 
In general, the distinct gapped boundaries of this TO are classified via the data provided in Sec.~\ref{sec:gapped_abelian_spt}. Suppose now though that we are considering a \staco, i.e., we started with the TO given by $\mathcal{C}$, condensed anyons in accordance with some condensable subalgebra $\mA$ in a middle region, and then performed the fold along the mirror line to obtain the folded boundary of the \staco. In this case, the indicator function $M$ corresponding to the folded boundary obeys the following constraints:
\begin{equation}
\label{eq:m_constraints_2}
    M_{a,b}=1 \implies 
    \begin{array}{ll}
         & M_{b,a} = 1 \, ,\\
         & M_{a \times b,1}=1.
    \end{array}
\end{equation}
The first constraint arises from the layer exchange symmetry of the \staco, which descends from the reflection symmetry about the fold line in the unfolded theory. The second constraint can likewise be understood from the unfolded picture: if $M_{a,b}=1$, in the unfolded theory this implies that if we bring an anyon $a$ from the left region (with $\mathcal{C}$ TO) and an anyon $b$ from the right region (with $\mathcal{C}$ TO) towards the fold line, either (i) both $a,b$ condense at the gapped boundaries between $\mathcal{C}$ and $\mathcal{C}/\mA$, or (ii) they both pass into the region $\mathcal{C}/\mathcal{A}$ and annihilate each other. Suppose now that we instead bring both $a$ and $b$ from the left region with $\mathcal{C}$ TO and bring them towards the middle region with $\mathcal{C}/\mA$ TO. Then, either (i) both condense at the gapped boundary to $\mathcal{C}/\mathcal{A}$ or, (ii) they annihilate each other after passing through this gapped boundary. In the folded theory, this implies that $M_{a \times b,1}=1$. Note that, by combining Eq.~\eqref{eq:m_constraints_2} with Eq.~\eqref{eq:m_constraints_1}, we can also derive from $M_{a,b} = 1$ that $M_{1,a \times b}=1$ and $M_{a,\bar{a}}=M_{b,\bar{b}}=1$.

In fact, the constraints in Eq.~\eqref{eq:m_constraints_2} provide a \textit{sufficient} condition for a general gapped boundary of $\mD$ to also define a consistent folded boundary. In other words, given a $\mathcal{C}\boxtimes\overline{\mathcal{C}}$ TO with a gapped boundary whose indicator function $M$ satisfies Eq.~\eqref{eq:m_constraints_2}, we can derive a condensable subalgebra $\mA$ of $\mathcal{C}$ such that this boundary can equivalently be realized as a folded boundary of the corresponding \staco. Indeed, we can define the set of condensed anyons $\mathcal{A}$ such that an anyon $c\in\mathcal{C}$ is in $\mA$ if $M_{c,1} = 1$. That $\mathcal{A}$ is a condensable subalgebra follows immediately from the fact that $(c,1)$ are elements of a Lagrangian subalgebra, and thus all $c \in \mathcal{A}$ must have trivial topological spin and trivial braiding with all other elements of $\mA$. Given this, we can straightforwardly construct a folded boundary by taking a $\mathcal{C}$ TO, picking the condensable subalgebra $\mA$ we have constructed above, condensing it in a subregion, and folding along the mirror symmetric line; this procedure automatically realizes a folded boundary of $\mathcal{C}\boxtimes\overline{\mathcal{C}}$ to vacuum. One may check that this folded boundary is precisely the original gapped boundary we started with.

From the above discussion, we see that a gapped boundary of $\mathcal{C}\boxtimes\overline{\mathcal{C}}$ can be realized as a folded boundary if and only if Eq.~\eqref{eq:m_constraints_2} holds. This constraint can be more compactly expressed as imposing a positivity condition on the indicator function $M$. First, by combining Eqs.~\eqref{eq:m_constraints_1} and \eqref{eq:m_constraints_2}, we can derive the following inequality:
\begin{equation}
\label{eq:m_constraints_3}
    M_{a, \bar{b}}+M_{a,\bar{c}}\leq M_{a,\bar{a}}+M_{b,\bar{c}} \,.
\end{equation}
Conversely, if this inequality holds, then so does Eq.~\eqref{eq:m_constraints_2} (see Appendix~\ref{app:positivity} for the proof). We now make use of the following fact: a symmetric binary matrix $M$ is positive semi-definite if and only if Eq.~\eqref{eq:m_constraints_3} holds~\cite{Letchford2012}. Interpreting the indicator function $M_{a, \bar{b}}$ as such a symmetric binary matrix, it follows from Eq.~\eqref{eq:m_constraints_3} that the indicator function $M_{a, \bar{b}}$ is positive semi-definite. Therefore, a gapped boundary of $\mathcal{C}\boxtimes\overline{\mathcal{C}}$ can be realized as a folded boundary of the \staco\, if and only if the corresponding indicator function $M$ is a symmetric, positive semi-definite matrix\footnote{Positivity also played an important role in the derivation by Ref.~\cite{Duivenvoorden2017}, which established a similar correspondence between folded boundaries of quantum doubles and $G\times G$-SPTs.}.

To illustrate the constraint that the requirement of being a folded boundary imposes on general gapped boundaries, consider the boundary of $\mD(\mbb{Z}_2) \times \mD(\mbb{Z}_2) \cong \mD(\mbb{Z}_2 \times \mbb{Z}_2)$ that corresponds to the nontrivial $\mbb{Z}_2 \times \mbb{Z}_2$ $1+1$d SPT. The anyons condensed at the boundary are $1$, $e_1 m_2$, $m_1 e_2$, and $f_1 f_2$ (see Section~\ref{sec:gapped_abelian_spt}). We can immediately see that the condition of Eq.~\eqref{eq:m_constraints_3} is not satisfied: we have $M_{e,m} = M_{m,e} = 1$ but $M_{e \times m, 1} = 0$. Therefore, the positivity condition is violated and this boundary cannot be obtained via folding, i.e., while this is valid gapped boundary to vacuum of a general $\mD(\mbb{Z}_2\times \mbb{Z}_2)$ TO, it cannot describe a gapped boundary to vacuum for a $\mD(\mbb{Z}_2\times \mbb{Z}_2)$ TO that is obtained by folding a $\mD(\mbb{Z}_2)$ TO along a subregion where some condensable subalgebra $\mA$ is condensed.

\paragraph*{Non-Abelian TOs:} While the above constraint on the indicator function completely characterizes folded boundaries of Abelian TOs, it does not generalize to the non-Abelian setting since a matrix $M$ satisfying Eq.~\eqref{eq:m_constraints_1} is insufficient for completely specifying gapped boundaries in this case. Nevertheless, we are able to obtain useful partial information from a suitably generalized indicator function $M$. We define $M$ as 
\begin{equation}
    M : \mathcal{D} \to \mbb{N}
\end{equation}
such that $M_a$ is the multiplicity of fusion channels of the anyon $a$ when fused to the boundary. The anyon $a$ is confined (i.e. not condensed) at the boundary if $M_a = 0$. This reduces to the definition provided above in the case of Abelian TOs--Abelian anyons can only have multiplicity $0$ or $1$~\cite{kaidi2022higher}. 

We now turn to a more general treatment, which includes the non-Abelian case, by using the \emph{tunneling matrix} to construct the indicator function $M$ (which reduces to the definition given above in the Abelian case). Specifically, the tunneling matrix of the unfolded theory in Fig.~\ref{fig:main} can be used to construct the $M$-matrix of the folded theory. While $M$ describes the condensed anyons at a gapped boundary between a TO and vacuum, the tunneling matrix $W$ describes how anyons tunnel across a gapped boundary between two general topological orders~\cite{lan2015}. Consider an arbitrary anyon theory $\mathcal{C}$ with a finite set of anyons $\{a\}$ and, as before, condense some condensable subalgebra $\mA$ in a middle region to obtain the reduced TO $\mathcal{C}' = \mathcal{C}/\mA$ in this subregion. We can then label the anyons in $\mathcal{C}'$ as $\{\alpha\}$. In general, $\alpha$ can be identified with a direct sum of anyons in $\mathcal{C}$ as 
\begin{equation}
    \alpha = \bigoplus_{a} W_{\alpha,a} a \, ,
\end{equation}
which defines the tunneling matrix $W^T$. This matrix is fully determined, albeit indirectly, by the condensable algebra $\mathcal{A} \in \mathcal{C}$. The converse is generally not true since the $\mA$ contains more data than the tunneling matrix. The transpose $W_{a, \alpha}$ provides the fusion multiplicities for the process in which $a \in \mathcal{C}$ tunnels through a domain wall to turn into $\alpha \in \mathcal{C}'$. 

We now apply the folding trick to obtain the \staco, as shown in Fig.~\ref{fig:main}, which results in a gapped boundary to vacuum for the $\mathcal{C}\boxtimes\overline{\mathcal{C}}$ TO. Importantly, anyons from this TO undergo a sequence of two condensations to get to the boundary: first, an anyon $a_+ b_- \in \mathcal{C}\boxtimes\overline{\mathcal{C}}$ tunnels into the region with $\mathcal{C}'\boxtimes\overline{\mathcal{C}}'$ TO and turns into $\bigoplus_{\alpha\beta} W_{a \alpha} W_{b \beta} \, \alpha_+  \beta_-$. Next, anyons from $\mathcal{C}'\boxtimes\overline{\mathcal{C}}'$ can condense at the gapped boundary to vacuum specified by condensation of the Lagrangian subalgebra
\begin{equation}
\label{eq:alpha_alphabar}
    \mathcal L' = \bigoplus_\alpha \alpha_+ \bar{\alpha}_- \, ,
\end{equation}
which corresponds to the canonical gapped boundary of the condensed TO $\mathcal{C}' \boxtimes \overline{\mathcal{C}}'$. In other words, the corresponding condensation matrix for this gapped boundary to vacuum is given by $M'_{\alpha, \bar{\beta}} = \delta_{\alpha \beta}$. Hence, the gapped boundary to vacuum of the \staco, i.e., of the $\mathcal{C}\boxtimes\overline{\mathcal{C}}$ TO is specified by the data describing this sequence of condensations. The condensation matrix $M$ describing the folded boundary of the \staco\, is therefore given by 
\begin{equation}
    M_{a, \bar{b}} = W_{a, \alpha} W_{\bar{b}, \bar{\beta}} \delta_{\alpha \bar{\beta}} = W_{a, \alpha} W_{\bar{b}, \alpha} = (W W^T)_{a \bar{b}} \, ,
\end{equation}
from which we see that $M_{a, \bar{b}}$ is a positive semidefinite matrix. Thus, the general condition for when a gapped boundary of an arbitrary TO $\mathcal{C}$ provides an admissible folded boundary of the corresponding \staco\, is encoded in the positivity of the condensation matrix $M$, which is constructed through the intermediate step of the tunneling matrix $W$ (which is in turn specified by the condensable subalgebra $\mA$ that is part of the data of the \staco\, as defined in this paper).


\subsection{Folded Boundaries of Quantum Doubles}
\label{sec:folddouble}

Following the preceding general characterization of folded boundaries in $2+1$d TOs described by some UMTC $\mathcal{C}$, let us now specialize to the case when $\mathcal{C} = \mD(G)$. We will now translate the constraints that determine when a gapped boundary is an allowed folded boundary into constraints on the $1+1$d gapped phases that can be realized on the folded boundary. Stated physically, we wish to translate these constraints into conditions on the subgroup $H \subset G \times G$ and 2-cocycle class $[\omega] \in H^2(H,U(1))$ (which characterize general gapped boundaries of $\mD(G\times G)$, see Sec.~\ref{sec:genreview}) which are compatible with the boundary being a folded boundary of the \staco. These conditions will in turn restrict the gapped $1+1$d $G\times G$-symmetric phases which can be realized through the \staco\, construction and will also allow us to make clear the correspondence between this class of $G\times G$-symmetric gapped phases and $G$-symmetric igSPTS.

As before, we start with the simpler case when $G$ is Abelian. In this case, after folding, we obtain a $\mD(G \times G) \cong \mD(G) \times \overbar{\mD(G)}$ TO with a gapped boundary to vacuum. We can then use the general discussion in the previous subsection to find the conditions under which this gapped boundary is a consistent folded boundary. First, we need to constrain the subgroup $H \subset G \times G$ which, physically, consists of the different flux labels for every anyon that is condensed at the boundary. From the second constraint of Eq.~\eqref{eq:m_constraints_2}, we see that if an anyon $a_+ b_- \in \mD(G) \times \overbar{\mD(G)}$ carrying flux $(g_1, g_2) \in H$ is condensed, then the anyon $b_+ a_-$ must also be condensed, implying that $(g_2,g_1)$ is also in $H$. Therefore, if the gapped boundary specified by $H$ defines a consistent folded boundary, we can write $H=K\times L$, where $K$ consists of elements of the form $(k,k)$, $L$ consists of elements of the form $(l,e)$, and $L\subset K$. Note that if the condensable subalgebra $\mathcal{A}$ (which is condensed in the middle subregion before folding) contains no pure charges, then $K=G$~\cite{Beigi2011,cong2016top}. Furthermore, if the condensable algebra $\mathcal{A}$ contains anyons that carry flux in a normal subgroup $A \subset G$, then $L = A$. 

Next, we constrain the two-cocycle $\omega$. Recall that, when $H$ is Abelian, the cohomology class of $\omega$ is completely fixed by the function $\Omega(g,h)=\omega(h,g)/\omega(g,h)$ defined in Eq.~\eqref{eq:Omegadefinition}, satisfying the properties listed in Eq.~\eqref{eq:Omegaproperties}. Moreover, as discussed after Eq.~\eqref{eq:Omegaproperties}, the set of anyons in the Lagrangian subgroup specified by $H$ and $\omega$ take the form $(h,\pi_h)$, where $h \in H$ and $\pi_h$ is a $1$d representation of $H$ determined by the function $\Omega(h,\cdot)$. Using the constraints on anyon condensation arising from folded boundaries derived in the previous subsection, we now derive a corresponding constraint on the form of the function $\Omega$.

Let us understand how the condensed dyons are constrained by Eq.~\eqref{eq:m_constraints_2}. Consider a dyon whose flux lives entirely within one layer of the \staco. By the folding construction, the attached charge also only lives within the same layer and thus acts trivially on any dyon whose flux lives entirely within the other layer. This means that $\Omega((g,e), (e,h)) = 1$ for all $g, h \in L$. Next, consider a dyon $(a,1)$ which lives only in one layer and which condenses at the gapped boundary to vacuum. If $(a,1)$ condenses, then $(1,a)$ also condenses. Recall that anyons in the second layer are understood to have opposite braiding as anyons in the first layer. This means that the charge associated to a flux $(g,e)$ in one layer is the complex conjugate (i.e. inverse) charge associated to a flux $(e, g)$ in the other layer. This means that $\Omega((g,e), (h,e)) = \Omega((e,g), (e,h))^*$ for all $g, h \in A$. 

Suppose that no pure charges condense in $\mA$, such that elements of $H$ have the form $(gl,g)$ for $g\in G$ and $l\in A \subset G$. For two elements $(gl,g),(g'l',g')\in H$, we can use the above constraints to write
\begin{equation}
    \Omega((gl,g),(g'l',g')) = \frac{\Omega((g,e),(l',e))}{\Omega((g',e),(l,e))} \Omega((l,e),(l'e)) \, .
\end{equation}
If we define the functions $\eta: A \times A \rightarrow U(1)$ such that $\eta(l,l')=\omega((l,e),(l',e))$, and $\epsilon: A \times G\rightarrow U(1)$ such that $\epsilon(l,g) = \Omega((g,e),(l,e))$, then we can write
\begin{equation}
\label{eq:Omega}
    \Omega((gl,g),(g'l',g')) = \frac{\epsilon(l',g)}{\epsilon(l,g')}\frac{\eta(l,l')}{\eta(l',l)} \, .
\end{equation}
Therefore, $G\times G$-symmetric folded boundaries of $\mD(G\times G)$ can be specified by the functions $\epsilon$ and $\eta$. These functions satisfy the following relations: (a) $\epsilon$ is bilinear, (b) $\eta$ is a 2-cocycle on $L$, and (c) $\epsilon(l,l') = \eta(l,l')/\eta(l',l)$ for all $l,l'\in L$. A normal subgroup $L$ and a pair of functions $\eta$, $\epsilon$ satisfying these conditions are exactly the data that specify a condensable subalgebra $\mathcal{A}$ which does not contain any pure charges, as described in Ref.~\cite{Davydov2009} (specializing to the case of Abelian $G$). We see that this same data can be packaged cleanly into a subgroup $H\subset G\times G$ and a 2-cocycle class $[\omega]\in H^2(H,U(1))$, which together specify a gapped $1+1$d phase with $G\times G$ symmetry. Thus, we have shown that folded boundaries of the \staco\, are in one-to-one correspondence with gapped $G\times G$-symmetric $1+1$d phases which are specified by the subgroup $H$ and the 2-cocycle class $[\omega]$ satisfying the constraints derived above.

The above discussion unfortunately does not generalize straightforwardly to the non-Abelian setting. The difficulty primarily arises from the fact that the full data characterizing a gapped boundary cannot be reconstructed given only the set of condensed anyons on the boundary (along with their multiplicities)~\cite{davydov2014}. The subgroup $H \subset G \times G$ is obtained in the same manner as in the Abelian case discussed above, and consists of elements of the form $(lg,g)$ for $l \in A$ and $g \in G$. This is the subgroup $A \rtimes G$, where the action of $G$ on $A$ is given by conjugation, $l \mapsto glg^{-1} \in A$. However, the 2-cocycle class $[\omega]$ cannot in general be obtained given only the set of condensed anyons and their multiplicities. One must instead resort to the \textit{full center} construction introduced in Ref.~\cite{davydov2010centrealgebra} to obtain the data characterizing a folded boundary in the non-Abelian case. 

Instead of taking this formal mathematical approach to deriving the fractionalization data for a folded SPT from the corresponding folded boundary, we will utilize a different correspondence made possible by the \staco: first, in the following Section, we will use the \staco\, construction to argue that folded boundaries of $\mD(G \times G)$ TOs correspond to $1+1$d $G$-symmetric mixed-state (or average) SPT phases--we will argue this explicitly for $G$ Abelian and conjecture it also holds for $G$ non-Abelian. Next, we will analyze the symmetry fractionalization data characterizing such mixed-state SPTs and use the aforementioned correspondence to obtain the 2-cocycle class $[\omega]$. The result, whose derivation we delay to the next Section, is that
\begin{equation}
\label{eq:nonabelian_omega}
    \omega((kg, g), (k'g', g')) = \epsilon(k', g) \eta( k, gk' g^{-1}) \, ,
\end{equation}
for a folded boundary of $\mD(G \times G)$ arising from partial condensation of $\mathcal A \subset \mD(G)$ described by $\epsilon$ and $\eta$. Under the correspondence to be established between mixed-state SPTs and folded SPTs, this implies that a folded SPT corresponding to the same anyon condensate determined by $(\eta,\epsilon)$ is characterized by the 2-cocycle given by Eq.~\eqref{eq:nonabelian_omega}. This (indirectly) implies that folded SPTs have fractionalization data determined by a folded boundary of the corresponding $\mathcal{D}(G\times G)$ SymTaco. Remarkably, this data is identical to the data characterizing $1+1$d $G$-symmetric igSPTs (see Sec.~\ref{sec:gapless}), which underlines the utility of the \staco\, construction in obtaining non-trivial correspondences between gapped and gapless phases (albeit with different symmetries), and their corresponding fractionalization data.

In fact, the correspondence established by the \staco\, between folded SPTs and igSPTs goes beyond a formal identification of the symmetry fractionalization data and provides a direct method for generating folded SPTs from bilayers of igSPTs via suitable Hamiltonian deformations which can be inferred from the \staco\, construction (as we will discuss in Sec.~\ref{sec:subsystem}, we can use a similar procedure for generating mixed-state SPTs from igSPTs via decoherence that is specified by the \staco\, construction). Suppose we consider a $G$-symmetric igSPT and stack it with its inverse, i.e., with a $G$-symmetric igSPT which carries the opposite $H^3(G_{low},U(1))$ anomaly class (where $G_{low} = G/A$ is the low-energy symmetry group and $A \lhd G$ is a normal subgroup, see Eq.~\eqref{eq:groupext}). This stack realizes a bilayer $G\times G$-symmetric $1+1$d system, whose corresponding SymTFT is simply a decoupled stack with $\mD(G) \times \mD(G)$ on two independent layers. We can then realize a \staco\, from this bilayer SymTFT by allowing the tunneling of anyons between the two layers, thereby allowing us to gap out the gapless degrees of freedom that remain in the decoupled igSPTs. This corresponds to explicitly breaking the $G\times G$ symmetry down to $A \rtimes G$, consisting of elements of the form $(lg,g)$ for $l \in A$ and $g \in G$. In particular, as discussed in Section~\ref{sec:gapped_abelian_spt}, given a reference boundary, the set of anyons condensed on the dynamical boundary of the SymTFT correspond to specific operators in the $1+1$d lattice model; thus, given a Hamiltonian realizing the decopuled igSPT bilayer, the \staco, through the diagonal interlayer anyon condensation, suggests the appropriate interlayer couplings to add to the bilayer igSPT Hamiltonian to realize the corresponding folded SPT. We expect that this construction can always be done explicitly on the lattice when $G$ is Abelian. In the following subsection, we illustrate this protocol through the simple example of the $\mathbb{Z}_4$ igSPT.


\subsection{Example: $\mbb{Z}_4 \times \mbb{Z}_4$ Folded SPT}
\label{sec:z4xz4}

To illustrate the general principles discussed in the previous two subsections, let us consider the case when $G = \mbb{Z}_4$ and show how we recover the condensed subalgebra $\mA$ corresponding to the folded SPT that is in one-to-one correspondence with the $\mbb{Z}_4$ igSPT reviewed in Sec.~\ref{sec:gapless}.

Recall that the $\mbb{Z}_4$ igSPT can be understood within the SymTFT construction via the condensation of the subalgebra generated by the $e^2 m^2$ boson in the bulk anyon theory $\mD(\mbb{Z}_4)$, with the fractionalization data provided by Eq.~\eqref{eq:z4_eta_epsilon}. 
Here, we label anyons $a_A b_B \in \mD(G) \times \mD(G)$ rather than $\mD(G) \times \overline{\mD(G)}$.
Let us now obtain the folded SPT that corresponds to this igSPT. In this case, we have the \staco\, defined by the $\mD(\mbb{Z}_4 \times \mbb{Z}_4)$ TO with a gapped boundary to vacuum and we would like to determine the Lagrangian subgroup that corresponds to this gapped boundary (note that we are given the condensable subalgebra $\mA$ that determines the intermediate TO as part of the data of the \staco). 

Following the steps described in the prior subsection, we find that the subgroup characterizing the folded SPT of interest is given by $H = \mbb{Z}_4 \times \mbb{Z}_2 \subset \mbb{Z}_4 \times \mbb{Z}_4$, corresponding to elements $(gl, g)$ for $g \in \mbb{Z}_4$, $l \in \mbb{Z}_2$. Using Eq.~\eqref{eq:Omega}, we can further determine the $2$-cocycle characterizing the folded SPT: 
\begin{equation}
    \Omega((gl, g), (g'l', g')) = \frac{\epsilon(l', g)}{\epsilon(l, g')} \, .
\end{equation}
Here, we used the fact that $[\eta] \in H^2(\mathbb{Z}_2,U(1)) = \mathbb{Z}_1$ is necessarily trivial, and so we can set $\eta(l, l') \equiv 1$. We can now reinterpret this data in terms of the condensed anyons at the folded boundary of the \staco\, as follows: since the symmetry is broken to the subgroup $H$, the electric anyons $e_A^2 e_B^2$ are condensed. To determine the charges bound to fluxes, it suffices to consider $\Omega((gl,g), \cdot)$ for the generators of $H$. For the diagonal subgroup generated by $(1,1)$, we have that
\begin{equation}
    \Omega_{(1,1)} (g' l', g') = \epsilon(l', 1) = (-1)^{l'/2} \, .
\end{equation}
Therefore, $\Omega_{(1,1)}$ sends $(g', g') \mapsto 1$ and $(l', 1) \mapsto (-1)^{l'/2}$. This corresponds to the representation $e_A e_B^3$, which implies that the anyon $e_A m_A e_B^3 m_B$ is condensed.\footnote{Note that this is the image of $e_+ m_+ \bar{e}_- \bar{m}_-$ under the isomorphism $\overline{\mD(G)} \to \mD(G \times G)$.}
Similarly,
\begin{equation}
    \Omega_{(2,0)} (g'l', g') = \epsilon^*(2, g') = (-1)^{g'}
\end{equation}
corresponds to the condensation of $e_A^2$. Therefore, the condensed anyons at the folded boundary are precisely generated by the Lagrangian subgroup in Eq.~\eqref{eq:L_SSBSPT} and the folded SPT in this case corresponds to the third example we had previously encountered in Sec.~\ref{sec:gapped_abelian_spt}. Thus, we see that the \staco\, provides an exact correspondence between the symmetry fractionalization data of this gapped, folded SPT with that of the $1+1$d $\mbb{Z}_4$-symmetric igSPT.

Let us now show how we can use the data provided by the \staco\, (specifically, the set of condensed anyons) to explicitly obtain a folded SPT by gapping out two copies of the corresponding igSPT in a manner which preserves the fractionalization data. Let us stack together two copies of the gapless $\mbb{Z}_4$ SPT (more precisely, the gapless SPT and its ``inverse"), such that the total symmetry is $\mbb{Z}_4^A \times \mbb{Z}_4^B$. We allow tunneling between the two systems, which preserves the low energy subspace defined by $\mbb{Z}_2^A \times \mbb{Z}_2^B \subset \mbb{Z}_4^A \times \mbb{Z}_4^B$. Within the low energy subspace, the symmetry acts as $\mbb{Z}_2^{A, low} \times \mbb{Z}_2^{B,low}$. We then allow tunneling between the two systems, designed to (spontaneously or explicitly) break the symmetry within the low energy subspace to the diagonal $\mbb{Z}_2^{AB, low}$. This operation corresponds to the usual operation of stacking two systems with the same symmetry. Since the emergent anomaly is $\mbb{Z}_2$ valued under stacking, the emergent anomaly is trivialized by the tunneling. The full symmetry group is thus reduced (spontaneously or explicitly) to $\mbb{Z}_2^A \times \mbb{Z}_4^{AB}$ by this procedure. Recall that $\mbb{Z}_2^A$ corresponds to the \emph{gapped} symmetry which fractionalizes on the gapless SPT. The tunneling between the two systems thus opens a gap while still preserving the fractionalization data of the original gapless SPT. 

Let us briefly discuss the explicit procedure for generating the folded SPT by stacking two $\mathbb{Z}_4$ igSPTs and choosing the interlayer tunneling in accord with the \staco. The lattice model for the $\mathbb{Z}_4$ gapless SPT is reviewed in Sec.~\ref{sec:z4eg}. To realize the folded SPT from this igSPT, we consider two copies of the $\mbb{Z}_4$ igSPT and label the two copies with the index $\alpha = A,B$. The total symmetry of the system is then $\mathbb{Z}_4^A \times \mathbb{Z}_4^B$. Note that the $\mbb{Z}_4$ igSPT carries a non-trivial emergent $\mbb{Z}_2$ anomaly class and is hence its own inverse. Guided by the \staco, we can now deduce the coupling which will open a gap and realize the folded SPT. Explicitly, after adding this coupling, the Hamiltonian for the folded SPT is given by
\begin{equation}
\begin{aligned}
    H = -\sum_i \Big( &\cZ_{A,i-1}^2 \cX_{A,i-1}^2 \cZ_{A,i}^2 + \cZ_{B,i-1}^2 
    \cX_{B,i-1}^2 \cZ_{B,i}^2 \\
    + &\cZ^2_{A,i-1} \cZ^2_{A,i} \cZ^2_{B,i-1}  \cZ^2_{B,i} \\
    + &\cZ_{A,i-1} \cZ^\dagger_{A,i} \cX_{A,i} \cZ_{B,i-1} \cZ^\dagger_{B,i} \cX_{B, i} + \text{ h.c.}  \Big) \,.
\end{aligned}
\end{equation}
The first two terms impose the energetic constraints Eq.~\eqref{eq:HDelta} on the two copies of the system and correspond to the condensation of $e^2 m^2$ and $\bar{e}^2 \bar{m}^2$ on each copy in the corresponding $\mD(\mbb{Z}_4)$ and $\overbar{\mD(\mbb{Z}_4)}$ SymTFTs. The second line constructs the \staco\, from the decoupled layers and corresponds to tunneling $e^2$ between the two copies (equivalently, condensing  $e^2 \bar{e}^2$), and spontaneously breaks the original $\mbb{Z}_4 \times \mbb{Z}_4$ symmetry to $\mbb{Z}_2^{A} \times \mbb{Z}_4^{AB}$. The final term fully gaps out the system and corresponds to tunneling $em$ and $e^3 m$ between the two copies. The ground of this Hamiltonian realizes the non-trivial $1+1$d gapped $\mbb{Z}_4 \times \mbb{Z}_2$ SPT that was discussed as Example 3 in Sec.~\ref{sec:gapped_abelian_spt}.


\section{Mixed-State SymTFT: Folded Boundaries as Mixed-State SPTs}
\label{sec:mixed}

\begin{figure}
    \centering
    \includegraphics[width=\linewidth]{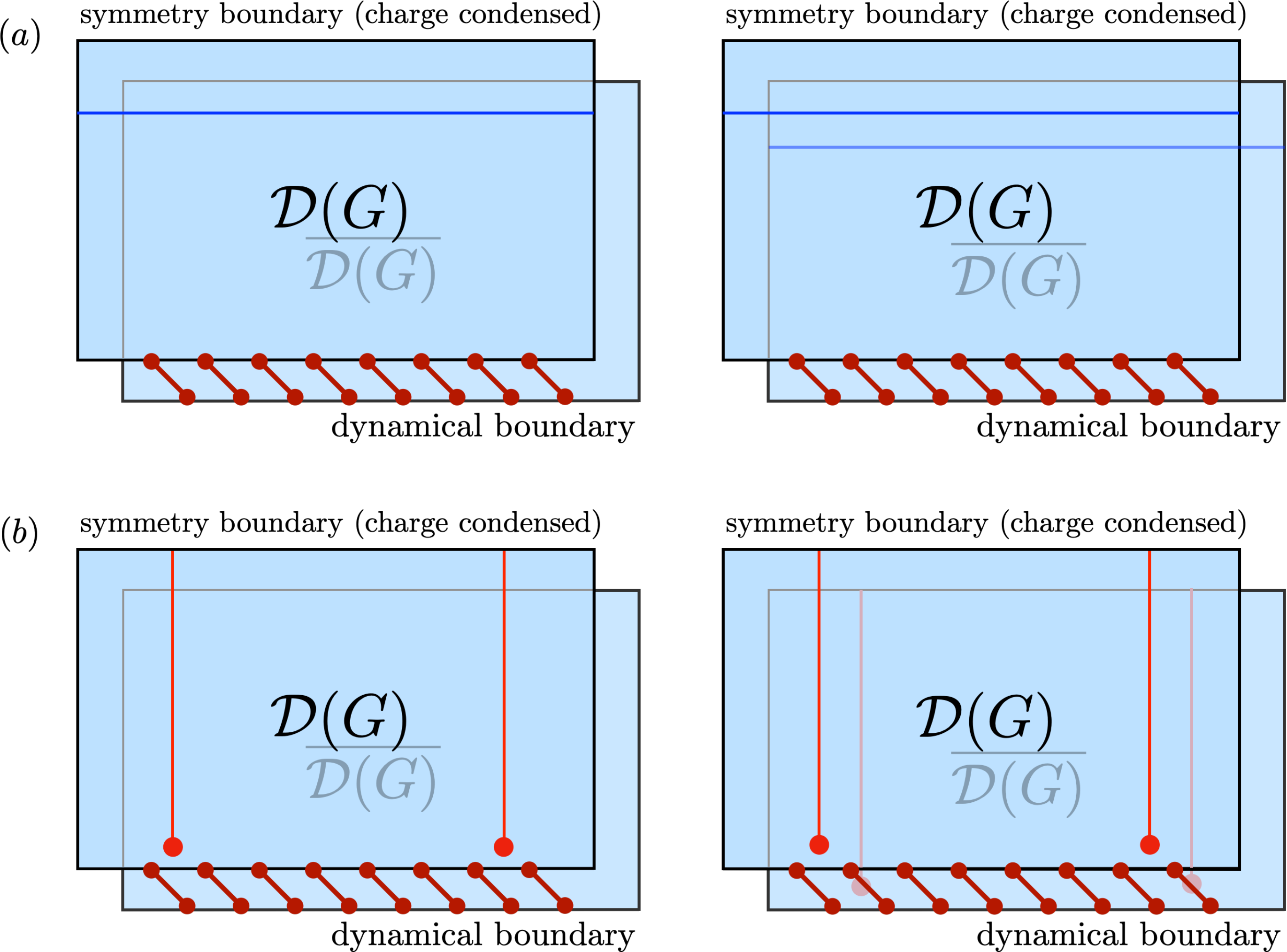}
    \caption{SymTFT for the Choi state. (a) Distinction between strong and weak symmetry operators. (b) Distinction between strong and weak order parameters. The bold red lines connecting the layers represent decoherence ($a_+ \bar{a}_-$ anyon condensation), which has the effect of coupling the ket and bra spaces of $\kket{\rho}$.}
    \label{fig:choi_symtft}
\end{figure}

Having established an exact correspondence between $G$-igSPTs and folded $G\times G$-SPTs, we now proceed to reveal a striking connection between these phases and \emph{mixed-state} SPTs. Recent work has unveiled the existence of \textit{intrinsically} mixed-state phases of matter, which have no pure-state analogues; examples include intrinsically average SPT (iASPT) order~\cite{ma2024,Ma2025,xue2024,guo2024} as well as intrinsically mixed topological order (imTO)~\cite{Sohal2025,Ellison2025}. The central ingredient that allows these novel phases to appear in the mixed-state setting is that symmetries may be \textit{strong/exact} or \textit{weak/average}. Physically, for a density matrix describing the reduced dynamics of a system coupled to a bath, strong symmetries correspond to those under which the system on its own is invariant, whereas weak symmetries are those under which only the system and bath together are invariant. Symmetry charges are thus conserved for the system only when it respects a strong symmetry. 

A natural question then is whether the SymTFT picture can be extended to mixed states, accommodating their aforementioned novel symmetry structure and describing intrinsically mixed-state phases. In this Section, we answer this in the affirmative: given a $1+1$d $G$-symmetric mixed state $\rho$, we propose the \staco\, as the SymTFT for the corresponding Choi state, a vectorized version of $\rho$. Using this framework, we arrive at an exact correspondence between iASPTs, igSPTs, and the folded SPTs discussed in previous Sections. We emphasize that, while this correspondence between different classes of SPTs builds on each of the prior Sections, the only background required for readers solely interested in the SymTFT description for mixed-state SPTs is provided in Section~\ref{sec:genreview}. 


\subsection{Strong and Weak Symmetries}
\label{sec:swsym}

Let us begin by first reviewing the symmetry structure of mixed states~\cite{buca2012,albert2014sym}. Consider a mixed state $\rho$ invariant under a symmetry group $G$ acting in the adjoint as 
\begin{equation}
\label{eq:Gweak}
    U_g \rho U_g^\dagger = \rho \, ,
\end{equation}
for all $g \in G_{weak}$. Such a symmetry is called a \emph{weak} or \emph{average} symmetry. Physically, $\rho$ describes a stochastic ensemble of states, 
which is symmetric ``on average" under $G_{weak}$, but in which each state is not necessarily symmetric. For a group $K$, the state $\rho$ may instead satisfy for $k \in K$
\begin{equation}
\label{eq:Kstrong}
    U_k \rho = e^{i \theta_k} \rho \, .
\end{equation}
The group $K$ is referred to as a \emph{strong} or \emph{exact} symmetry of $\rho$. This condition implies that $\rho$ can be expressed as an ensemble of states in which each state is individually symmetric under $K$. While one possibility is that the strong and weak symmetry groups commute such that the total symmetry group is a tensor product $G = G_{weak} \times K$, the most general symmetry obeyed by a density matrix is a group extension of $G_{weak}$ by $K$, as encapsulated in the short, exact sequence,
\begin{equation}
\label{eq:mixedgroupext}
    1 \to K \to G \to G_{weak} \to 1 \, ,
\end{equation}
Here, $K$ is a normal subgroup of $G$ and $G_{weak} = G / K$. This presentation makes the similarity to gapless SPTs manifest (see Eq.~\eqref{eq:groupext}). Hereafter, we will refer to the symmetry $G$ as the total symmetry group, even though it acts as a weak (or average) symmetry to reminder the reader that it is obtained via the above group extension.

The possibility of a density matrix $\rho$ obeying a strong symmetry, as in Eq.~\eqref{eq:Kstrong}, motivates the consideration of a larger $G_+ \times G_-$ symmetry action on $\rho$. The factor $G_+$ acts on the left of $\rho$ as $U_g \rho$, while the factor $G_-$ acts on the right as $\rho U_g^\dagger$. We denote elements of $G_+ \times G_-$ as $(g_1, g_2)$, for $g_1,g_2 \in G$. For a density matrix obeying Eq.~\eqref{eq:Gweak}, $\rho$ is invariant under the action of $(g, g)$, i.e., the diagonal subgroup of $G_+ \times G_-$. If a normal subgroup $K \subset G$ acts as a strong symmetry on $\rho$, then $\rho$ is invariant (up to a phase) under the action of $(k, e)$. Together, the strong and weak symmetries generate a subgroup $K \rtimes G \subset G_+ \times G_-$ (where the action of $G$ on $K$ is by conjugation, $k \mapsto gkg^{-1}$) whose elements are of the form $(kg, g)$. Recall that this is precisely the subgroup of $G \times G$ that we encountered in Section~\ref{sec:gapped} in our analysis of folded $G\times G$-symmetric SPTs (note that the role played by $K$ here is played by the subgroup $A \subset G$ in that discussion). Thus, we see that the symmetry structure of mixed states mirrors the symmetry structure of folded SPTs, suggesting a potential correspondence.

While there are evident parallels, in order to establish the tetraptych of Fig.~\ref{fig:main}, we must quantitatively establish that the data characterizing iASPTs is in exact correspondence with that characterizing folded SPTS (and thus also igSPTs). This should already seem plausible, since the symmetry and positivity conditions described in Sec.~\ref{sec:foldedbdy} are reminiscent of the symmetry and positivity conditions that characterize physical density matrices. In the following, we will make precise this connection in the Abelian case and conjecture that it extends as well to the non-Abelian case. 


\subsection{Choi Isomorphism and Mixed-State Correlations}
\label{sec:choi}

In the analysis of mixed states, it is convenient to express the operators $O$ acting on $\mathcal{H}$ as a vectors $\kket{O}$ in a ``doubled" Hilbert space using the Choi-Jamio\l{}kowski isomorphism\footnote{We will follow the standard convention in the mixed-state literature and refer to this vectorization of $\rho$ as the ``Choi state" $\kket{\rho}$. Strictly speaking, the Choi state refers to the vectorized version of a quantum channel, not a density matrix.}~\cite{choi1975,jamio1972}. This allows us to map the problem of characterizing density matrices to that of characterizing pure states (subject to the constraints of Hermiticity and positive semi-definiteness). For instance, the density matrix $\rho$ is mapped to a state $\kket{\rho} \in \mathcal{H}_+ \otimes \mathcal{H}_-$ as 
\begin{align}
    \rho &= \sum_{i,j} \rho_{i,j} \ket{i}\bra{j} \mapsto \kket{\rho} = \sum_{i,j} \rho_{i,j} \ket{i}_+ \ket{j}_-^* \,,
\end{align}
where $\pm$ denote the ``ket" and ``bra" space, respectively, and the asterisk denotes complex conjugation. Operators acting on the left of $\rho$ are mapped to operators acting on $\mathcal{H}_+$, and operators acting on the right of $\rho$ are mapped to their transpose acting on $\mathcal{H}_-$. In this presentation, the weak and strong symmetry conditions, Eqs. \eqref{eq:Gweak} and \eqref{eq:Kstrong}, respectively, translate into
\begin{align}
    U_{g+}U_{g-}^* \kket{\rho} = \kket{\rho}
\end{align}
and
\begin{align}
    U_{k+}\kket{\rho} = U_{k-}^*\kket{\rho} = e^{i\theta}\kket{\rho} \,.
\end{align}
Hence, $\kket{\rho}$ describes a bilayer state that is symmetric under the diagonal action of $G \subset G_+ \times G-$ as well as the action of $K \subset G$ on either layer. The action of the symmetries, in the SymTFT picture described in the following Section, is depicted in Fig.~\ref{fig:choi_symtft}.

The trace inner product on operators in $\mathcal H$ maps, under the Choi-Jamio\l{}kowski isomorphism, to the usual inner product of states on the doubled space, 
\begin{equation}
    \text{Tr}(A^\dagger B) = \langle \bra{A} B \rangle\rangle \, ,
\end{equation}
while physical expectation values of an operator $O$ are computed by 
\begin{equation}
    \langle O \rangle = \text{Tr}(O \rho) = \bbra{I} O_+ \kket{\rho} \, .
\end{equation}
On the other hand, usual expectation values of operators within the doubled state correspond to R\'enyi-2-like quantities in the physical Hilbert space--these quantities probe properties of the second moment of the density matrix. For example, 
\begin{equation}
    \bbra{\rho} A_+ B_-^* \kket{\rho} = \text{Tr}(\rho A \rho B^\dagger). 
\end{equation}
While such nonlinear functions of the density matrix are difficult to access in experiment~\cite{ippoliti2025}, as a theoretical tool such R\'enyi-2 correlation functions still provide a a useful diagnostic for distinguishing different symmetry breaking and SPT orders in mixed states~\cite{lee2023,sala2024}.

Explicitly, suppose we have a system with a strong symmetry $K$ and a local operator (i.e. an order parameter) $O_i$ supported at site $i$ which is charged under $K$. We may then compute the correlator,
\begin{align}
    C_I(i,j) = \frac{\Tr[\rho O_i^\dagger O_j \rho]}{\Tr[\rho^2]}  = \frac{\bbraket{\rho|O_{i+}^\dagger O_{i+}|\rho}}{\bbraket{\rho | \rho} }\label{eq:CI_correlator}
\end{align}
and the R\'enyi-2 correlator,
\begin{align}
    C_{II}(i,j) & =  \frac{\Tr[\rho O_i^\dagger O_j \rho  O_j^\dagger O_i]}{\Tr[\rho^2]} \\
    & = \frac{\bbraket{\rho|O_{i+}^\dagger O_{j+}O_{j-} O_{i-}^T O_{j-}^*|\rho}}{\bbraket{\rho|\rho}}\, .
    \label{eq:CII_correlator}
\end{align}
The standard and R\'enyi-2 correlators are useful to characterize distinct patterns of symmetry breaking in mixed states.
If the standard correlator exhibits long-range correlations, i.e., $C_I(i,j) \to O(1)$ as $|i-j|\to \infty$, then we have the usual spontaneous symmetry breaking (SSB) of the symmetry $G$; equivalently, this is ``strong-to-nothing" symmetry breaking\footnote{In place of $C_I$, we can alternatively consider the standard correlator $\tilde{C}_I(i,j) = \bbraket{I|O_{i+}^\dagger O_{i-}|\rho} = \mathrm{Tr}[\rho O_{i}^\dagger O_j]$ as a diagnostic of strong symmetry breaking~\cite{ma2024,zhang2025}. This coincides with $C_I$ for ``fixed-point" states, where $\rho \propto \rho^2$ is a projector, which describe a maximally mixed state within some subspace. Since local quantum channels cannot generate long-range correlations in finite-depth, on physical grounds we expect that the qualitative behavior of $C_I$ and $\tilde{C}_I$ is the same within some region of phase space around such fixed points. $\tilde{C}_I$ has the advantage that it allows strong symmetry breaking to be described via a correlation function in Choi space.}. Conversely, if the standard correlator does not exhibit long-range order, i.e., $C_I(i,j) \sim \exp(-|i-j|)$, but the R\'enyi-2 correlator exhibits long-range order, we say the system exhibits \emph{strong-to-weak spontaneous symmetry breaking} (SWSSB). If, instead, we consider a system with only a \emph{weak} $G$-symmetry, then we say it exhibits weak-to-nothing spontaneous symmetry breaking if $C_{I}$ exhibits long-range order (since, if $O_i$ is charged under the strong symmetry, it is necessarily also charged under the corresponding weak symmetry). This combination of standard and R\'enyi-2 correlators may also be employed to characterize e.g., mixed-state topological orders where, instead of local charged operators, one needs to study correlators of non-local string operators (so-called 1-form symmetry operators)~\cite{zhang2025}.

It is important to note, however, that these correlators are not necessarily ideal probes of SWSSB. Alternative probes that have been introduced include the fidelity correlator~\cite{lessa2025ssb} and R\'enyi-1, or Wightman, correlator (in which correlators are computed via the canonical purification of the mixed state)~\cite{weinstein2025,liu2024diagnosing}. These alternative probes satisfy certain information theoretic properties that the R\'enyi-2 correlators do not. In particular, the R\'enyi-2 correlators may incorrectly diagnose the positions (e.g. the critical decoherence strength) of certain mixed-state phase transitions. Nevertheless, we expect that for ``fixed point states" deep within a given mixed-state phase, all sets of correlators exhibit the same qualitative behavior and thus the R\'enyi-2 correlators are sufficient for diagnosing distinct mixed-state phases. Stated differently, if $\rho_1$ and $\rho_2$ are such ``fixed point states" that belong to different (symmetry protected) Choi state phases, it is expected that they correspond to distinct (symmetry protected) mixed state phases of matter (see related discussions in Refs.~\cite{ma2024, zhang2025,li2024replica}). We therefore posit that the SymTFT we propose should accurately capture all $G$-symmetric mixed-state phases. 

In the following subsection, we propose a SymTFT for describing ``gapped" mixed-state phases. For pure states, the properties of the ground state of a local Hamiltonian are strongly constrained by the latter's spectral gap. In $1+1$d, gapped, non-degenerate ground states of local Hamiltonians exhibit exponentially decaying correlations--with the correlation length set by the gap--and are short-range entangled~\cite{hastings2007}. The existence of a gap allows for well-defined notions of phases of matter, such that gapped ground states within the same phase are related by quasi-adiabatic Hamiltonian deformations. In principle, one can also define phases of matter based solely on states rather than Hamiltonians, with the appropriate equivalence relation provided by quasi-local finite-depth unitary evolution (locality, as encoded in a Lieb-Robinson bound, is the crucial ingredient for distinguishing phases based on their long-range correlations). In the mixed-state setting, there is no canonical notion of a ``gap" and hence no canonical notion of a phase. Despite this, there is considerable ongoing effort in making precise the notion of a mixed-state phase, primarily by exploring different equivalence relations on states~\cite{sang2023mixed,sang2024def,rakovszky2023stable} and using (emergent) anomalies to characterize mixed-state phases~\cite{Sohal2025,Ellison2025,zhang2025,hsin2025lre}. 

In what follows, we will adopt a perspective analogous to that employed in e.g., Refs.~\cite{ma2024,zhang2025}, and use the Choi state $\kket{\rho}$, to define the properties of the mixed-state $\rho$. In particular, we will say that $\rho$ is short-range entangled if $\kket{\rho}$ is short-range entangled; likewise, we will define $\rho$ as a ``gapped" mixed state if $\kket{\rho}$ is a gapped pure state (i.e., has exponentially decaying correlations of local observables or is the ground state of a--fictitious--gapped Hamiltonian in the doubled Hilbert space). While this simplification appears dramatic, it has nonetheless proven extremely useful in diagnosing mixed-state phases as it permits one to leverage known results regarding pure state phases to characterize mixed states. An important caveat, though, is that under this simplification, we do not expect the characterization of mixed-state phases to coincide with other proposals (see Ref.~\cite{ma2024} for a discussion regarding this point)\footnote{For instance, the $\mathbb{Z}_2$ SWSSB state discussed in Sec.~\ref{sec:swssb} is manifestly separable, but would be considered long-range entangled as per this definition, since the corresponding Choi state is a GHZ state in the doubled space.}. That being said, defining mixed-state phase via their Choi states constitutes a perfectly valid, consistent definition of mixed-state phases and is one that is naturally suited in the context of the SymTFT correspondence we develop in the following subsection, where we propose the Choi states of pure-state TOs as the SymTFTs for symmetric mixed states. 

To recap, the key message of this Section is that the characterization of mixed-state phases of matter can largely mapped to the characterization of pure-state phases \textit{via} the Choi isomorphism. As discussed above, SWSSB can for instance be understood as conventional SSB in the Choi state. We will thus use the properties of the Choi state to define the mixed-state phase of matter to which $\rho$ belongs. Crucially, however, not all states in the doubled Hilbert space map back to valid density matrices under the Choi isomorphism: by definition, a density matrix is constrained to be Hermitian ($\rho = \rho^\dagger$) and positive ($\rho \geq 0$). In the following subsection, we will show how accounting for this subtlety allows us to extend the SymTFT paradigm to the mixed-state setting, with the SymTaco appearing as the appropriate bulk object that captures the symmetry structure intrinsic to mixed states.


\subsection{Mixed-State SymTFT and the Symmetry Taco}
\label{sec:mixed-symtaco}

\begin{figure}
    \centering
    \includegraphics[width=\linewidth]{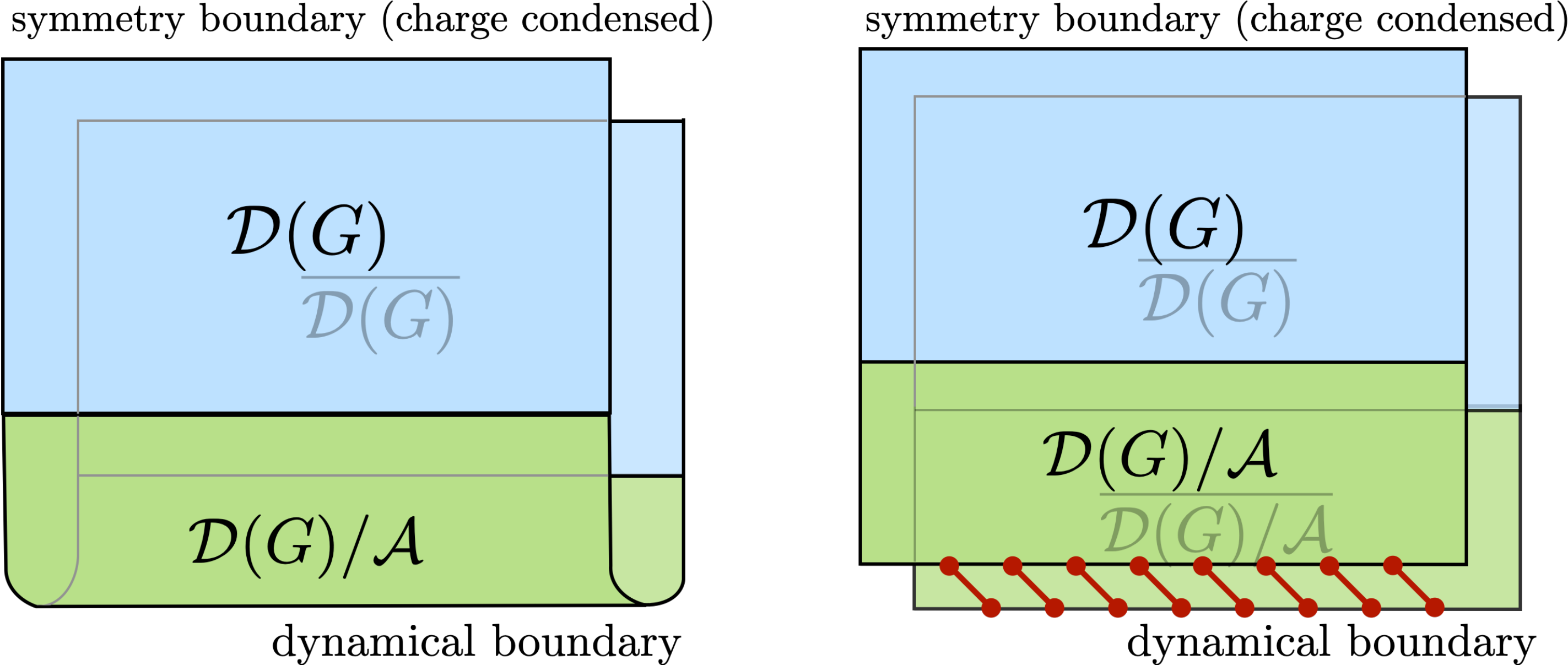}
    \caption{\staco\, correspondence between folded SPTs and mixed-state SPTs. Folding along the $\mathcal{D}(G)/\mathcal{A}$ TO corresponds to maximal decoherence of $\mathcal{D}(G)/\mathcal{A}$ in the Choi space. Folded SPTs with $G \times G$ symmetry can be interpreted as mixed-state SPTs with total $G_+ \times G_-$ symmetry. The bold red lines connecting the layers represent decoherence, which is implemented via condensation of anyons $a_+ \bar{a}_-$ in $\mD(G)/\mA \boxtimes \overline{\mD(G)/\mA}$.}
    \label{fig:folded_mixed}
\end{figure}

In this subsection, we propose a SymTFT framework for one-dimensional gapped phases of mixed states with total symmetry group $G$. Recall that we take a mixed state $\rho$ to be gapped if its Choi state $\kket{\rho}$ describes a gapped pure state. We will argue that phases of mixed states with $G_+ \times G_-$ symmetry are classified by the data governing allowed anyon condensations in a bulk topological order $\mathcal{D}(G) \boxtimes \overline{\mathcal{D}(G)}$ that satisfy a symmetry and positivity constraint. These constraints implement the Hermiticity and positivity conditions that mixed states are required to satisfy. We find that these constraints are \textit{identical} to the constraints characterizing folded SPTs (see Sec.~\ref{sec:gapped}), which is central to the tetraptych of correspondences shown in Fig.~\ref{fig:main} and establishes the SymTaco as the appropriate bulk object for studying $G$-symmetric mixed states. 
Indeed, folding the $\mD(G)$ TO along $\mD(G)/\mA$ can physically be interpreted as maximal decoherence of the anyons in $\mD(G)/\mA$, which corresponds a diagonal condensate of anyons between the ket and bra layers of $\mD(G)/\mA \boxtimes \overline{\mD(G)/\mA}$, as shown in Fig.~\ref{fig:folded_mixed}.
This perspective reproduces the classification of mixed-state SPTs in one dimension and provides a transparent method for extracting the relevant string order parameters. 

We begin by considering the case where the $1+1$d mixed state has a strong $G$ symmetry. The bulk TO encoding the symmetry data of the boundary ASPT is given by the bilayer TO $\mathcal{D}(G) \boxtimes \overline{\mathcal{D}(G)}$. This is to be understood as a topological order in the doubled space, arising from the density matrix of a pure state $\mathcal{D}(G)$ TO. We will label anyons of the bilayer TO as $a_+ b_-$, where $a,b \in \mathcal{D}(G)$; here,  the second anyon is always understood to have opposite (complex conjugate) spin and braiding. Similarly to the pure state case discussed in Sec.~\ref{sec:genreview}, we realize the $1+1$d mixed-state SPT as a gapped boundary to vacuum of the $\mathcal{D}(G) \boxtimes \overline{\mathcal{D}(G)}$ using the thin slab construction. As before, the thin slab has two boundaries, one of which we take to be the symmetry boundary and the other the dynamical boundary. On the symmetry boundary, we condense all pure charges of the TO. This corresponds to the Lagrangian algebra $\text{Rep}(G_+ \times G_-)$. The line operators for the confined anyons form the category $\text{Vec}(G_+ \times G_-)$, which label the symmetry operators acting on the mixed state\footnote{As in the previous subsection, we have used $G_+$ and $G_-$ to label the ``ket" and ``bra" symmetries of the mixed state.}. The strong symmetries are given by the confined fluxes that live entirely within a single layer, while the weak symmetries are given by diagonal fluxes which have support on both layers, as we will see in later examples (see Fig.~\ref{fig:choi_symtft}(a)).

Just as in the pure state case, the $1+1$d state in the Choi space is specified by a pattern of anyon condensation on the dynamical boundary. However, the allowed anyon condensates are now constrained by the fact that the Choi state must describe a density matrix, which is Hermitian and positive semi-definite. We claim that these constraints restrict the allowed anyon condensates precisely to those describing the folded boundaries discussed in Section~\ref{sec:foldedbdy}. We will show this explicitly in the Abelian case and hypothesize that this correspondence holds as well in the non-Abelian case.

Let us begin with the Abelian case. Suppose that $\rho_0 = \ket{\psi_0}\bra{\psi_0}$ describes a $2+1$d pure state hosting Abelian topological order, $\mathcal{C}$. Its corresponding Choi state $\kket{\rho_0}$ then describes a bilayer topological order, $\mathcal{C}\times \overline{\mathcal{C}}$. Suppose now that we subject $\rho_0$ to a local finite-depth quantum channel $\mathcal{E}[\cdot]$, which can be understood as forcing the incoherent proliferation of anyon excitations. The resulting density matrix $\rho \equiv \cE[\rho_0]$ then describes a mixed-state topological order, as discussed in Refs.~\cite{Sohal2025,Ellison2025}. The Choi state $\kket{\rho}$ of this decohered density matrix again describes a topological order in the doubled Hilbert space.

In the doubled space, the effect of the decoherence on the original state $\kket{\rho_0}$ is to induce the conventional condensation of bulk anyons~\cite{bao2023mixed}. 
Let us define the indicator function $M_{a,b} \equiv M_{a_+ b_-}$ as in Eq.~\eqref{eq:indicator-function} to keep track of which anyons from the original TO are condensed, where $a_+$ ($b_-$) lives in the ket (bra) space: $a_+ b_- \in \mathcal{C}\times \overline{\mathcal{C}}$. Now, for an anyon $a$, we can write down a Wilson line operator $W_a^\gamma$ on an open string $\gamma$, which creates $a$ and its conjugate $\bar a$ at its endpoints. An anyon $a$ is condensed if $W_a^\gamma$ exhibits long-range correlation as a function of the distance between the end-points of the open string $\gamma$. For ease of presentation, we restrict ourselves to ``fixed-point" states in which the anyons have been maximally decohered~\cite{Sohal2025}--in this case an anyon $a$ is condensed if the application of the open string operator $W_a^\gamma$ leaves $\rho$ unchanged. Suppose
$a_+ \bar{b}_-$
is condensed; then,
\begin{align}
    W_a^\gamma \rho (W_b^\gamma)^\dagger = \rho \implies W_b^\gamma \rho (W_a^\gamma)^\dagger = \rho \, ,
\end{align}
by Hermiticity, so 
$b_+ \bar{a}_-$
is also condensed. Thus, if $M_{a,b} = 1$, then we must have $M_{\bar{b},\bar{a}} =  M_{b,a} =1$, where the second equality follows since the condensed anyons form a group. Hence, we see that Hermiticity leads to the condition that anyon condensation is symmetric between the two layers.

Next, the fact that $\rho$ is positive semi-definite implies that we may write it in the form $\rho = \sqrt{\rho} \sqrt{\rho}$ for a unique, positive operator $\sqrt{\rho}$. Then, we have (dropping the explicit dependence on $\gamma$)
\begin{align}
\begin{split}
    \Tr[W_a \rho W_b^\dagger \rho] &= \Tr[W_a \sqrt{\rho} \sqrt{\rho} W_b^\dagger \sqrt{\rho} \sqrt{\rho}] \\
        &= \Tr[\sqrt{\rho} W_a \sqrt{\rho} \sqrt{\rho} W_b^\dagger \sqrt{\rho}] \\
        &= \bbraket{\sqrt{\rho} W_a^\dagger \sqrt{\rho} | \sqrt{\rho} W_b^\dagger \sqrt{\rho}},
\end{split}
\end{align}
where we used cyclicity of the trace and the definition of the inner product on the doubled space. Now, using the Cauchy-Schwarz inequality, we find that, if
$a_+ \bar{b}_-$
is condensed, 
\begin{align}
\begin{split}
    \Tr[\rho^2] &= |\bbraket{\sqrt{\rho} W_a^\dagger\sqrt{\rho} | \sqrt{\rho} W_b^\dagger \sqrt{\rho}} |^2 \\
    &\leq \bbraket{\sqrt{\rho} W_a^\dagger \sqrt{\rho} | \sqrt{\rho} W_a^\dagger \sqrt{\rho}} \bbraket{\sqrt{\rho} W_b^\dagger \sqrt{\rho} | \sqrt{\rho} W_b^\dagger \sqrt{\rho}} \\
    &= \Tr[(\sqrt{\rho} W_a \sqrt{\rho}) \sqrt{\rho} W_a^\dagger \sqrt{\rho}] \Tr[(\sqrt{\rho} W_b \sqrt{\rho}) \sqrt{\rho} W_b^\dagger \sqrt{\rho}] \\
    &= \Tr[\rho W_a \rho W_a^\dagger] \Tr[\rho W_b \rho W_b^\dagger] \, .
\end{split}
\end{align}
The two traces in the final line must both be non-zero in order to satisfy the inequality, implying that both 
$a_+ \bar{a}_-$ and $b_+ \bar{b}_-$
must also be condensed. Using the fact that the condensed anyons form a group, this in turn implies that 
$(a\times\bar{b})_+1_-$
must also be condensed. Hence, if $M_{a,\bar b} = 1$, we have that $M_{a\times \bar b , 1} = 1$ as well.

Thus, we have shown that the indicator functions characterizing the allowed anyon condensations in an Abelian mixed-state topological order \textit{exactly} correspond to those characterizing the folded condensates studied in Section~\ref{sec:foldedbdy}! It is this correspondence that allows us to leverage the results in the preceding Sections to establish correspondences between mixed-state SPTs and certain classes of gapped and gapless pure states. 

Let us note that, in this Abelian case, the allowed Choi state anyon condensations can always be decomposed into a two step process. Since the condensation matrix $M$ satisfies Eqs.~\eqref{eq:m_constraints_1}-\eqref{eq:m_constraints_3}, the subset of anyons that are supported on a single layer form condensable subgroups of anyons on the ket and bra layers of $\rho$. The full anyon condensation then arises from condensing these anyons on each layer separately, followed by condensation of $a_+ \bar{a}_-$ of anyons in the residual TO. 

We will not attempt to derive similar constraints in the more general non-Abelian case. However, based on the above analysis, we believe it is reasonable to conjecture that the allowed anyon condensates for non-Abelian Choi states should also follow a two step condensation process: first, condensing a Lagrangian subalgebra on each layer, followed by the condensation of anyons of the form $a_+ \bar{a}_-$ in the residual bilayer TO. From this assumption, it follows that anyon condensates in non-Abelian Choi states should also be in exact correspondence with the folded condensates discussed in Section~\ref{sec:foldedbdy}.

Finally, let us turn to the case where $\rho$ has a strong symmetry $K \lhd G$ and the total symmetry group $G$ is an extension of the weak symmetry by $K$, as in Eq.~\eqref{eq:mixedgroupext} (we do not restrict to Abelian $K$ or $G$ here). 
This is to contrast with the previous situation where $\rho$ has a $G_+ \times G_-$ symmetry, which could be \emph{spontaneously} broken down to such a subgroup after decoherence. Here, we wish to realize the symmetry breaking explicitly, via condensation of electric anyons in the bulk Choi state TO.
As noted in Sec.~\ref{sec:swsym}, the full symmetry group in the Choi space is $K \rtimes G$, consisting of elements of the form $(kg, g)$. In this case, the bulk TO in the Choi space will be isomorphic to $\mathcal{D}(K \rtimes G)$. To obtain this TO, we can begin with the original $\mathcal{D}(G_+ \times G_-)$ TO and and condense charges to break the gauge group down to $K \rtimes G$. 
Let $\bigoplus_i \pi_i \subset \mathcal{D}(G)$ be the set of charges condensed such that the resulting TO is given by $\mathcal{D}(K)$, for $K \subset G$. These charges correspond to those that are uncharged under $K$. Then, condensing the charges $\bigoplus_i \pi_i \pi^*_i \subset \mathcal{D}(G_+ \times G_-)$ leads to a bulk topological order $\mathcal{D}(K \rtimes G)$ in the Choi space. This condensation is positive, since the anyons condensed are of the form $a_+ \bar{a}_-$. Physically, this condensation corresponds to starting from a pure state $\mathcal{D}(G)$ TO and decohering the $\pi_i$ charges, which can be done by applying maximal decoherence composed of the short string operators for these charges (see Refs.~\cite{Sohal2025,Ellison2025} for the explicit procedure). Hence, the bulk TO of a $1+1$d theory with a $K \subset G$ strong symmetry is given by a mixed state TO whose Choi state topological order is isomorphic to $\mathcal{D}(K \rtimes G)$. 


\subsection{Example: Strong-to-weak SSB}
\label{sec:swssb}

As a simple example illustrating the positivity constraints on the SymTaco, let us consider the possible gapped phases of mixed states with strong $\mbb{Z}_2$ symmetry. Such a system can exhibit three distinct gapped phases: a trivial and symmetric phase, a SWSSB phase, and a complete (strong-to-nothing) SSB phase. Working with qubits placed on sites of a one-dimensional chain, with the usual Pauli operators $X_i$ and $Z_i$, such that the $\mathbb{Z}_2$ symmetry is generated by $X \equiv \prod_i X_i$, representative states for these three phases in the computational basis are given by:
\begin{align}
    \ket{+} &= \ket{+}^{\otimes L} \, ,\\
    \rho_{\mathrm{SW}} &= (1 + X)/2^L \, , \\
    \ket{\mathrm{GHZ}} &= \frac{1}{\sqrt{2}}\left( \ket{0}^{\otimes L} + \ket{1}^{\otimes L} \right) \, ,
\end{align}
where $L$ is the length of the system.

It is clear that $C_I(i,j)$ and $C_{II}(i,j)$ (defined in Eqs.~\eqref{eq:CI_correlator} and~\eqref{eq:CII_correlator} respectively) vanish for all $i$ and $j$ for the operator $O_i = Z_i$, which is charged under $X$, for the symmetric state $\ket{+}$. In contrast, $C_I(i,j) = 1$ for all $i$ and $j$ for the state $\ket{\mathrm{GHZ}}$, indicating complete SSB. Meanwhile, $C_{II}(i,j) = 1$ while $C_I(i,j) = 0$ for the state $\rho_{\mathrm{SW}}$, indicating SWSSB of the strong $\mathbb{Z}_2$ symmetry. Note that the Choi state, $\kket{\rho_{\mathrm{SW}}} \propto \ket{0}_+^{\otimes L}(\ket{0}^*_-)^{\otimes L} + \ket{1}_+^{\otimes L}(\ket{1}^*_-)^{\otimes L}$ is simply a GHZ state, indicating SSB of the diagonal $\mathbb{Z}_2$ subgroup of the full symmetry $\mathbb{Z}_2^+ \times \mathbb{Z}_2^-$. In contrast, the Choi state of $\ket{\mathrm{GHZ}}$ is simply two decoupled copies of GHZ states on the ket and bra spaces, indicating SSB of the full symmetry $\mathbb{Z}_2^+ \times \mathbb{Z}_2^-$.

Let us see how this physics is reproduced in the SymTaco. The bulk TO in the Choi space for a strong $\mathbb{Z}_2$ symmetry is givne by $\mathcal{D}(\mbb{Z}_2^+ \times \mbb{Z}_2^-)$, which is the Choi state of the pure state $\mathbb{Z}_2$ Toric Code. The anyons in the Choi space form a $\mbb{Z}_2^4$ group under fusion, generated by the charges $e_+, e_-$ and the fluxes $m_+, m_-$, with a $-1$ braiding between $e_+$ and $m_+$, as well as between $e_-$ and $m_-$. The reference boundary of the SymTaco is chosen to be the charge condensed boundary, with the Lagrangian subgroup given by $\mathcal{L} = \langle e_+, e_- \rangle$. With this choice, the $m_+$ and $m_-$ generate the $\mathbb{Z}_2^+ \times \mathbb{Z}_2^-$ symmetry, respectively. The weak symmetry, corresponding to the diagonal subgroup $\mathbb{Z}_2 \subset \mathbb{Z}_2^+ \times \mathbb{Z}_2^-$ is then generated by the composite flux $m_+ m_-$. Due to the nontrivial braiding of the charges with fluxes, we see that the local order parameters for breaking of the full symmetry $\mathbb{Z}_2^+ \times \mathbb{Z}_2^-$ correspond to the Wilson lines of $e_+$ and $e_-$ stretching from the reference boundary to the dynamical boundary. The order parameter for breaking down to the diagonal subgroup $\mathbb{Z}_2$ is then given by the Wilson line for $e_+ e_-$ which, indeed, has trivial braiding with the generator of this preserved symmetry, $m_+ m_-$. In the original physical space, this corresponds to the order parameter for SWSSB.

Having fixed the symmetry algebra dictated by the SymTaco, we can now map out the possible mixed state phases by identifying the allowed anyon condensates on the reference boundary. We are restricted to anyon condensations which obey the positivity constraint of Eq.~\eqref{eq:m_constraints_3}. As argued in Sec.~\ref{sec:folddouble}, such boundaries may equivalently be obtained by considering condensable algebras within a single layer $\mathcal{D}(\mbb{Z}_2)$ and folding. There are three condensable algebras of $\mathcal{D}(\mbb{Z}_2)$: the trivial condensation $\mathcal{L} = 1$, and the Lagrangian subgroups $\mathcal L_e = \langle e \rangle$ and $\mathcal L_m = \langle m \rangle$. Upon folding, these become 
\begin{equation} \label{eq:z2mixed}
\begin{aligned}
    \mathcal{L}_1 &= \langle e_+ e_-, m_+ m_-  \rangle , \\
    \mathcal{L}_e &= \langle e_+, e_- \rangle \, ,\\
    \mathcal{L}_m &= \langle m_+, m_- \rangle \, .
\end{aligned}
\end{equation}
Using the identification of the order parameters above, we see that the boundaries $\mathcal L_e$ and $\mathcal L_m$ correspond to Choi space representations of the SSB and trivial pure states, respectively. On the other hand $\mathcal L_1$ is \textit{intrinsically mixed}, and has no pure state counterpart--in the pure-state context, there is no condensable subgroup that contains both $e$ and $m$. Indeed, since $e_+ e_-$ is condensed on this boundary, we see from the above description of the order parameters that this choice corresponds to the SWSSB phase. We thus see that we can indeed reproduce the mixed-state phases of a $\mathbb{Z}_2$-symmetric system within our SymTaco description. 


\subsection{Symmetry Fractionalization and the Tetraptych Correspondence}
\label{sec:symfrac}

Having introduced the SymTaco and illustrating that it captures the correct symmetry properties via the simple example of $\mbb{Z}_2$ SWSSB, we now turn to a systematic analysis of gapped $1+1$d mixed-state SPTs with strong symmetry $K \lhd G$, where $G$ is the total symmetry group obtained via the group extension Eq.~\eqref{eq:mixedgroupext}. Our goal in this Section is to explicitly derive the data characterizing the symmetry fractionalization of mixed-state SPTs (including iASPTs). We will find that this data exactly corresponds to that characterizing the pure-state folded SPTs discussed in Section~\ref{sec:gapped}. This is consistent with our argument that the valid set of anyon condensates for our mixed-state SymTFT is in one-to-one correspondence with the set of valid condensates for the pure-state SymTFT for folded states, in the case where the symmetry group is Abelian. We will also comment on the non-Abelian case and, doubling back to Sec.~\ref{sec:folddouble}, argue that this correspondence allows us to indirectly obtain the 2-cocycle characterizing folded SPTs in the general non-Abelian case (which we were unable to directly derive in  Section~\ref{sec:folddouble}). Taken together, the results of this subsection complete the correspondences of phases in the tetraptych of Fig.~\ref{fig:main}.

Let us begin by deriving the symmetry fractionalization data of $1+1$d mixed-state SPTs (see also Refs.~\cite{ma2024,guo2024,xue2024}). Recall from Section~\ref{sec:choi} that we say that a mixed state is short-range entangled (and hence, in $1+1$d, an SPT) if its Choi state is short-range entangled. For our purposes, the essential consequences of assuming that $\rho$ is short-range entangled are that the strong and weak symmetries, when mapped to operators in the doubled space, \emph{fractionalize} on the short-range entangled pure state $\kket{\rho}$~\cite{else2014}. Explicitly, if $U_k$ is the unitary representation of a strong symmetry $k \in K$, then its restriction to a sufficiently large interval $I$ on the short-range entangled state $\rho$ fractionalizes as
\begin{equation}
    U_{k,I} \rho = U_k^L U_k^R \rho \, ,
\end{equation}
where $U_k^{L,R}$ are unitary operators with support at the left and right endpoints of $I$. Similarly, under the assumption that $\kket{\rho}$ is SRE in the Choi space, the restriction of a weak symmetry $g \in G$ implemented by $V_g$ to the region $I$ fractionalizes as 
\begin{align}
    V_{g,I} \rho V^\dagger_{g,I} = V_g^L V_g^R \rho (V_g^L V_g^R)^\dagger \, .
\end{align}

As in the case of pure-state SPTs, these fractionalized symmetry operators may form a projective representation of the full symmetry group. Since the $U_k$ form a linear representation of $K$, the fractionalized operators may in fact form a projective representation,
\begin{equation}
    U_k^{L} U_{k'}^{L} = \eta(k, k') U_{k k'}^{L} \, , \label{eq:Uk_fractionalization}
\end{equation}
where $\eta(k,k')$ is a 2-cocycle, $\eta \in Z^2(K,U(1))$. On the other hand, since $\rho$ transforms in the conjugate representation of $G$, the conjugate action of the fractionalized operators forms a linear representation 
\begin{align}
    (V^L_g V^L_{g'}) (\cdot) (V^L_g V^L_{g'})^\dagger = V^L_{gg'}  (\cdot) V^{L\dagger}_{gg'} \, , \label{eq:Vg_fractionalization}
\end{align}
since any projective phases associated with $V^L_g$ will cancel out with the opposite projective phase associated with $V^{L\dagger}_g$. Physically, these relations tell us that the mixed state $\rho$ may form an ensemble of states such that each state is an SPT (with the same SPT index) with respect to the strong symmetry $K$. On the other hand, $\rho$ cannot form an SPT protected solely by the weak symmetry $G_{\mathrm{weak}}$~\cite{Ma2023,Ma2025}.

The weak symmetry still has a role to play, as there may be a non-trivial interplay between the weak and strong symmetries--it is precisely this interplay that opens the possibility for intrinsically mixed-state phases. Let us consider acting with a strong symmetry $k \in K$, followed by acting with the weak symmetry $g \in G$, on a large finite interval $I$:
\begin{align}
        V_{g,I} U_{k,I} \rho V_{g,I}^\dagger
        &= V_{g,I} U^L_k U^R_k \rho V^\dagger_{g,I}\\
        &= U^L_{gkg^{-1}} U^R_{gkg^{-1}} V_{g,I} \rho V^\dagger_{g,I}\, ,
\end{align}
where in the second equality, we used the fact that the combination $U_g^L U_g^R$ still forms a linear representation. Alternatively, we again fractionalize $U_k$ on $\rho$ but instead write,
\begin{align}
        V_{g,I} U_{k,I} \rho V_{g,I}^\dagger &= V_{g,I} U_k^L U_k^R \rho V_{g,I}^\dagger \\
        &= (V_{g,I} U_k^L V_{g,I}^{\dagger}) (V_{g,I} U_k^R V_{g,I}^\dagger) V_{g,I} \rho V^\dagger_{g,I} \,.
\end{align}
Comparing the above two expressions, we see that, by locality, $U^L_{gkg^{-1}}$ and $V_{g,I} U^L_k V^{\dagger}_{g,I}$ are related by a phase, 
\begin{equation}
    V_{g,I} U^L_k V^{\dagger}_{g,I} = \epsilon(k, g) U^L_{g k g^{-1}} \,. \label{eq:VgUk_fractionalization}
\end{equation}
On the other hand, we can repeat the above derivation in a slightly modified setting, where we act with a strong symmetry $k \in K$ on an interval $I$, followed by the action of an average symmetry $g \in G$ on on the entire system. We then find that, 
\begin{align}
    V_g U^L_k V^\dagger_g = \epsilon'(k,g) U^L_{gkg^{-1}}.
\end{align}
To see that $\epsilon$ and $\epsilon'$ are the same phase, notice that we could repeat the argument on a system with open boundary conditions, and take $I$ to be the entire system. Indeed, the function $\epsilon(k,g) \in U(1)$, with $k \in K$ and $g \in G$, gives the $G$ charge  carried by the fractionalized $U_k^{L,R}$ operators. 

We see that the symmetry fractionalization data $(\epsilon, \eta)$ of (intrinsically) ASPTs exactly parallels that of igSPTs reviewed in Sec.~\ref{sec:symigspt}. In an igSPT, the gapped degrees of freedom, charged under $A$, can form an $A$-SPT, while in an iASPT, the degrees of freedom charged under the strong symmetry $K$ can form an SPT; this data is captured by the 2-cocycle $\eta$. In the igSPT, the degrees of freedom charged under the low-energy symmetry group do not fractionalize as they are gapless while, in an iASPT, the weak symmetry simply cannot fractionalize. Finally, in an igSPT (iASPT), the fractionalized high-energy (strong) symmetry operators can carry charge under the low-energy (weak) symmetry; this data is captured by the function $\epsilon$. Repeating the same analysis as in Ref.~\cite{Wen2023}, one can check that the pair $(\eta,\epsilon)$ characterizing an iASPT satisfy the same consistency conditions as those for an igSPT (see Eq.~\eqref{eq:condensationdata}.) This correspondence suggests that it may be possible to generate iASPTs from igSPTs--for the $\mbb{Z}_4$ iASPT, this process was discussed in Ref.~\cite{Ma2025}. We will return to this relation in Section~\ref{sec:apps} from a SymTaco perspective and argue that a general protocol exists for generating iASPTs from igSPTs in $1+1$d.

From the SymTaco, we observe that the Choi state of an iASPT should also correspond to a $G \times G$ (folded) SPT. The 2-cocycle which characterizes this SPT in the doubled space is fully determined by $\eta$ and $\epsilon$. Defining $\omega((kg,g), (k'g',g'))$ through,
\begin{align}
\begin{split}
    U^L_k U^L_g U^L_{k'} U^L_{g'} \rho (U^L_g U^L_{g'})^\dagger = \omega((kg, g), (k'g', g')) \\ \times U^L_{k g k' g^{-1}} U^L_{gg'} \rho U^{L \dagger}_{gg'} \, ,
\end{split}
\end{align}
a straightforward computation using Eqs.~\eqref{eq:Uk_fractionalization}, \eqref{eq:Vg_fractionalization}, and Eq.~\eqref{eq:VgUk_fractionalization} gives 
\begin{equation}
\omega((kg, g), (k'g', g')) = \epsilon(k', g) \eta( k, gk' g^{-1}) \,. \label{eq:mixed-state-coycle}
\end{equation}
One can verify explicitly that $\omega$ as defined above is indeed a $2$-cocycle provided that $\epsilon$ and $\eta$ satisfy Eq.~\eqref{eq:condensationdata}. Conversely, the symmetry fractionalization data $\eta$ and $\epsilon$ can be read off from the 2-cocycle for the mixed state SPT $\omega$ by setting $\epsilon(k',g) = \omega((g,g), (k',e))$ and $\eta(k,k') = \omega((k,e),(k',e))$. 
Note that this reduces to the 2-cocycle found via Abelian anyon condensation of Eq.~\eqref{eq:Omega}. Recall now that in the non-Abelian case, we have only conjectured that there is an exact correspondence between the SymTaco for mixed-state SPTs and gapped SPTs. However, we expect that this correspondence extends in general given that we have directly argued that the correspondence holds in the Abelian case and that the 2-cocycles for mixed-state and folded SPTs match in this case. We therefore conjecture that the 2-cocycle for folded SPTs, corresponding to a folded boundary in the SymTaco characterized by the data $(\epsilon,\eta)$, is identical to that of the corresponding mixed-state SPT, Eq.~\eqref{eq:mixed-state-coycle} above--this provides the justification for Eq.~\eqref{eq:nonabelian_omega}.

Let us consolidate our results thus far and map out the correspondences advertised in the tetraptych of Fig.~\ref{fig:main} for $G$-symmetric systems. In this subsection, we have established an exact correspondence between the symmetry fractionalization data of iASPTs and igSPTs. In Section~\ref{sec:gapped}, we established an analogous correspondence between igSPTs and folded SPTs. Further, we have demonstrated that the correspondences between these three families of phases can be unified through a common holographic bulk description as encapsulated by the SymTaco. As reviewed in Section~\ref{sec:gapless}, the SymTFT description of $G$-igSPTs is given by the $\mD(G)$ quantum double, with the dynamical boundary described by a condensable subalgebra $\mA$. In Section~\ref{sec:gapped}, we showed that such SymTFTs are in one-to-one correspondence with $\mD(G) \times \overline{\mD(G)}$ quantum doubles, specifically those whose Lagrangian subalgebras satisfy a certain ``folded" constraint--we dubbed such a SymTFT the SymTaco. These SymTacos, in turn, provide the holographic bulk for certain $G\times G$ SPTs, which we termed folded SPTs. Finally, in this Section, we proposed the Choi state of the $\mD(G)$ quantum double, which is the $\mD(G) \times \overline{\mD(G)}$ anyon theory in the doubled Hilbert space. We showed that for Abelian $G$ (and conjectured for non-Abelian $G$), the set of Lagrangian subalgebras of this Choi state--consistent with the Hermiticity and positivity constraints on a density matrix--precisely correspond to the folding constraints for the SymTaco description of $G \times G$ folded SPTs. We thus conclude that the SymTaco provides a unified holographic bulk description for pure-state igSPTs, pure-state folded SPTs, and mixed-state SPTs, as advertised.


\subsection{Example: $\mbb{Z}_4$ Intrinsically Average SPT}
\label{sec:z4iaspt}

The positivity condition describing folded SPTs precisely maps to the positivity condition that density matrices must satisfy as states in the doubled Hilbert space. With this understanding in mind, we can return to the $\mbb{Z}_4 \times \mbb{Z}_4$ folded SPT and view it as a mixed-state SPT in the doubled space, which corresponds to an intrinsically average SPT. Recall that the condensed anyons characterizing the folded SPT are 
\begin{align}
\label{eq:choi_spt_condensation}
\begin{split}
    \mathcal{L} &= \langle e_A^2 e_B^2, e_A^2 m_A^2, e_A e_B m_A m_B^3 \rangle  \\
    &\mapsto \langle e_+^2 e_-^2, e_+^2 m_+^2, e_+ \overline{e}_- m_+ \overline{m}_- \rangle \, .
\end{split}
\end{align}
We can identify the $A$ and $B$ $\mbb{Z}_4$ symmetries with the ket (+) and bra (-) $\mbb{Z}_4$ symmetries, respectively. 
Note that, in the first line of Eq.~\eqref{eq:choi_spt_condensation}, we are labeling the anyons as elements of $\mD(G \times G)$, while the anyons in the second line are labeled as elements of $\mD(G_+) \boxtimes \overline{\mD(G_-)}$ where the $-$ anyons have spin opposite to the $+$ anyons.
For instance, note that $e^3_B m_B$ has opposite braiding compared to $e_A m_A$. 
In the above set of anyons, we thus identify $e_A m_A$ with $e_+ m_+$ and $e^3_B m_B$ with $e_- m_-$, so that $e_A m_A e_B m^3_B$ is identified with $e_+ m_+ \overline{e}_- \overline{m}_-$ in the second line. 

We now provide an explicit density matrix realizing this pattern of anyon condensation. 
We consider a chain of four dimensional qudits $\cX_i$, $\cZ_i$ such that the extended $\mbb{Z}_4$ symmetry is generated by $\cX \equiv \prod_i \cX_i$, and the strong $\mbb{Z}_2$ subgroup is generated by $\cX^2 \equiv \prod_i \cX_i^2$.
We claim that the density matrix corresponding to this pattern of anyon condensation is given by 
\begin{equation}
    \rho \propto \left( \prod_i (1 + \cZ_i^2 \cX_i^2 \cZ_{i+1}^2) \right) \, .
\end{equation}
In the following Section, we will validate this claim and provide a general prescription to write down Pauli stabilizer models of Abelian mixed-state SPTs. For now, we content ourselves with verifying that $\rho$ satisfies the expected universal properties. Observe that $\rho$ satisfies 
\begin{align}
    \cZ_i^2 \cX_i^2 \cZ_{i+1}^2 \rho &= \rho \, , \label{eq:shortstring_strong} \\
    \cZ^\dagger_i \cZ_{i+1} \cX_{i+1} \rho \cX^\dagger_{i+1} \cZ_{i+1}^\dagger \cZ_i &= \rho \, . \label{eq:shortstring_weak}
\end{align}
These correspond via Eq.~\eqref{eq:anyontostabilizer} to the $e_+^2 m_+^2$ and $e_+ m_+ \overline{e}_- \overline{m}_-$ condensed anyons in the Lagrangian subgroup of Eq.~\eqref{eq:choi_spt_condensation}. 

\begin{figure}
    \centering
    \includegraphics[width=\linewidth]{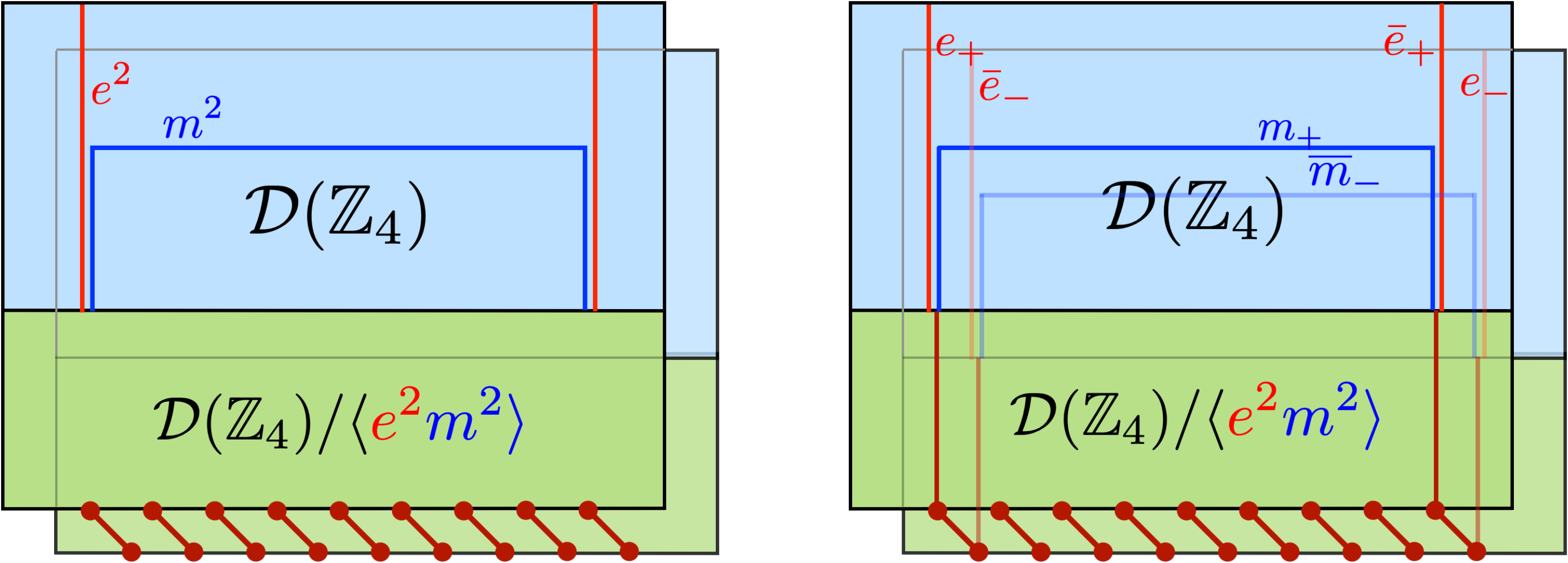}
    \caption{Strong (left) and weak (right) string order parameters characterizing the $\mbb{Z}_4$ iASPT, encoding the relations Eq.~\eqref{eq:longstring_strong} and Eq.~\eqref{eq:longstring_weak}.}
    \label{fig:z4mspt_stringorder}
\end{figure}

The state $\rho$ clearly commutes with $\prod_i \cX_i$, so is invariant under the total (average) $\mbb{Z}_4$ symmetry. To see that it is invariant under the strong $\mbb{Z}_2$ symmetry given by $\prod_i \cX_i^2$, observe that the strong symmetry operator is obtained by a product of the stabilizers $\cZ_i^2 \cX_i^2 \cZ_{i+1}^2$ over the entire lattice (assuming periodic boundary conditions). 
Using the relations above, we can derive the following behavior of the string order parameters, 
\begin{align}
    \cZ_i^2 \cX_i^2 \ldots \cX_{j}^2 \cZ_{j+1}^2 \rho &= \rho \label{eq:longstring_strong} \\
    \cZ^\dagger_i  \cX_{i+1} \ldots \cX_j \cZ_j \rho (\cZ^\dagger_i  \cX_{i+1} \ldots \cX_j \cZ_j)^\dagger &= \rho \, . \label{eq:longstring_weak}
\end{align}
The string order parameters are illustrated in the \staco\, in Fig.~\ref{fig:z4mspt_stringorder}.
We find that the truncated strong symmetry operator $\prod_{i=1}^L \cX^2_i$ fractionalizes on $\rho$ as $\cZ_0^2 \cX_0^2 \cZ_{L+1}^2$, and the weak symmetry operator  $\prod_{i=1}^L \cX_i$ fractionalizes on $\rho$ as the adjoint action of $\cZ_0^\dagger \cZ_{L+1}\cX_{L+1}$. The endpoint of the strong symmetry operator carries charge $2$ under the $\mbb{Z}_4$ symmetry, while the endpoint of the weak symmetry operator carries charge under the strong $\mbb{Z}_2 \subset \mbb{Z}_4$ symmetry. 

\section{Applications of the Mixed-State SymTaco}
\label{sec:apps}

In the preceding Section, we introduced the Choi state of the quantum double $\mD(G)$--the \staco--as a mixed-state SymTFT for $1+1$d $G$-symmetric systems, where $G$ contains both strong and weak symmetries. Using the \staco, we established one of the main results of the present work: a one-to-one correspondence between intrinsically average SPTs, intrinsically gapless SPTs, and ``folded" gapped SPTs. Now, beyond providing a formal tool for classifying states, the SymTFT formalism in the pure-state setting also sheds light on other \emph{kinematic} properties and manipulations of $G$-symmetric systems. As discussed in Section~\ref{sec:genreview}, this includes the anomalies and gauging of a symmetry. In this Section, as a step towards understanding the full implications of the \staco\, on the properties of mixed states, we study the process of \emph{weak} or \emph{classical} gauging and, in doing so, uncover a novel \emph{mixed-state anomaly} in strongly and/or weakly $\mathbb{Z}_2$ symmetric systems. As another application of the \staco, we also demonstrate how our correspondence between igSPTs and iASPTs can be used to motivate explicit decoherence channels to generate the latter from the former. Finally, for Abelian symmetries $G$, we also develop a description of mixed-state SPTs in terms of Pauli subsystem codes, which may be of independent interest.  


\subsection{Gauging Weak Symmetries and a New Mixed-State ``Anomaly"}
\label{sec:mixed-gauging}

As reviewed in Section~\ref{sec:gauging}, the process of gauging a symmetry $G$ in pure states can be described within the SymTFT picture by changing the reference boundary from the ``electric" condensate to the boundary which describes the spontaneous breaking of the full symmetry group $G$ (assuming such a boundary exists). It is thus natural to ask what implications the finer symmetry structure of mixed states has for gauging and anomalies in this context. While we defer a comprehensive exploration to a forthcoming work~\cite{classgauge}, in this Section we show that, even in the relatively simple case of $\mbb{Z}_2$ symmetric mixed states, the \staco\, leads us to new insights regarding gauging and anomalies.

As first observed in Ref.~\cite{Ellison2025}, the refined symmetry structure of mixed states opens the possibility for \emph{classically} or \emph{weakly} gauging a symmetry, i.e., gauging a weak symmetry. In brief, when gauging a symmetry in a pure state $\ket{\psi}$, one projects the state to the subspace satisfying a certain Gauss' law (see Refs~\cite{williamson2016mpo,williamson2019gauge} for a discussion). For instance, for a spin-$1/2$ chain with spin-flip symmetry $X \equiv \prod_i X_i$, one adds ancilla degrees of freedom (the gauge fields) and enforces the Gauss' law $G_i = Z_{i-1/2} X_i Z_{i+1/2} \equiv 1$, where the original qubits live on integer valued sites and the gauge fields live on the half-integer valued sites. The gauged state, $\ket{\tilde{\psi}}$ satisfies $G_i \ket{\tilde{\psi}} = \ket{\tilde{\psi}}$ for all $i$. Gauging promotes the symmetry generated by $X$ to a local symmetry, in the sense that we can apply spin-flips to a finite region, at the expense of adding the gauge degrees of freedom. In contrast, when we \emph{classically} gauge a symmetry of a mixed state $\rho$, we only demand that the gauged state, $\tilde{\rho}$, is invariant under \textit{conjugation} by the Gauss' law: $G_i \tilde{\rho} G_i^\dagger = \tilde{\rho}$. Physically, classical gauging may be understood as incoherently proliferating symmetry defects of the symmetry being weakly gauged.

At the level of states, this process may be understood in the doubled Hilbert space via the Choi isomorphism. Suppose $\rho$ is strongly symmetric under $K$, such that $U_g\rho = e^{i\theta_g} \rho$ for $g \in K$. Then, the vectorized density matrix $\kket{\rho}$ is symmetric under $K\times K$. Classical or weak gauging then corresponds to applying the usual gauging procedure to the diagonal subgroup $K^{\mathrm{diag}}\subset K\times K$, rather than the full symmetry group $K\times K$ (which would describe the familiar ``strong" gauging). This makes clear how we should implement classical gauging at the level of the \staco--one simply needs to exchange the electric reference boundary with that corresponding to spontaneous breaking down to $K^{\mathrm{diag}}$ which, in the original Hilbert space, describes strong-to-weak spontaneous symmetry breaking of the strong symmetry $K$. Note that, if the system only has a \emph{weak} symmetry $G_{weak}$ to begin with, classically gauging it would correspond to exchanging the charge condensed boundary with that corresponding to weak-to-nothing breaking of the $G_{weak}$ symmetry, i.e., replacing the charge-condensed boundary with the original symmetry preserving boundary. 

To illustrate this procedure, let us consider the example of
$1+1$d states invariant under a strong $\mbb{Z}_2$ symmetry, as discussed in Section~\ref{sec:swssb}. Recall that the corresponding bulk TO in the Choi space is $\mD(\mbb{Z}_2^+ \times \mbb{Z}_2^-)$, a bilayer Toric Code. The reference boundary of the \staco\, is chosen to be the charge condensed boundary, with the corresponding Lagrangian subgroup given by $\mathcal{L}_e = \langle e_+, e_- \rangle$. With this choice, the strong $\mathbb{Z}_2$ symmetry is generated by the magnetic anyons, $m_+$ and $m_-$. Now, to \emph{classically} gauge the $\mbb{Z}_2$ symmetry within the \staco, we exchange the charge condensed Lagrangian subgroup $\mathcal{L}_e$ on the reference boundary with the subgroup corresponding to SWSSB of the $\mathbb{Z}_2$ symmetry, namely to the Lagrangian subgroup, $\mathcal{L}_1 = \langle e_+ e_-, m_+ m_-  \rangle$.

Let us now analyze the symmetry structure of the gauged theory.
As usual, the symmetry operators are given by the anyons that are confined at the reference boundary, which is described by $\mathcal{L}_1$: these anyons are $e_+$ and $m_+$, which are identified with $e_-$ and $m_-$, respectively. At first glance, it appears that the anyons $e_\pm$ and $m_\pm$ should generate strong symmetries, as they only act on either one of the ket or bra space at one time. However, acting with $m_+$ in the \staco\, is \emph{equivalent} to acting with $m_-$ due to the identification $m_+ \sim m_-$ at the reference boundary. We refer to such symmetries as \emph{strong modulo weak} symmetries. 

To be more explicit, if we start with a strongly $G$-symmetric state $\rho$ and weakly gauge $G$, the new state $\tilde{\rho}$ will also be strongly symmetric under $G$. That is, if $U_g \rho = e^{i\theta_g} \rho$, then $U_g \tilde{\rho} = e^{i\theta_g} \tilde{\rho}$. After weakly gauging $G$, we project to those states which are weakly symmetric: $U_g \rho U_g^\dagger = \rho$. Thus, on one of these states, $U_g^\dagger \rho = U_g^\dagger (U_g \rho U_g^\dagger) = \rho U_g^\dagger$. Hence, the left and right strong symmetries are in fact no longer independent and we hence claim that they only generate a ``strong modulo weak" symmetry. We emphasize that the distinction between a strong symmetry and a strong modulo weak symmetry is only apparent when considering the full space of gauge-invariant states, and as such is largely immaterial at the level of analyzing individual states. 

In the $\mathbb{Z}_2$ example, we identify the symmetry generated by $m_+ \sim m_-$ as the original strong $\mathbb{Z}_2$ symmetry modulo the gauged weak symmetry as, prior to gauging, the anyons $m_+$ and $m_-$ generated the strong $\mbb{Z}_2$ symmetry. It is only after gauging, which introduces $m_+ m_-$ to the reference boundary condensate, that $m_+$ and $m_-$ become identified. On the other hand, the anyon $e_+ \sim e_-$ generates the dual symmetry obtained after gauging. Interestingly, this implies that the dual symmetry obtained by gauging a weak symmetry is a strong modulo weak symmetry, rather than another weak symmetry. A heuristic way to see why this should be the case is as follows: upon gauging, the operators charged under the new dual symmetry are the domain wall operators of the original symmetry. Since the original symmetry was a weak symmetry, the domain wall operators act simultaneously on the bra and ket spaces. The dual symmetry operators which detect a domain wall act on either the bra or the ket space; however, their action on the domain wall operators is the same in either case. Therefore, the symmetry actions on the bra and ket spaces are identified, and the dual symmetry is also a strong modulo weak symmetry. We will systematically study the symmetry structure obtained after weak gauging more generally in forthcoming work~\cite{classgauge}.

For now, we comment on what appears to be an ``anomaly" in the weakly gauged theory. At first glance, the strong modulo weak symmetries generated by $m_+ \sim m_-$ and $e_+ \sim e_-$ do not appear to have an anomaly in the usual sense, since it is possible to dress each line at the dynamical boundary in a way that removes their mutual braiding statistics. Explicitly, while $m_+$ and $e_+$ have non-trivial braiding, we can use the reference boundary to convert $m_+$ into $m_-$ which does braid trivially with $e_+$. Nevertheless, we find that there is no gapped dynamical boundary that is symmetry preserving \textit{and} satisfies the positivity constraint that must hold for density matrices. Instead, we find that the only allowed choices of gapped boundary for the reference boundary are given in Eq.~\eqref{eq:z2mixed}, each of which spontaneously breaks at least one of the strong modulo weak symmetries, since the reference boundary shares at least one condensed anyon with each of the boundaries in Eq.~\eqref{eq:z2mixed}. In the language of Ref.~\cite{zhang2024_categorical}, there is no Lagrangian subgroup that is both magnetic and also satisfies the positivity constraint. Thus, we refer to this symmetry as being ``anomalous," in the sense that there is an obstruction to a trivially gapped symmetric state (note although that these states need not be long-range entangled as is the case for states invariant anomalous strong symmetries~\cite{Lessa:2024wcw,li2024anyon,lessa2025higher,hsin2025lre}). Remarkably, this anomaly is unique to mixed states, as its existence crucially relies on the positivity constraint on density matrices.


\subsection{Fermionic SymTacos}
\label{sec:fstaco}

Having built some intuition for bosonic systems with $\mathbb{Z}_2$ symmetry, let us briefly consider fermionic systems with only a \emph{weak} $\mbb{Z}_2^F$ fermion parity symmetry. In the pure-state setting, the SymTFT for fermionic systems has recently been developed in Refs.~\cite{wen2024fermion,huang2025,tiwari2024fermion}. For a $1+1$d pure state with only $\mathbb{Z}_2^F$ fermion parity symmetry, the relevant SymTFT is the usual $2+1$d Toric Code with the reference boundary taken to be $\mathcal{L}_f = \langle 1, f \rangle$; here, the emergent fermion $f$ is understood to be condensed by adding an ancilliary fermionic chain (only along the reference boundary and \emph{not} the dynamical boundary or the bulk) and pair condensing the $f$ anyons with the physical fermions. The ancilla fermions along the reference boundary encode the fact that the SymTFT describes a system composed of microscopic fermions. On the dynamical boundary however, we are still only permitted to condense bosonic anyons since, on physical grounds, only bosonic operators can condense. We can choose the fermion parity symmetry to be generated by the confined $m$ anyon. The two choices of gapped boundary for the dynamical boundary--the $e$ and $m$ condensates--then correspond to the same phases as the Kitaev chain and a trivially gapped chain, respectively. We can also gauge the fermion parity symmetry (which corresponds to bosonizing the system) by exchanging the $\mathcal{L}_f$ reference boundary with $\mathcal{L}_e$; this yields a bosonic system with $\mathbb{Z}_2$ symmetry generated by the $m$ line. In particular, the Kitaev chain and trivial phases, corresponding to the $e$ and $m$ condensed dynamical boundaries, map to the $\mathbb{Z}_2$-SSB and trivial symmetric phases, as expected under the familiar Jordan-Wigner transformation.

Let us now consider a mixed state with only a \emph{weak} $\mbb{Z}_2^F$ fermion parity symmetry. We propose that the relevant SymTFT for such a system is provided by the $f$-decohered $2+1$d Toric Code discussed in Refs.~\cite{Wang2025intrinsic,Sohal2025,Ellison2025}. In the Choi-state representation, this corresponds to starting with a bilayer Toric Code and condensing $f_+ f_-$, such that the bulk anyons are generated by $\langle f_+ \sim f_-, e_+ e_- \sim m_+ m_- \rangle$. On the reference boundary, we then condense $\mathcal{L}_{f_\pm} = \langle f_+ \sim f_- \rangle$ by introducing ancilla fermions. This leaves a single confined anyon on the boundary, $e_+ e_- \sim m_+ m_-$, which we take to be the generator of the weak $\mathbb{Z}_2^F$ fermion parity symmetry. This leaves only a single choice of Lagrangian subgroup for the dynamical boundary that respects positivity (recall that this must be bosonic): condensing $e_+ e_- \sim m_+ m_-$. This choice reflects the fact that, in the presence of a weak $\mbb{Z}_2^F$ symmetry, there is only a single short-range-entangled phase given by the maximally mixed state. In particular, the distinction between the trivial phase and the Kitaev chain phase is removed when the strong $\mbb{Z}_2^F$ is broken to the weak $\mbb{Z}_2^F$~\cite{kawabata2023sym}.

As in the pure-state setting, we can classically gauge the weak $\mbb{Z}_2^F$ symmetry via the corresponding SymTaco. This is implemented by changing the reference boundary from $\mathcal{L}_f$ to the boundary described by the condensate $\mathcal{L}_{e_+ e_-} = \langle e_+ e_- \sim m_+ m_- \rangle$. Note that after gauging, there are no more physical fermions remaining in the system. The new symmetry that appears is a strong modulo weak symmetry, given by the $f_+ \sim f_-$ line. In contrast to the pure-state example, where gauging the fermion parity symmetry (i.e. bosonizing) yields a bosonic system with non-anomalous $\mathbb{Z}_2$ symmetry, here the dual symmetry appears to be ``anomalous" in the following sense. 
While the $f_+ \sim f_-$ symmetry line has non-trivial braiding with itself, it can be dressed by $e_+ e_-$ at the reference boundary to give $m_+ e_-$, which does not have nontrivial spin. Hence the $f_+ \sim f_-$ symmetry does not have an anomaly in the usual sense. Nevertheless, there is no choice of gapped boundary that preserves the corresponding  symmetry--the only option for the dynamical boundary is $\mathcal{L}_{e_+ e_-}$, which spontaneously breaks the symmetry. This indicates an ``anomaly" in the sense described in the previous subsection.


\subsection{Pauli Subsystem Codes for Abelian Mixed-State SPTs}
\label{sec:subsystem}

A natural setting in which nontrivial mixed states arise is in the context of quantum error-correcting codes, which are engineered to stabilize quantum information against environmental noise (up to certain thresholds). Indeed, there is a deep conceptual connection between mixed-state phases of matter and quantum error correction, since any resource-theoretic property of a specific mixed state is expected to persist throughout the entire mixed-state phase~\cite{sang2023mixed}. For topologically ordered phases in $2+1$d, the correspondence with quantum error correction is particularly well established. In the pure-state setting, for example, any $2+1$d Abelian anyon theory that admits a gapped boundary to vacuum can be fully described in terms of Pauli stabilizer codes~\cite{Ellison2023}. More recently, it has been shown that any Abelian anyon theory (even non-modular theories) can be captured by topological subsystem codes \cite{Ellison2023} and that these are in correspondence with mixed-state topological orders~\cite{Sohal2025,Ellison2025}. Motivated by these developments, we use the intuition provided by the SymTaco construction to develop a Pauli subsystem code description of mixed-state SPTs. 
Indeed, in the same way that the boundary anyon condensates in a pure-state SymTFT motivate explicit Hamiltonians, we find that the condensates in the SymTaco naturally correspond to subsystem codes.
This yields an explicit form for the density matrices of $1+1$d mixed-state SPTs, from which their universal data can be extracted. Moreover, this approach naturally suggests a protocol for preparing $1+1$d mixed-state SPTs with Abelian symmetries, which we briefly discuss in Sec.~\ref{sec:prepare}.

We begin by reviewing the language of Pauli topological subsystem codes (TSSCs)~\cite{poulin2013,Bombin2010,bombin2012,Bombin2014}. Consider a system with $n$-dimensional qudits placed on vertices of a $d$-dimensional lattice and let $\mathcal{H}$ be the corresponding Hilbert space ($\mathcal{H} = \otimes_j \mathcal{H}_j$, $\mathcal{H}_j = \mathbb{C}^n$). We denote by $\mathcal P$ the group generated by all the Pauli clock and shift operators, $\cZ_i$ and $\cX_i$, acting on this lattice. A subsystem code is a particular class of a \emph{quantum error correcting code}, which is simply a decomposition of the Hilbert space as $\cH = \cH_C \oplus \cH_C^\perp$, where $\cH_C$ is the \emph{code space} and $\cH_C^\perp$ its complement. Quantum information is stored by preparing a state in the code space. Errors, i.e., operators which take this state out of the code space, can be detected by appropriate measurement operations without destroying the encoded information. For a subsystem code, the code space further decomposes as $\cH_C = \cH_L \otimes \cH_{\cG}$, where $\cH_L$ is the \emph{logical subsystem} and $\cH_{\cG}$ is the \emph{gauge subsystem}. The code space is determined by a \emph{stabilizer group} $\mathcal{S}$, which is generated by elements within $\mathcal{P}$. Explicitly, 
\begin{align}
    H_C = \{ \ket{\psi}  \text{ such that }  S\ket{\psi} = \ket{\psi} \, \forall S \in \cS \} \, .
\end{align}
For a TSSC, we demand that there are no nonlocal stabilizers (which act on a number of qudits that increases with system size) on the infinite plane\footnote{Note that nonlocal stabilizers are permitted on the torus or on other topologically non-trivial manifolds.}.

Unlike a topological stabilizer code, in which logical information is stored in the entire code space, logical information in a TSSC is only stored in $\cH_L$. This is made manifest by a choice of \emph{gauge group} $\cG$, which is generated by a set of local elements of $\mathcal{P}$; note that the elements of $\cG$ need not commute. The gauge group is chosen such that its center corresponds to the stabilizer group, up to the inclusion of $U(1)$ phase factors (which we implicitly include in the following): $Z(\cG) \propto \cS$. One can also proceed in the reverse direction by first choosing $\cG$ and then taking the stabilizer group to be $Z(\cG)$. The gauge operators therefore, by definition, do not take states out of the code space and do not act on the logical space--in this sense, they may be thought of as redundant errors. The group of operators which \emph{do} act on the logical space are called the (bare) \emph{logical} operators, and are given the centralizer of $\cG$ in $\mathcal{P}$, i.e., $\cZ_{\mathcal{P}}(\cG)$. 

Now, in the same way that topological stabilizer codes can be used to construct stabilizer Hamiltonians (see e.g. Ref.~\cite{Ellison2022}), TSSCs can be used to construct fixed-point mixed states~\cite{Sohal2025,Ellison2025}. To see this, let us consider a lattice system with $n$-level qudits placed on the vertices as above. Suppose that the gauge group $\cG$ is generated (up to roots of unity) by local operators $O_i^\alpha$, where $i$ labels the lattice site and $\alpha$ indexes different operators. For each of these operators, we construct the local quantum channels
\begin{align}
    \cE_{i, \alpha} [\rho] = \frac{1}{n} \sum_{m=0}^{n-1} O_{i,\alpha}^m \rho O^{m\dagger}_{i,\alpha} \, ,
\end{align}
which correspond to maximal decoherence of the operators $O_{i,\alpha}$. We then construct a quantum channel from these local channels as
\begin{align}
\label{eq:noise}
    \cE_{\cG} = \bigotimes_{i,\alpha} \cE_{i,\alpha} \, .
\end{align}
This channel may physically be understood as the process of measuring all the gauge operators and then forgetting the measurement outcomes. This results in a state that is maximally mixed within the gauge subsystem.
Now, let $\ket{\psi}$ be an arbitrary state in the code space. The TSSC code naturally describes the mixed state
\begin{align}
\label{eq:Gchannel}
    \rho_{\cG} \equiv \cE_{\cG}[\ket{\psi}\bra{\psi}] \, .
\end{align}
By construction, this state satisfies the following relations:
\begin{subequations} \label{eq:rhoG_relations}
\begin{align}
    S \rho_{\cG} = \rho_{\cG} S^\dagger &= \rho_{\cG} \, , \qquad \forall S \in \cS \\
    O \rho_{\cG} O^\dagger &= \rho_{\cG} \, , \qquad \forall O \in \cG \, .
\end{align}
\end{subequations}
We see that the stabilizer and gauge groups are intimately related to the strong and weak symmetries of a mixed state; precisely this structure was exploited in Refs.~\cite{Sohal2025,Ellison2025} to show that $2+1$d TSSCs can be used to describe mixed-state topological orders. Note that if the logical space is empty, we can write
\begin{align}
    \rho_{\cG} \propto \prod_i \left( \sum_{m=0}^{n-1}  \, S_i^m \right) \, ,
\end{align}
which simply expresses the fact that, after maximal decoherence via the channel $\mathcal{E}_\mathcal{G}$, $\rho_{\cG}$ is the maximally mixed state within the code space.

We now show that a slightly modified TSSC construction proves convenient for describing symmetric mixed states. Given a symmetry group $G$ and a (normal) subgroup of strong symmetries $K \lhd G$, we define a $(G,K)$-symmetric TSSC by specifying a gauge group whose gauge operators commute with the strong symmetry $K$. We define the symmetric stabilizer group as the set of Pauli operators commuting both with elements of $\cG$ \emph{and} the total symmetry $G$: $\cS_{G} = \cZ_{\mathcal{P}}(\cG \cup G)$ (see Eq.~\eqref{eq:mixedgroupext} for the definition of the total symmetry).

Now, in the same vein that the SymTFT for a pure-state $G$-symmetric system can be used to motivate a stabilizer Hamiltonian (for Abelian $G$) which realizes all phases determined by the symmetry group $G$ (see Sec.~\ref{sec:genreview}), we can use the SymTaco for a $G$-symmetry to motivate a $G$-symmetric TSSC which describes all mixed-state phases determined by the $G$ symmetry. Following the above discussion, this in turn allows us to construct explicit channels to generate these states. The relevant information is again provided by the set of condensed anyons, which correspond to the elements of the gauge group $\cG$ in the same manner that they corresponded to terms in the Hamiltonian in the pure-state setting. In particular, we can see from Eq.~\eqref{eq:rhoG_relations}, that the gauge operators are associated to \emph{interlayer} anyon condensation in the SymTaco (i.e. the condensed anyons with support on both the ket and bra space), while the stabilizers are associated to \emph{intralayer} anyon condensation (i.e. the condensed anyons with support on either the ket or the bra space).

We illustrate this subsystem code formalism for $1+1$d mixed-state phases discussed above using three examples of increasing complexity: (i) the SWSSB state, (ii) the decohered cluster state, and (iii) the $\mbb{Z}_4$ intrinsically mixed SPT state. 

\paragraph*{Example 1: $\mbb{Z}_2$ SWSSB.}
The first example is the SWSSB of a strong $\mbb{Z}_2$ symmetry. This phase, together with a representative density matrix, was discussed in Sec.~\ref{sec:swssb}. We consider a system with a strong $\mbb{Z}_2$ symmetry, for which the bulk SymTaco is given by $\mD(\mbb{Z}_2^+ \times \mbb{Z}_2^-)$. From the SymTaco perspective, the SWSSB phase arises from the Choi state condensation
\begin{align}
    \mathcal L_1 = \langle e_+ e_- \,,  m_+ m_- \rangle \, .
\end{align}
Using the heuristic identifications in Eq.~\eqref{eq:anyontostabilizer}, we add $Z_i Z_{i+1}$ (from $e_+ e_-$) and $X_i$ (from $m_+ m_-$) to the gauge group $\mathcal G = \langle Z_i Z_{i+1}, X_i \rangle$. The center $\cZ(\cG) = \cS = \langle \prod_i X_i \rangle$ is generated by the operator $\prod_i X_i$. 
Following the protocol described above, we find that the resulting density matrix is the maximally mixed state within the code subspace, 
\begin{align}
    \rho_\cG \propto  1 + \prod_i X_i \, ,
\end{align}
which exhibits SWSSB of the $\mbb{Z}_2$ symmetry as reviewed in Sec.~\ref{sec:swssb}.

\paragraph*{Example 2: $\mbb{Z}_2^{strong} \times \mbb{Z}_2^{weak}$ Cluster State.} Next, consider a system with $G = \mbb{Z}_2^A \times \mbb{Z}_2^B$ symmetry, with a strong subgroup $K = \mbb{Z}_2^B$. This system is defined on a one-dimensional chain with two qubits per unit cell, labeled by $A$ and $B$. The two symmetries are generated by $\prod_i X_i^{A,B}$. Then, the bulk SymTaco is given by the Choi state TO $\mD(\mbb{Z}_2^{A+} \times \mbb{Z}_2^{B+} \times \mbb{Z}_2^{A-} \times \mbb{Z}_2^{B-})$, and the dynamical boundary condition is specified by the condensation of
\begin{equation}
\begin{aligned}
    \mathcal{L} = \langle & e_+^A m_+^B \, , e_-^A m_-^B \, , 
     \\ &e^A_+ e^A_- \, ,  (m^A e^B)_+ (m^A e^B)_- \rangle  \, .
\end{aligned}
\end{equation}
It can be checked that this anyon condensation satisfies Eqs.~\eqref{eq:m_constraints_1}-\eqref{eq:m_constraints_3}, and is therefore positive. This pattern of anyon condensation can be obtained by condensing $e^A m^B$ within a single layer of $\mathcal{D}(\mbb{Z}_2^A \times \mbb{Z}_2^B)$ and then folding to obtain the SymTaco. 

From the set of condensed anyons, we can again read off the generators of the gauge group $\cG$ using the heuristic correspondence in Eq.~\eqref{eq:anyontostabilizer}. We obtain
\begin{align}
    \mathcal{G} = \langle Z_i^A Z_{i+1}^A X^B_{i}   \, , Z^B_i Z^B_{i+1} X^A_{i+1} \, , Z^A_i \,  \rangle \, .
\end{align}
The first generator corresponds to the condensed anyons $e^A_+ m^B_+$ and $e^A_- m^B_-$, while the second generator corresponds to $(m^A e^B)_+ (m^A e^B)_-$. The third generator corresponds to $e^A_+ e^A_-$, which reduces the $\mbb{Z}_2^A$ symmetry group to a weak symmetry\footnote{Here, we end up with three generators of $\cG$ rather than four, since the generator $Z_i^A Z_{i+1}^A X^B_i$, which ends up being a stabilizer, accounts for two of the generators of $\mathcal L$.}. We choose $Z_i$ in the gauge group to break the symmetry explicitly rather than spontaneously. This set of gauge operators can be understood as starting from the familiar $\mbb{Z}_2 \times \mbb{Z}_2$ cluster state stabilizer code and adding $Z^A_i$ decoherence to break $\mbb{Z}_2^A$ to a weak symmetry. The set of gauge operators defined above commutes with the strong symmetry $K = \mbb{Z}_2^B$.

On a one-dimensional chain with periodic boundary conditions, the center is
\begin{align}
    \mathcal{S}_G = \mathcal{S} = \langle Z_i^A Z_{i+1}^A X^B_i \rangle \, ,
\end{align}
which commutes with the total symmetry $G = \mbb{Z}_2^A \times \mbb{Z}_2^B$. After decohering a state in the code-space via the channel in Eq.~\eqref{eq:noise}, we obtain the corresponding density matrix of the mixed-state SPT:
\begin{equation}
    \rho_\cG \propto \prod_i (1 + Z_i^A Z_{i+1}^A X^B_i ) \, .
\end{equation}
This is precisely the state studied in Ref.~\cite{Ma2023,Ma2025}. Let us see how the symmetry leads to a symmetry-protected degeneracy when the system is placed on open boundary conditions. For a chain with $L$ sites labeled $0 \ldots L-1$ with open boundary conditions, the symmetric stabilizer group is 
\begin{equation}
    \mathcal{S}_G^{o.b.c} = \langle Z_i^A Z_{i+1}^A X^B_i\rangle, \; i \in \{0, \ldots L-2\} \, ,
\end{equation}
while the full stabilizer group is
\begin{equation}
    \cS^{o.b.c} = \cS_G^{o.b.c} \cup \langle Z_0^A,  Z_{L-1}^A X_{L-1}^B \rangle \, .
\end{equation}
The non-symmetric stabilizers $Z_0^A$ and $Z^A_{L-1} X^B_{L-1}$ in $\cS$ that are not in $\cS_G$ become protected classical bits, unaffected by the application of the channel Eq.~\eqref{eq:noise}. The fixed-points (steady states) of \eqref{eq:Gchannel} are 
\begin{multline}
    \rho_\cG^{\pm \pm} \propto  \left( \prod_{i=0}^{L-2} (1 + Z^A_i Z^A_{i+1} X^B_i) \right) \\ \times  (1 \pm Z^A_0)  (1 \pm  Z^A_{L-1} X^B_{L-1}) \, ,
\end{multline}
together with their convex sums. The extremal points above correspond to the distinct values of the classical bits $Z_0^A$ and $ Z^A_{L-1} X^B_{L-1}$. This generalizes the notion of locally indistinguishable mixed states from the topologically ordered case.

\paragraph*{Example 3: $\mbb{Z}_4$ Intrinsically Mixed SPT}

We now return to the main example exemplifying the results of this paper: the $\mbb{Z}_4$ iASPT, reinterpreted through the lens of symmetric subsystem codes. We work on a periodic chain with four-dimensional qudits at each site and impose the total $\mbb{Z}_4$ symmetry generated by $\prod_i \cX_i$ and a strong subgroup $\prod_i \cX_i^2$. Let us consider the TSSC specified by the gauge operators 
\begin{equation}
    \mathcal{G} = \langle \cZ_i^2, \cX_i^2, \cZ^\dagger_{i-1} \cZ_i \cX_i \rangle \, . \label{eq:z4_gauge}
\end{equation}
The generators $\cZ_i^2$, $\cX_i^2$, and $\cZ^\dagger_{i-1} \cZ_i \cX_i$ correspond to the condensed anyons in the Choi space $e^2_+ e_-^2$, $m_+^2 m_-^2$, and $e_+ m_+ \overline{e}_- \overline{m}_-$, respectively. The operators in the gauge group are chosen to ensure that they commute with the strong $\mbb{Z}_2$ symmetry, but they can violate the total $\mbb{Z}_4$ symmetry. The stabilizer group, which is given by the center of the group of gauge operators, is 
\begin{equation} 
    \cS = Z(\mathcal{G}) = \langle \cZ_{i}^2 \cX_{i}^2 \cZ_{i+1}^2 \rangle \, .
\end{equation}
Each stabilizer commutes with the total $\mbb{Z}_4$ symmetry. 

The mixed-state SPT, obtained via the decoherence procedure provided earlier in this Section, is the maximally mixed state within the subspace where all of the stabilizers are $+1$. This is given by the density matrix 
\begin{equation} \label{eq:z4mspt}
    \rho_\cG \propto \prod_i (1 + \cZ^2_{i-1} \cX^2_{i-1} \cZ^2_i ) \, ,
\end{equation}
which is the projector onto the $+1$ eigenspace of all of the stabilizers. This density matrix was previously introduced in Sec.~\ref{sec:z4iaspt}, where we also saw how the universal data of symmetry fractionalization was realized on the density matrix $\rho_\cG$. Here, we us explore a different feature of this intrinsically mixed SPT: its properties on a system with open boundary conditions. 

Suppose the model is defined on a chain of length $L+1$, with sites ranging from $0$ to $L$. The gauge operators are 
\begin{equation}
    \mathcal G^{o.b.c} = \langle \cZ_i^2, \cX_i^2, \cZ_i^\dagger \cZ_{i+1} \cX_i, \cZ_L^2, \cX_L^2 \rangle \, ,
\end{equation}
for $i$ ranging from $0$ to $L-1$. We have only included gauge operators from Eq.~\eqref{eq:z4_gauge} whose support is completely contained within $0$ and $L$. Due to the removal of certain generators, the center becomes
\begin{equation}
    Z(\mathcal G^{o.b.c}) = \langle \cZ^2_i \cX^2_i \cZ^2_{i+1}, \cZ_0^2 \cX_0^2, \cZ_L^2 \rangle \, ,
\end{equation}
for $i$ ranging from $0$ to $L-1$. Compared to the case of periodic boundary conditions, the center contains the Pauli operators $\cZ_0^2 \cX_0^2$ and $\cZ_L^2$, which are localized at the boundaries and are not symmetric under the total $\mbb{Z}_4$ symmetry.

We take our stabilizers to be the subgroup of $Z(\mathcal G^{o.b.c})$ respecting the weak symmetry,
\begin{equation}
    \mathcal{S}_G = \langle \cZ_i^2 \cX_i^2 \cZ_{i+1}^2 \rangle \, , \quad i \in \{0, \ldots L-1 \} \, .
\end{equation}
Here, $\cZ^2_0 \cX_0^2$ and $\cZ_L^2$ are non-symmetric stabilizers which commute with the stabilizers and preserve the code subspace. On open boundary conditions, this subsystem code falls short of encoding any logical qubits however, since there is no operator in $Z(\mathcal G^{o.b.c})$ which anti-commutes with $\cZ_L^2$. Instead, this symmetric subsystem code protects two classical bits, given by the eigenvalues of $\cZ_0^2 \cX_0^2$ and $\cZ_L^2$. This manifests as a degenerate manifold of steady state density matrices of \eqref{eq:Gchannel}, with extremal points 
\begin{equation}
    \rho^{\pm\pm} \propto \left( \prod_{i=0}^{L-1} (1+\cZ^2_i \cX_i^2 \cZ_{i+1}^2) \right) (1  \pm \cZ_0^2\cX_0^2) (1 \pm \cZ_L^2) \, 
\end{equation}
corresponding to $\pm 1$ eigenvalues of the non-symmetric stabilizers $\cZ_0^2 \cX_0^2$ and $\cZ_L^2$.


\subsection{Preparing iASPTs from igSPTs}
\label{sec:prepare}

As a final application, we discuss how iASPTs can be directly generated via local decoherence of igSPTs. This was first discussed in Ref.~\cite{Ma2025}--here we emphasize the SymTFT perspective on this process. Intuitively, it should be clear that such a protocol exists since $G$-symmetric igSPTs and iASPTs share the same fractionalization data, as we have made clear via the \staco, even though they define distinct phases of matter--igSPTs have algebraic correlations while iASPTs are short-range correlated, meaning that they cannot be connected via locality-preserving finite-depth quantum channels. Nevertheless, the correspondence established via the \staco\, suggests that iASPTs arise from maximal decoherence within the low-energy subspace of the intrinsically gapless SPT. Physically, this follows from the fact that mixed-state phases in the \staco\, are given by the anyon condensation of $a_+ \bar{a}_-$ in the Choi state of a daughter TO, which is obtained by partial anyon condensation within a parent TO. The first step--the initial partial anyon condensation--determines the fractionalization data of an igSPT, and the second step--maximal decoherence--implements the $a_+ \bar{a}_-$ anyon condensation in the Choi state to generate the iASPT. This process of maximal decoherence eliminates the algebraic correlations present in the bulk of the igSPT in a manner that leaves the fractionalization data intact (parallel to our discussion of generating folded SPTs from igSPTs in Sec.~\ref{sec:folddouble}). In the Abelian case, described in the preceding subsection, the explicit form of the operators used to induce this decoherence on the low-energy subspace are provided using Eq.~\eqref{eq:anyontostabilizer}. While the correspondence between mixed states and symmetric subsystem codes may not hold, 
based on the SymTaco correspondence we expect that a similar set of decoherence operators can be obtained even in the non-Abelian setting to generate iASPTs from igSPTs.


\section{Conclusions and Future Directions}
\label{sec:cncls}

In this work, we have taken the first step towards extending the SymTFT framework to encompass $1+1$d mixed-state phases of matter. Building on the holographic correspondence central to the SymTFT, we have introduced the \textit{symmetry taco}--a folded, bilayer $2+1$d topological order constructed from a generic TO $\mathcal{C}$ and its condensable subalgebras $\mA$--which provides a unified bulk object for encoding gapped, gapless, and mixed-state symmetry-protected topological (SPT) phases in $1+1$d systems. Here, we have focused in particular on $1+1$d systems with finite $G$-symmetries and developed the tetraptych of correspondences between condensable subalgebras in the bulk $\mD(G)$ TO, folded $G\times G$ SPTs, intrinsically gapless $G$-SPTs, and intrinsically average $G$-SPTs (see Fig.~\ref{fig:main}).

Strikingly, the \staco\, naturally encodes the positivity and Hermiticity constraints on physical density matrices, opening a new route towards a categorical description of mixed-state phases and revealing new connections between ostensibly distinct topological states. As we have argued, igSPTs and iASPTs can be understood as partially gapped or decohered boundaries of the same \staco, tied to the same condensable algebra. This shared origin provides not only a formal and conceptual identification of these phases, but also a recipe for obtaining new mixed-state phases from familiar pure-states. Our construction thus illustrates how folded SPTs can be generated via controlled deformations of igSPT Hamiltonians, and how iASPTs can be generated via the local decoherence of igSPTs.

Alongside establishing the tetraptych of Fig.~\ref{fig:main}, as applications of the \staco\, we have also obtained a new mixed-state anomaly, provided a general prescription for generating iASPTs from igSPTs, and, for Abelian $G$, provided a description of iASPTs as symmetric Pauli subsystem codes. The \staco\, opens many exciting avenues for further exploration, particularly for discovering new mixed-state phases and transitions between them. As a first step, we expect that generalizing the \staco\, to $1+1$d mixed states invariant under categorical symmetries $\mathcal{C}$ will shed light on the general algebraic structure of strong and weak non-invertible symmetries, and reveal new mixed-state SPTs as well as new mixed-state anomalies. For example, invariance of a density matrix under a strong non-invertible symmetry $\mathcal{C}$ \textit{does not} generally imply that the state is also invariant under a weak non-invertible symmetry $\mathcal{C}$. It would be interesting to understand the physical consequences of this symmetry structure in full via the \staco.

Here, we have restricted our discussion to mixed-state SPTs that appear in the context of decohered or open quantum systems, but another rich arena is that of mixed-state SPTs arising within disorder ensembles. However, the distinct physical setting of disordered systems imposes additional constraints which would need to be taken as further input to the SymTaco. Specifically, when considering decohered systems, any basis decomposition of the density matrix is valid when computing correlation functions; in contrast, for disordered systems, such calculations must be performed in the basis of Hamiltonian ground states since only this choice corresponds to physical ground states of some random Hamiltonian in the ensemble~\cite{Ma2025}. It remains to be seen how this additional data governing ground states of disordered quantum systems enters into the SymTaco perspective and restricts our classification to that previously obtained in Ref.~\cite{Ma2025}.

A particularly intriguing direction for future research is investigating $G$-symmetric mixed-state phases which provide analogues of gapless phases or critical points which, in the pure-state context, have been investigated from a SymTFT perspective. One instance in which such mixed states could appear is on the boundaries of intrinsically mixed-state TOs. However, it is not yet understood how the bulk anomalous 1-form symmetries characterizing mixed-state TOs manifest on their boundaries; for instance, we suspect the boundary of the $f$-decohered Toric Code is likely described by a random-singlet Majorana phase. More generally, it remains to be seen what physical consequences, if any, can be inferred for the boundary states of intrinsically mixed-state TOs from the SymTaco picture and whether there are explicit connections to be made with infinite randomness fixed-points of anyon chains~\cite{bonesteel2007,fidkowski2008prb}. From the opposite direction, an exciting open question is also to understand the mixed-state phases or mixed-state critical points for which intrinsically mixed-state TOs provide the appropriate bulk SymTFT. We remark that although we have focused on $1+1$d $G$-symmetric phases here, several elements of our construction extend naturally to higher dimensions, which we leave to future work. We expect that the subsystem code formalism we have developed for $G$-symmetric mixed states generalizes mutatis mutandis to higher dimensions and could shed light on igSPTs in higher dimensions, for which a comprehensive framework is lacking.

The mixed states we have considered here belong to well-defined phases of matter only in the context of finite-time noisy evolution. From a practical and fundamental perspective however, the most natural setting in which mixed-state phases appear are as steady-states of open system dynamics. While there has been partial progress in characterizing topologically non-trivial steady-state phases of locally generated non-unitary dynamics~\cite{liu2024,rakovszky2023stable,chirame2024spt,chirame2024to,davydova2024local,shah2024instability}, classifying steady-state phases even in the restricted setting of local Lindbladian dynamics remains a challenge. It is thus imperative to extend the SymTFT perspective to the dynamical realm.

Finally, the realization of intrinsically mixed-state phases, especially iASPTs, in NISQ-era devices remains a pressing goal. Our identification of large families of such phases with Pauli subsystem codes suggests a concrete path forward. This is particularly true in systems with engineered noise channels, where these phases could be obtained via local decoherence of igSPTs or by using the explicit quantum channels that our prescription in Sec.~\ref{sec:subsystem} provides.


\stoptoc
\begin{acknowledgements}
We thank Michael Hermele for collaboration during the initial stages of the project,  Ruochen Ma for numerous insightful discussions, and Ho Tat Lam for a related collaboration. A. P. is grateful to Corey Jones, Po-Shen Hsin, and Ryohei Kobayashi for stimulating discussions on related topics. 
The research of M.Q. (at University of Colorado Boulder, prior to August
2024) is supported by the U.S. Department of Energy, Office of Science, Basic Energy Sciences (BES) under Award number DE-SC0014415. Research of M.Q. after August 2024 and prior to February 2025 was supported, in part, by the Physical Science Division at the University of Chicago. Research of M.Q. (after February 2025) and D.T.S. was supported by the Simons Collaboration on Ultra-Quantum Matter, which is a grant from the Simons Foundation [Nos. 651442 (M.Q.), 651440 (D.T.S.)]. 
The work of R.S. was in part supported by a Simons Investigator Grant from the Simons Foundation (Grant No. 566116) awarded to Shinsei Ryu.
X.C. is supported by the Simons collaboration on `Ultra-Quantum Matter'' (grant number 651438), the Simons Investigator Award (award ID 828078), the Institute for Quantum Information and Matter at Caltech, and the Walter Burke Institute for Theoretical Physics at Caltech. This material is based upon work supported by the Sivian Fund and the Paul Dirac Fund at the Institute for Advanced Study and the U.S. Department of Energy, Office of Science, Office of High Energy Physics under Award Number DE-SC0009988 (A.P.). Part of this work was conducted while M. Q. and A. P. were visiting the Okinawa Institute of Science and Technology (OIST) through the Theoretical Sciences Visiting Program (TSVP).
\end{acknowledgements}

\emph{Note.---}We thank the authors of Ref.~\cite{sakuramixed} and Ref.~\cite{yinanmixed} for coordinating submission of their related independent works, which will appear on the arXiv in the same posting.


\appendix
\resumetoc

\section{Notes on Notation}
\label{app:notes}

Depending on the context, we will find it more convenient to view the bulk TO in the symmetry taco equivalently as $\mD(G \times G)$, $\mD(G) \boxtimes \mD(G)$, or $\mD(G) \boxtimes \overline{\mD(G)}$, all of which are isomorphic to each other. When deriving constraints on folded SPTs and when considering Choi states, the $\mD(G) \boxtimes \overline{\mD(G)}$ viewpoint is natural; when making the correspondence to $1+1$d pure state SPTs, the $\mD(G\times G)$ viewpoint turns out to be more natural. Unfortunately, this can sometimes lead to confusion when it is unclear which perspective is being used. We therefore take a moment to clarify the notation that is used throughout the paper. 

The first two quantum doubles, $\mD(G \times G)$ and $\mD(G) \boxtimes \mD(G)$, are canonically isomorphic. In either case, we will adopt one of the following anyon labeling schemes. Let $a$ and $b$ denote anyons of $\mD(G)$. We will denote anyons of $\mD(G \times G)$ and $\mD(G) \boxtimes \mD(G)$ as an ordered pair of anyons $(a,b)$, or with a ``flavor" $A/B$ index $a_A b_B$. 

On the other hand, $\mD(G) \boxtimes \overline{\mD(G)}$ is isomorphic, but not canonically so, to $\mD(G \times G)$ and $\mD(G) \boxtimes \mD(G)$. 
For anyons $a, b \in \mD(G)$, we label anyons of $\mD(G) \boxtimes \overline{\mD(G)}$ as $a_+ b_-$. Anyons in $\overline{\mD(G)}$, with the $-$ label, have opposite (complex conjugate) braiding to their counterparts in $\mD(G)$, with the $+$ label. 

Choosing an isomorphism between these two amounts to choosing an isomorphism between $\mD(G)$ to $\overline{\mD(G)}$, sending $a$ to $a_-$. Such an isomorphism must preserve the fusion rules while all of the braiding phases are complex conjugated. Let $a = ([k], \pi)$ be an anyon of $\mD(G)$. There are two possibilities for the isomorphism $a \to a_-$. The first is to assign $a_- = ([k^{-1}], \pi)$, and the second is to assign $a_- = ([k], \pi^*)$. 

\section{Details on Condensable Algebras and Condensed Anyons}
\label{app:nonabelian}

The complete data describing a condensable algebra $\mathcal A$ in a quantum double $\mD(G)$ is given by Eq.~\eqref{eq:condensationdata}, for both Abelian and non-Abelian $G$. In many cases, however, it is more helpful and intuitive to simply keep track of which anyons have been condensed, along with their multiplicities. In other words, we want the decomposition 
\begin{equation}
\label{eq:anyondecomp}
    \mathcal A = \bigoplus_a n_a a \, ,
\end{equation}
of the condensable algebra $\mathcal A$ into anyons $a$ together with their multiplicities $n_a$.

In general, the data describing the set of condensed anyons is coarser than the full data of the algebra $\mathcal A$ since the multiplication map is not manifest in Eq.~\eqref{eq:anyondecomp}~\cite{davydov2014}. Nevertheless, this decomposition makes transparent certain aspects of the condensation, such as which anyons can be moved freely into the boundary from the bulk TO. Moreover, in certain instances the full data of the algebra can indeed be derived entirely from Eq.~\eqref{eq:anyondecomp} and other knowledge of $\mD(G)$; the case when $G$ is Abelian is one such instance. In this Appendix, we describe how to obtain the decomposition Eq.~\eqref{eq:anyondecomp} starting from the data in Eq.~\eqref{eq:condensationdata}. To do so, we will review the description of $\mD(G)$ as a category of representations of a certain Hopf algebra called $DG$. A central tool will be the theory of characters, which we will outline below, following Ref.~\cite{Beigi2011}.

Let us recall the mathematical description of $\mD(G)$ as a unitary modular tensor category. We will view $\mD(G)$ as the category of representations of a Hopf algebra $DG$, known as the quantum double of $G$. As a vector space, $DG = \mbb{C}[G] \otimes \mbb{C}[G]^*$ is $|G|^2$-dimensional and is spanned by elements $gh^*$. The multiplication on $DG$ is
\begin{equation}
\label{eq:DGmult}
    (g_1 h_1^*) (g_2 h_2^*) = \delta_{h_2, g_2^{-1} h_1 g_2} (g_1 g_2) h_2^* \,.
\end{equation}
$DG$ contains both $\mbb{C}[G]$ and $\mbb{C}[G]^*$ as subalgebras. The natural inclusion of $\mbb{C}[G] \to DG$ is $g \mapsto \sum_{h \in G} g h^*$, and the natural inclusion of $\mbb{C}[G]^* \to DG$ is $h^* \mapsto e h^*$. For convenience, we will identify $g \in DG$ and $h^* \in DG$ with their images under the respective inclusions. One can then check that 
\begin{equation}
    h^* g = g (g^{-1} h g)^* \,.
\end{equation}
The multiplicative unit of $DG$ is $e = \sum_h eh^*$. 

$\mD(G)$ is the category of finite-dimensional representations of $DG$. A representation $(V, \rho)$ of $DG$ can be succinctly described as a $G$-graded vector space with a compatible $G$ action. In other words, $V$ is equipped with a $G$-grading
\begin{equation}
V = \bigoplus_{k \in G} V_k
\end{equation}
and a $G$-action $\rho$ satisfying $\rho(g) V_k = V_{g k g^{-1}}$. To extend $\rho$ to a $DG$ action, we define $\rho(h^*)$ to be the projector onto the $V_h$ subspace and $\rho(gh^*) = \rho(g) \rho(h^*)$. We see that the compatibility condition ensures that $(V, \rho)$ is indeed a representation of $DG$. 

Representations of $DG$ decompose as direct sums into irreducible representations. These irreducible representations are the simple objects of the category $\mD(G)$, and label the distinct anyon types (or superselection sectors) in $\mD(G)$ topological order. Irreducible representations of $DG$ correspond to pairs $([k], \pi)$, with $[k]$ a conjugacy class and $\pi$ an irreducible representation of the centralizer $Z(k)$ of $k$. For such a pair, there is a $DG$ representation $(V_{([k], \pi)}, \rho)$
given by the induced representation of $G$. It can be defined in a basis-independent way as 
\begin{equation}
    V_{([k],\pi)} = \mbb{C}[G] \zotimes W \, ,
\end{equation}
where $W$ is the space on which $\pi$ acts. The tensor product $\zotimes$ means that elements of $\mbb{C}[Z(k)]$ can be moved between the factors to act on $\mbb{C}[G]$ or $W$. In other words, we have the relation
\begin{equation} \label{eq:tensor_over_z}
    g z \zotimes w = g \zotimes \pi(z) w
\end{equation} 
for any $z \in Z(k)$. The $G$-action is given by left-multiplication as 
\begin{equation}
    \rho(g) (\tau \otimes w) = g \tau \zotimes w \, ,
\end{equation}
while the $G$-grading is given by
\begin{equation}
    \rho(h^*) (\tau \otimes w) = \delta_{h,\tau k \tau^{-1}} \tau \zotimes w \, .
\end{equation}

To specify a convenient basis for $V_{([k], \pi)}$, fix a basis $\{w_i \}$ of $W$, and an element $\tau_a \in G$ for every element of the conjugacy class $a \in [k]$ such that $a = \tau_a k \tau_a^{-1}$ and $\tau_k = e$. Then a basis for $V_{([k], \pi)}$ is given by
\begin{equation}
\label{eq:repbasis}
    | \tau_a, w_i \rangle \coloneqq \tau_a \zotimes w_i \, .
\end{equation}
In this basis, the action of $g$ is given by 
\begin{equation}
\begin{aligned}
    \rho(g) | \tau_a, w_i \rangle &= g \tau_a \zotimes w_i \\
    &= \left(\tau_{g a g^{-1}} \tau^{-1}_{g a g^{-1}} \right) g \tau_a \zotimes w_i \\
    &= \tau_{g a g^{-1}} \zotimes \pi(\tau^{-1}_{g a g^{-1}} g \tau_a) w_i \\
    &= | \tau_{g a g^{-1}}, \pi(\tau^{-1}_{g a g^{-1}} g \tau_a) w_i  \rangle \, ,
\end{aligned}
\end{equation}
where the third line follows from the relation Eq.~\eqref{eq:tensor_over_z} and the fact that $\tau^{-1}_{g a g^{-1}} g \tau_a \in Z(k)$. The grading is 
\begin{equation}
    \rho(h^*) | \tau_a , w_i \rangle = \delta_{h,a} | \tau_a, w_i \rangle \, .
\end{equation}
The dimension of the irreducible representation is $|[k]| |Z(k)|$, which specifies the the quantum dimension of the anyon $([k], \pi)$. 

Let us now discuss the character theory of representations of $DG$. Characters provide an efficient method of determining the decomposition of a representation into irreducibles. Given a representation $(V, \rho)$, its character $\chi_V: DG \to \mbb{C}$ is 
\begin{equation}
    \chi_V (g h^*) = \text{Tr}_V (\rho(gh^*)) \,.
\end{equation}
It can be shown, due to the compatibility condition between the $G$-grading and the $G$-action, that $\chi_V(g h^*)$ vanishes unless $gh = hg$. Direct computation using the basis \eqref{eq:repbasis} gives 
\begin{equation} \label{eq:anyon_character}
    \chi_{([k], \pi)}(g h^*) = \delta_{h \in [k]} \delta_{gh,hg} \text{Tr}_\pi (\tau_h^{-1} g \tau_h) \,.
\end{equation}
Characters are useful in determining the direct sum decomposition of a particular representation into irreducible representations. This is because characters of irreducible representations satisfy an orthogonality relation. Define the inner product on characters to be
\begin{equation} \label{eq:character_innerproduct}
    \langle \chi_1, \chi_2 \rangle = \frac{1}{|G|} \sum_{g,h} \overline{\chi_1(g h^*)} \chi_2 (g h^*)
\end{equation}
for representations $(V_1, \rho_1)$ and $(V_2, \rho_2)$. If $V_1$ and $V_2$ are irreducible, the orthogonality relation for characters guarantees that the inner product of $\chi_1$ and $\chi_2$ gives $1$ if $(V_1, \rho_1) \cong (V_2, \rho_2)$ as irreducible representations and $0$ otherwise. Similarly, if $V_1$ is reducible while $V_2$ is irreducible, then the inner product of characters $\chi_1$ and $\chi_2$ computes the multiplicity of $V_2$ within $V_1$. 

Let us compute the irrep decomposition for the condensable algebras encountered in the main text. Following Ref.~\cite{Davydov2009}, a general condensable algebra is specified by: 
\begin{enumerate}
    \item a subgroup $H \subset G$,
    \item a normal subgroup $A \subset H$, 
    \item a $2$-cocycle $\omega: A \times A \to U(1)$, and 
    \item a function $\epsilon: A \times H \to U(1)$ which satisfies the conditions given in Eqs.~\eqref{eq:condensationdata}.
\end{enumerate}
From the above data, an algebra $\mathcal A$ with a $G$-action $\rho(g)$ and $G$-grading $\rho(h^*)$ is defined as follows: $\mathcal A$ is generated by a set $\{e_{x,a}\}$ (for $x \in G$, $a \in A$) subject to the relations
\begin{align}
    e_{xy, a} = \epsilon_a(y) e_{x, yay^{-1}}, \; \epsilon_a(y) \coloneqq \epsilon(a, y) 
\end{align}
for $y \in H$. This expression makes sense since $yAy^{-1} = A$, given that $A$ is a normal subgroup of $H$.  
Picking a set of (left) coset representatives $\{x_i\}$ for $G/H$ (recall that $H$ is not necessarily a normal subgroup of $G$), we can define an orthonormal basis $e_{x_i, a}$. As a vector space $\mathcal A$ is $|G/H| |F|$-dimensional.
The $G$-grading and $G$-action on $\mathcal A$ are given by 
\begin{align}\label{eq:A_g_grading_action}
\begin{split}
    \rho(h^*) e_{x_i, a} &= \delta_{h, x_i a x_i^{-1}} e_{x_i, a} \, , \\
    \rho(g) e_{x_i, a} &= e_{gx_i, a} \, .
\end{split}
\end{align}
Since $G = \bigcup_i x_i H$, we can find an $x_{i'}$ and $y \in H$ such that $gx_i = x_{i'} y$. The $G$-action then becomes  
\begin{equation} \label{eq:A_g_action}
    \rho(g) e_{x_i, a} = e_{g x_i, a} = \epsilon_a(y) e_{x_{i'}, y a y^{-1}}.
\end{equation}
The character of the condensable algebra $\mathcal A$ is given by 
\begin{widetext}
\begin{equation} \label{eq:algebra_character}
\begin{aligned}
    \chi_{\mathcal A}(gh^*) &= \sum_{x_i} \sum_{a \in A} \langle e_{x_i, a} | \rho(gh^*) | e_{x_i, a} \rangle \\
    &= \sum_{x_i}  \sum_{a \in A} \delta_{h, x_i a x_i^{-1}} \delta_{x_i, x_{i'}} \delta_{a, yay^{-1}} \epsilon_a (y), \quad gx_i = x_{i'} y \text{  for  } y \in H \\
    &= \sum_{x_i} \delta_{x_i^{-1} h x_i \in A} \delta_{x_i^{-1} g x_i \in H} \delta_{hg,gh} \epsilon(x_i^{-1} h x_i, x_i^{-1} g x_i) \\
    &= \delta_{gh,hg} \frac{1}{|H|} \sum_{x \in G} \delta_{x h x^{-1} \in A} \delta_{x gx^{-1} \in H} \epsilon(x h x^{-1}, x g x^{-1}) \, .
\end{aligned}
\end{equation}
In going from the first line to the second, we applied the definition of the $DG$ action \eqref{eq:A_g_grading_action} on $\mathcal A$. In the third line, we performed the sum over $a \in A$. Finally, we replaced the sum over coset representatives with a sum over the entire group, compensating for the over counting by introducing the $1/|H|$ prefactor. The expression for the character simplifies nicely when $H = G$, i.e., when no charges are condensed. In this case, the character is 
\begin{equation}
    \chi_\mathcal{A} (gh^*) = \delta_{gh,hg} \delta_{h \in A} \epsilon_h(g) \,.
\end{equation}

The multiplicities of the condensed anyons can be found by using the orthogonality relation on characters. The multiplicity of the anyon $([k], \pi)$ in $\mathcal A$ is given by 
\begin{equation} \label{eq:anyon_multiplicity}
\begin{aligned}
    \langle \chi_\mathcal{A}, \chi_{([k], \pi)} \rangle &= \frac{1}{|G||H|} \sum_{g,h} \sum_x \delta_{gh,hg} \delta_{xhx^{-1} \in A} \delta_{xgx^{-1} \in H} \; \epsilon^*(xhx^{-1}, xgx^{-1}) \delta_{h \in [k]} \text{Tr}_\pi (\tau_h^{-1} g \tau_h) \\
    &= \frac{1}{|H|} \sum_{g,h} \delta_{gh,hg} \delta_{h \in A} \delta_{g \in H} \; \epsilon^*(h, g) \delta_{h \in [k]} \text{Tr}_\pi (\tau_h^{-1} g \tau_h) \\
    &= \frac{1}{|H|} \sum_{g \in Z(k)} \sum_{h \in [k] \cap A} \delta_{\tau_h g \tau_h^{-1} \in H} \; \epsilon^*(\tau_h k \tau_h^{-1}, \tau_h g \tau_h^{-1}) \text{Tr}_\pi (g) \\
    &= \frac{1}{|H|} \sum_{g \in Z(k)} \epsilon^*(k, g) \text{Tr}_\pi (g) \sum_{h \in [k] \cap A} \delta_{\tau_h g \tau_h^{-1}  \in  H} \, .
\end{aligned}
\end{equation}
\end{widetext}
Here, the first line comes from substituting Eqs.~\eqref{eq:anyon_character} and~\eqref{eq:algebra_character} into the inner product defined in Eq.~\eqref{eq:character_innerproduct}. To arrive at the second line, we perform a change of variables: $g \mapsto x^{-1} g x$, $h \mapsto x^{-1} h x$ to make the terms independent of $x$ before summing over $x \in G$. The third line arises from performing another change of variables: $g \mapsto \tau_h^{-1} g \tau_h$ and substituting $h = \tau_h k \tau_h^{-1}$. Finally, we rearrange the sum, making use of the fact that $\epsilon^*(\tau_h k \tau_h^{-1}, \tau_h g \tau_h^{-1}) = \epsilon^*(k, g)$ for commuting $k$, $g$ is independent of $h$. 

It is instructive to understand various limiting cases of this expression. First, let us consider the case where $A = H = G$, i.e., a Lagrangian subalgebra $\mathcal L$ where no charges are condensed. Eq.~\eqref{eq:anyon_multiplicity} then becomes
\begin{equation}
\begin{aligned}
    \langle \chi_\mathcal{L} , \chi_{([k], \pi)} \rangle &= \frac{|[k]|}{|G|} \sum_{g \in Z(k)} \epsilon^*(k, g) \text{Tr}_\pi(g) \\
    &= \frac{1}{|Z(k)|} \sum_{g \in Z(k)} \epsilon^*(k, g) \text{Tr}_\pi(g) \, ,
\end{aligned}
\end{equation}
which reduces to the overlap of $Z(k)$ characters for the representations $\epsilon(k, \cdot)$ and $\pi$. Since $\epsilon(k, \cdot)$ is a one-dimensional representation and $\pi$ is irreducible, their overlap can only be non-vanishing if $\pi$ is also a one-dimensional representation. Therefore, only one-dimensional charges $\pi$ can condense with the flux $[k]$. 

Next, consider the case where the condensable algebra $\mathcal A$ contains no charges but the condensation is non-maximal, which corresponds to the case that we encountered in Sec.~\ref{sec:gapless} in our analysis of gapless SPTs. Here, we take $H = G$ so that no charges are condensed, and $A \subset H$ a normal subgroup of $G$. The last sum in Eq.~\eqref{eq:anyon_multiplicity} is replaced by an overall constant factor $|[k] \cap A|$, and the expression becomes 
\begin{equation}
\begin{aligned}
    \langle \chi_{\mathcal{A}}, \chi_{([k], \pi)} \rangle &= \frac{|[k] \cap A|}{|G|} \sum_{g \in Z(k)} \epsilon^*(k, g) \text{Tr}_\pi(g) \\
    &= \begin{cases} 0 & k \notin A \\ \frac{1}{|Z(k)|} \sum_{g \in Z(k)} \epsilon^*(k, g) \text{Tr}_\pi (g) & k \in A\end{cases}.
\end{aligned}
\end{equation}
We see that an anyon $([k], \pi)$ can only condense if $k$ is contained in the normal subgroup $A$. Moreover, as before, $\pi$ must be a one-dimensional charge if it is condensed together with a flux $[k]$. 

Finally, we consider the case of a Lagrangian subalgebra $\mathcal L$ which describes spontaneous symmetry breaking to a normal subgroup $H \subset G$. Normality of $H$ simplifies the final sum in Eq.~\eqref{eq:anyon_multiplicity}, since $\delta_{\tau_h g \tau_h^{-1} \in H}$ is independent of $h \in [k] \cap A$. Performing the sum gives 
\begin{equation} \label{eq:normalH}
\begin{aligned}
    \langle \chi_\mathcal{A}, \chi_{([k], \pi)} \rangle &= \frac{|[k] \cap A|}{|H|} \sum_{g \in Z(k) \cap H } \epsilon^*(k, g) \text{Tr}_\pi(g) \\ 
    &= \frac{|Z(k) \cap H| \times |[k] \cap A|}{|H|} \\ & \quad \times \frac{1}{|Z(k) \cap H|} \sum_{g \in Z(k) \cap H } \epsilon^*(k, g) \text{Tr}_\pi(g) \, .
\end{aligned}
\end{equation}
The prefactor is an integer which counts the number of $H$-conjugacy classes contained in $[k] \cap A$. The second line computes the overlap of characters $\epsilon(k, \cdot)_{Z(k) \cap H}$ and $\pi|_H$ restricted to the subgroup $Z(k) \cap H$. Unlike the previous case, now an anyon $([k], \pi)$ can condense even when $\pi$ is not one-dimensional, as long as it splits into one-dimensional irreps containing $\epsilon(k, \cdot)$ upon restricting to $Z(k) \cap H$. 

This is the situation that we encounter in studying folded and mixed-state SPTs in Secs.~\ref{sec:gapped} and~\ref{sec:mixed}. There, we identify gapped boundaries of $\mD(G \times G)$ that arise from partial anyon condensation and folding. Given a condensable algebra $\mathcal A$ in $\mD(G)$ containing no charges, characterized by a normal subgroup $A \subset G$ and a charge assignment $\epsilon:A \times G \to U(1)$ (we take the 2-cocycle $\omega$ to be trivial for simplicity), we argued that the corresponding gapped boundary of $\mD(G \times G)$ is given by a certain subgroup $H = A \rtimes G$ and the 2-cocycle Eq.~\eqref{eq:nonabelian_omega}. Since $H$ is normal in $G \times G$, we can apply Eq.~\eqref{eq:normalH} to compute the anyons condensed in $\mathcal L$ with $H$ and $\omega$ as above. 


\section{Positivity Constraint on the Condensation Matrix}
\label{app:positivity}

In this Appendix, we will prove Eq.~\eqref{eq:m_constraints_3}. We split the proof into four cases: first, if $M_{a,\bar{b}}=M_{a,\bar{c}}=0$, then the inequality is trivially satisfied. Second, if $M_{a,\bar{b}}=1$ and $M_{a,\bar{c}}=0$, then Eq.~\eqref{eq:m_constraints_2} implies $M_{a,\bar{a}}=1$, and so the inequality reads $1\leq 1+M_{b,\bar{c}}$, which is always true. The same argument holds for the third case when $M_{a,\bar{b}}=0$ and $M_{a,\bar c}=1$. Finally, if $M_{a,\bar b}=M_{a,\bar c}=1$, the inequality reads $2\leq 1 + M_{b,\bar c}$, and we need to show that $M_{b, \bar c}=1$. Using Eq.~\eqref{eq:m_constraints_2}, $M_{a,\bar b}=1$ implies that $M_{a \times \bar b, 1} = 1$. We have $M_{a \times \bar c, 1} = 1$ for the same reason. Then, combining $a \times \bar b$ and $a \times \bar c$ using Eq.~\eqref{eq:m_constraints_1}, we have $M_{b \times \bar{c}, 1} = 1$. Since Eq.~\eqref{eq:m_constraints_2} implies $M_{c, \bar{c}} = 1$, we have $M_{b, \bar c} = 1$ by Eq.~\eqref{eq:m_constraints_1}, as desired. This completes the proof.

Conversely, if Eq.~\eqref{eq:m_constraints_1} and Eq.~\eqref{eq:m_constraints_3} hold, then we can show that Eq.~\eqref{eq:m_constraints_2} holds. When $b=c$, Eq.~\eqref{eq:m_constraints_2} reads $2M_{a, \bar b}\leq M_{a, \bar a}+M_{b, \bar b}$, which means that if $M_{a, \bar b}=1$, then $M_{a, \bar a}=M_{b, \bar b}=1$. Then Eq.~\eqref{eq:m_constraints_1} says that $M_{\bar{b},{b}}=1$, and $M_{a, \bar b}=M_{\bar{b},b}=1$ gives $M_{a \times \bar{b},1}=1$, which is precisely Eq.~\eqref{eq:m_constraints_2} after replacing $\bar{b}$ with $b$. 



\stoptoc
\bibliography{biblio}

\begin{thebibliography}{142}%
\makeatletter
\providecommand \@ifxundefined [1]{%
 \@ifx{#1\undefined}
}%
\providecommand \@ifnum [1]{%
 \ifnum #1\expandafter \@firstoftwo
 \else \expandafter \@secondoftwo
 \fi
}%
\providecommand \@ifx [1]{%
 \ifx #1\expandafter \@firstoftwo
 \else \expandafter \@secondoftwo
 \fi
}%
\providecommand \natexlab [1]{#1}%
\providecommand \enquote  [1]{``#1''}%
\providecommand \bibnamefont  [1]{#1}%
\providecommand \bibfnamefont [1]{#1}%
\providecommand \citenamefont [1]{#1}%
\providecommand \href@noop [0]{\@secondoftwo}%
\providecommand \href [0]{\begingroup \@sanitize@url \@href}%
\providecommand \@href[1]{\@@startlink{#1}\@@href}%
\providecommand \@@href[1]{\endgroup#1\@@endlink}%
\providecommand \@sanitize@url [0]{\catcode `\\12\catcode `\$12\catcode `\&12\catcode `\#12\catcode `\^12\catcode `\_12\catcode `\%12\relax}%
\providecommand \@@startlink[1]{}%
\providecommand \@@endlink[0]{}%
\providecommand \url  [0]{\begingroup\@sanitize@url \@url }%
\providecommand \@url [1]{\endgroup\@href {#1}{\urlprefix }}%
\providecommand \urlprefix  [0]{URL }%
\providecommand \Eprint [0]{\href }%
\providecommand \doibase [0]{https://doi.org/}%
\providecommand \selectlanguage [0]{\@gobble}%
\providecommand \bibinfo  [0]{\@secondoftwo}%
\providecommand \bibfield  [0]{\@secondoftwo}%
\providecommand \translation [1]{[#1]}%
\providecommand \BibitemOpen [0]{}%
\providecommand \bibitemStop [0]{}%
\providecommand \bibitemNoStop [0]{.\EOS\space}%
\providecommand \EOS [0]{\spacefactor3000\relax}%
\providecommand \BibitemShut  [1]{\csname bibitem#1\endcsname}%
\let\auto@bib@innerbib\@empty
\bibitem [{\citenamefont {Anderson}(1972)}]{anderson1972}%
  \BibitemOpen
  \bibfield  {author} {\bibinfo {author} {\bibfnamefont {P.~W.}\ \bibnamefont {Anderson}},\ }\bibfield  {title} {\bibinfo {title} {More is different},\ }\href {https://doi.org/10.1126/science.177.4047.393} {\bibfield  {journal} {\bibinfo  {journal} {Science}\ }\textbf {\bibinfo {volume} {177}},\ \bibinfo {pages} {393} (\bibinfo {year} {1972})},\ \Eprint {https://arxiv.org/abs/https://www.science.org/doi/pdf/10.1126/science.177.4047.393} {https://www.science.org/doi/pdf/10.1126/science.177.4047.393} \BibitemShut {NoStop}%
\bibitem [{\citenamefont {Chen}\ \emph {et~al.}(2012)\citenamefont {Chen}, \citenamefont {Gu}, \citenamefont {Liu},\ and\ \citenamefont {Wen}}]{chen2012spt}%
  \BibitemOpen
  \bibfield  {author} {\bibinfo {author} {\bibfnamefont {X.}~\bibnamefont {Chen}}, \bibinfo {author} {\bibfnamefont {Z.-C.}\ \bibnamefont {Gu}}, \bibinfo {author} {\bibfnamefont {Z.-X.}\ \bibnamefont {Liu}},\ and\ \bibinfo {author} {\bibfnamefont {X.-G.}\ \bibnamefont {Wen}},\ }\bibfield  {title} {\bibinfo {title} {Symmetry-protected topological orders in interacting bosonic systems},\ }\href {https://doi.org/10.1126/science.1227224} {\bibfield  {journal} {\bibinfo  {journal} {Science}\ }\textbf {\bibinfo {volume} {338}},\ \bibinfo {pages} {1604} (\bibinfo {year} {2012})},\ \Eprint {https://arxiv.org/abs/https://www.science.org/doi/pdf/10.1126/science.1227224} {https://www.science.org/doi/pdf/10.1126/science.1227224} \BibitemShut {NoStop}%
\bibitem [{\citenamefont {Chen}\ \emph {et~al.}(2013)\citenamefont {Chen}, \citenamefont {Gu}, \citenamefont {Liu},\ and\ \citenamefont {Wen}}]{chen2013SPT}%
  \BibitemOpen
  \bibfield  {author} {\bibinfo {author} {\bibfnamefont {X.}~\bibnamefont {Chen}}, \bibinfo {author} {\bibfnamefont {Z.-C.}\ \bibnamefont {Gu}}, \bibinfo {author} {\bibfnamefont {Z.-X.}\ \bibnamefont {Liu}},\ and\ \bibinfo {author} {\bibfnamefont {X.-G.}\ \bibnamefont {Wen}},\ }\bibfield  {title} {\bibinfo {title} {Symmetry protected topological orders and the group cohomology of their symmetry group},\ }\href {https://doi.org/10.1103/PhysRevB.87.155114} {\bibfield  {journal} {\bibinfo  {journal} {Phys. Rev. B}\ }\textbf {\bibinfo {volume} {87}},\ \bibinfo {pages} {155114} (\bibinfo {year} {2013})}\BibitemShut {NoStop}%
\bibitem [{\citenamefont {Kramers}\ and\ \citenamefont {Wannier}(1941)}]{kramerswannier}%
  \BibitemOpen
  \bibfield  {author} {\bibinfo {author} {\bibfnamefont {H.~A.}\ \bibnamefont {Kramers}}\ and\ \bibinfo {author} {\bibfnamefont {G.~H.}\ \bibnamefont {Wannier}},\ }\bibfield  {title} {\bibinfo {title} {Statistics of the two-dimensional ferromagnet. part i},\ }\href {https://doi.org/10.1103/PhysRev.60.252} {\bibfield  {journal} {\bibinfo  {journal} {Phys. Rev.}\ }\textbf {\bibinfo {volume} {60}},\ \bibinfo {pages} {252} (\bibinfo {year} {1941})}\BibitemShut {NoStop}%
\bibitem [{\citenamefont {{Shao}}(2023)}]{shao2023review}%
  \BibitemOpen
  \bibfield  {author} {\bibinfo {author} {\bibfnamefont {S.-H.}\ \bibnamefont {{Shao}}},\ }\bibfield  {title} {\bibinfo {title} {{What's Done Cannot Be Undone: TASI Lectures on Non-Invertible Symmetries}},\ }\bibfield  {journal} {\bibinfo  {journal} {arXiv e-prints}\ }\href {https://doi.org/10.48550/arXiv.2308.00747} {10.48550/arXiv.2308.00747} (\bibinfo {year} {2023}),\ \Eprint {https://arxiv.org/abs/2308.00747} {arXiv:2308.00747 [hep-th]} \BibitemShut {NoStop}%
\bibitem [{\citenamefont {{Sch{\"a}fer-Nameki}}(2024)}]{sakurareview}%
  \BibitemOpen
  \bibfield  {author} {\bibinfo {author} {\bibfnamefont {S.}~\bibnamefont {{Sch{\"a}fer-Nameki}}},\ }\bibfield  {title} {\bibinfo {title} {{ICTP lectures on (non-)invertible generalized symmetries}},\ }\href {https://doi.org/10.1016/j.physrep.2024.01.007} {\bibfield  {journal} {\bibinfo  {journal} {Physics Reports}\ }\textbf {\bibinfo {volume} {1063}},\ \bibinfo {pages} {1} (\bibinfo {year} {2024})}\BibitemShut {NoStop}%
\bibitem [{\citenamefont {Barkeshli}\ \emph {et~al.}(2019)\citenamefont {Barkeshli}, \citenamefont {Bonderson}, \citenamefont {Cheng},\ and\ \citenamefont {Wang}}]{barkeshli2014sym}%
  \BibitemOpen
  \bibfield  {author} {\bibinfo {author} {\bibfnamefont {M.}~\bibnamefont {Barkeshli}}, \bibinfo {author} {\bibfnamefont {P.}~\bibnamefont {Bonderson}}, \bibinfo {author} {\bibfnamefont {M.}~\bibnamefont {Cheng}},\ and\ \bibinfo {author} {\bibfnamefont {Z.}~\bibnamefont {Wang}},\ }\bibfield  {title} {\bibinfo {title} {Symmetry fractionalization, defects, and gauging of topological phases},\ }\href {https://doi.org/10.1103/PhysRevB.100.115147} {\bibfield  {journal} {\bibinfo  {journal} {Phys. Rev. B}\ }\textbf {\bibinfo {volume} {100}},\ \bibinfo {pages} {115147} (\bibinfo {year} {2019})}\BibitemShut {NoStop}%
\bibitem [{\citenamefont {{Gaiotto}}\ \emph {et~al.}(2015)\citenamefont {{Gaiotto}}, \citenamefont {{Kapustin}}, \citenamefont {{Seiberg}},\ and\ \citenamefont {{Willett}}}]{ggs2015}%
  \BibitemOpen
  \bibfield  {author} {\bibinfo {author} {\bibfnamefont {D.}~\bibnamefont {{Gaiotto}}}, \bibinfo {author} {\bibfnamefont {A.}~\bibnamefont {{Kapustin}}}, \bibinfo {author} {\bibfnamefont {N.}~\bibnamefont {{Seiberg}}},\ and\ \bibinfo {author} {\bibfnamefont {B.}~\bibnamefont {{Willett}}},\ }\bibfield  {title} {\bibinfo {title} {{Generalized global symmetries}},\ }\href {https://doi.org/10.1007/JHEP02(2015)172} {\bibfield  {journal} {\bibinfo  {journal} {Journal of High Energy Physics}\ }\textbf {\bibinfo {volume} {2015}},\ \bibinfo {eid} {172} (\bibinfo {year} {2015})}\BibitemShut {NoStop}%
\bibitem [{\citenamefont {{Cordova}}\ \emph {et~al.}(2022)\citenamefont {{Cordova}}, \citenamefont {{Dumitrescu}}, \citenamefont {{Intriligator}},\ and\ \citenamefont {{Shao}}}]{cordova2022}%
  \BibitemOpen
  \bibfield  {author} {\bibinfo {author} {\bibfnamefont {C.}~\bibnamefont {{Cordova}}}, \bibinfo {author} {\bibfnamefont {T.~T.}\ \bibnamefont {{Dumitrescu}}}, \bibinfo {author} {\bibfnamefont {K.}~\bibnamefont {{Intriligator}}},\ and\ \bibinfo {author} {\bibfnamefont {S.-H.}\ \bibnamefont {{Shao}}},\ }\bibfield  {title} {\bibinfo {title} {{Snowmass White Paper: Generalized Symmetries in Quantum Field Theory and Beyond}},\ }\bibfield  {journal} {\bibinfo  {journal} {arXiv e-prints}\ }\href {https://doi.org/10.48550/arXiv.2205.09545} {10.48550/arXiv.2205.09545} (\bibinfo {year} {2022}),\ \Eprint {https://arxiv.org/abs/2205.09545} {arXiv:2205.09545 [hep-th]} \BibitemShut {NoStop}%
\bibitem [{\citenamefont {{McGreevy}}(2023)}]{mcgreevy2023review}%
  \BibitemOpen
  \bibfield  {author} {\bibinfo {author} {\bibfnamefont {J.}~\bibnamefont {{McGreevy}}},\ }\bibfield  {title} {\bibinfo {title} {{Generalized Symmetries in Condensed Matter}},\ }\href {https://doi.org/10.1146/annurev-conmatphys-040721-021029} {\bibfield  {journal} {\bibinfo  {journal} {Annual Review of Condensed Matter Physics}\ }\textbf {\bibinfo {volume} {14}},\ \bibinfo {pages} {57} (\bibinfo {year} {2023})}\BibitemShut {NoStop}%
\bibitem [{\citenamefont {{Etingof}}\ \emph {et~al.}(2009)\citenamefont {{Etingof}}, \citenamefont {{Nikshych}}, \citenamefont {{Ostrik}},\ and\ \citenamefont {{Ehud Meir}}}]{etingof2009}%
  \BibitemOpen
  \bibfield  {author} {\bibinfo {author} {\bibfnamefont {P.}~\bibnamefont {{Etingof}}}, \bibinfo {author} {\bibfnamefont {D.}~\bibnamefont {{Nikshych}}}, \bibinfo {author} {\bibfnamefont {V.}~\bibnamefont {{Ostrik}}},\ and\ \bibinfo {author} {\bibfnamefont {w.~a. a.~b.}\ \bibnamefont {{Ehud Meir}}},\ }\bibfield  {title} {\bibinfo {title} {{Fusion categories and homotopy theory}},\ }\bibfield  {journal} {\bibinfo  {journal} {arXiv e-prints}\ }\href {https://doi.org/10.48550/arXiv.0909.3140} {10.48550/arXiv.0909.3140} (\bibinfo {year} {2009}),\ \Eprint {https://arxiv.org/abs/0909.3140} {arXiv:0909.3140 [math.QA]} \BibitemShut {NoStop}%
\bibitem [{\citenamefont {{Kapustin}}(2010)}]{kapustin2010}%
  \BibitemOpen
  \bibfield  {author} {\bibinfo {author} {\bibfnamefont {A.}~\bibnamefont {{Kapustin}}},\ }\bibfield  {title} {\bibinfo {title} {{Topological Field Theory, Higher Categories, and Their Applications}},\ }\bibfield  {journal} {\bibinfo  {journal} {arXiv e-prints}\ }\href {https://doi.org/10.48550/arXiv.1004.2307} {10.48550/arXiv.1004.2307} (\bibinfo {year} {2010}),\ \Eprint {https://arxiv.org/abs/1004.2307} {arXiv:1004.2307 [math.QA]} \BibitemShut {NoStop}%
\bibitem [{\citenamefont {{Kong}}\ and\ \citenamefont {{Wen}}(2014)}]{kong2014bfc}%
  \BibitemOpen
  \bibfield  {author} {\bibinfo {author} {\bibfnamefont {L.}~\bibnamefont {{Kong}}}\ and\ \bibinfo {author} {\bibfnamefont {X.-G.}\ \bibnamefont {{Wen}}},\ }\bibfield  {title} {\bibinfo {title} {{Braided fusion categories, gravitational anomalies, and the mathematical framework for topological orders in any dimensions}},\ }\href {https://doi.org/10.48550/arXiv.1405.5858} {\bibfield  {journal} {\bibinfo  {journal} {arXiv e-prints}\ ,\ \bibinfo {eid} {arXiv:1405.5858}} (\bibinfo {year} {2014})},\ \Eprint {https://arxiv.org/abs/1405.5858} {arXiv:1405.5858 [cond-mat.str-el]} \BibitemShut {NoStop}%
\bibitem [{\citenamefont {{Douglas}}\ and\ \citenamefont {{Reutter}}(2018)}]{douglas2018f2c}%
  \BibitemOpen
  \bibfield  {author} {\bibinfo {author} {\bibfnamefont {C.~L.}\ \bibnamefont {{Douglas}}}\ and\ \bibinfo {author} {\bibfnamefont {D.~J.}\ \bibnamefont {{Reutter}}},\ }\bibfield  {title} {\bibinfo {title} {{Fusion 2-categories and a state-sum invariant for 4-manifolds}},\ }\bibfield  {journal} {\bibinfo  {journal} {arXiv e-prints}\ }\href {https://doi.org/10.48550/arXiv.1812.11933} {10.48550/arXiv.1812.11933} (\bibinfo {year} {2018}),\ \Eprint {https://arxiv.org/abs/1812.11933} {arXiv:1812.11933 [math.QA]} \BibitemShut {NoStop}%
\bibitem [{\citenamefont {{Gaiotto}}\ and\ \citenamefont {{Johnson-Freyd}}(2019)}]{gaiotto2019}%
  \BibitemOpen
  \bibfield  {author} {\bibinfo {author} {\bibfnamefont {D.}~\bibnamefont {{Gaiotto}}}\ and\ \bibinfo {author} {\bibfnamefont {T.}~\bibnamefont {{Johnson-Freyd}}},\ }\bibfield  {title} {\bibinfo {title} {{Condensations in higher categories}},\ }\bibfield  {journal} {\bibinfo  {journal} {arXiv e-prints}\ }\href {https://doi.org/10.48550/arXiv.1905.09566} {10.48550/arXiv.1905.09566} (\bibinfo {year} {2019}),\ \Eprint {https://arxiv.org/abs/1905.09566} {arXiv:1905.09566 [math.CT]} \BibitemShut {NoStop}%
\bibitem [{\citenamefont {Johnson-Freyd}(2022)}]{Johnson_Freyd_2022}%
  \BibitemOpen
  \bibfield  {author} {\bibinfo {author} {\bibfnamefont {T.}~\bibnamefont {Johnson-Freyd}},\ }\bibfield  {title} {\bibinfo {title} {On the classification of topological orders},\ }\href {https://doi.org/10.1007/s00220-022-04380-3} {\bibfield  {journal} {\bibinfo  {journal} {Communications in Mathematical Physics}\ }\textbf {\bibinfo {volume} {393}},\ \bibinfo {pages} {989–1033} (\bibinfo {year} {2022})}\BibitemShut {NoStop}%
\bibitem [{\citenamefont {{D{\'e}coppet}}\ \emph {et~al.}(2024)\citenamefont {{D{\'e}coppet}}, \citenamefont {{Huston}}, \citenamefont {{Johnson-Freyd}}, \citenamefont {{Nikshych}}, \citenamefont {{Penneys}}, \citenamefont {{Plavnik}}, \citenamefont {{Reutter}},\ and\ \citenamefont {{Yu}}}]{yu2024f2c}%
  \BibitemOpen
  \bibfield  {author} {\bibinfo {author} {\bibfnamefont {T.~D.}\ \bibnamefont {{D{\'e}coppet}}}, \bibinfo {author} {\bibfnamefont {P.}~\bibnamefont {{Huston}}}, \bibinfo {author} {\bibfnamefont {T.}~\bibnamefont {{Johnson-Freyd}}}, \bibinfo {author} {\bibfnamefont {D.}~\bibnamefont {{Nikshych}}}, \bibinfo {author} {\bibfnamefont {D.}~\bibnamefont {{Penneys}}}, \bibinfo {author} {\bibfnamefont {J.}~\bibnamefont {{Plavnik}}}, \bibinfo {author} {\bibfnamefont {D.}~\bibnamefont {{Reutter}}},\ and\ \bibinfo {author} {\bibfnamefont {M.}~\bibnamefont {{Yu}}},\ }\bibfield  {title} {\bibinfo {title} {{The Classification of Fusion 2-Categories}},\ }\bibfield  {journal} {\bibinfo  {journal} {arXiv e-prints}\ }\href {https://doi.org/10.48550/arXiv.2411.05907} {10.48550/arXiv.2411.05907} (\bibinfo {year} {2024}),\ \Eprint {https://arxiv.org/abs/2411.05907} {arXiv:2411.05907 [math.CT]} \BibitemShut {NoStop}%
\bibitem [{\citenamefont {{Freed}}\ and\ \citenamefont {{Teleman}}(2012)}]{freed2012}%
  \BibitemOpen
  \bibfield  {author} {\bibinfo {author} {\bibfnamefont {D.~S.}\ \bibnamefont {{Freed}}}\ and\ \bibinfo {author} {\bibfnamefont {C.}~\bibnamefont {{Teleman}}},\ }\bibfield  {title} {\bibinfo {title} {{Relative quantum field theory}},\ }\bibfield  {journal} {\bibinfo  {journal} {arXiv e-prints}\ }\href {https://doi.org/10.48550/arXiv.1212.1692} {10.48550/arXiv.1212.1692} (\bibinfo {year} {2012}),\ \Eprint {https://arxiv.org/abs/1212.1692} {arXiv:1212.1692 [hep-th]} \BibitemShut {NoStop}%
\bibitem [{\citenamefont {{Kong}}\ \emph {et~al.}(2015)\citenamefont {{Kong}}, \citenamefont {{Wen}},\ and\ \citenamefont {{Zheng}}}]{kong2015bdry}%
  \BibitemOpen
  \bibfield  {author} {\bibinfo {author} {\bibfnamefont {L.}~\bibnamefont {{Kong}}}, \bibinfo {author} {\bibfnamefont {X.-G.}\ \bibnamefont {{Wen}}},\ and\ \bibinfo {author} {\bibfnamefont {H.}~\bibnamefont {{Zheng}}},\ }\bibfield  {title} {\bibinfo {title} {{Boundary-bulk relation for topological orders as the functor mapping higher categories to their centers}},\ }\bibfield  {journal} {\bibinfo  {journal} {arXiv e-prints}\ }\href {https://doi.org/10.48550/arXiv.1502.01690} {10.48550/arXiv.1502.01690} (\bibinfo {year} {2015}),\ \Eprint {https://arxiv.org/abs/1502.01690} {arXiv:1502.01690 [cond-mat.str-el]} \BibitemShut {NoStop}%
\bibitem [{\citenamefont {{Kong}}\ \emph {et~al.}(2017)\citenamefont {{Kong}}, \citenamefont {{Wen}},\ and\ \citenamefont {{Zheng}}}]{kong2017}%
  \BibitemOpen
  \bibfield  {author} {\bibinfo {author} {\bibfnamefont {L.}~\bibnamefont {{Kong}}}, \bibinfo {author} {\bibfnamefont {X.-G.}\ \bibnamefont {{Wen}}},\ and\ \bibinfo {author} {\bibfnamefont {H.}~\bibnamefont {{Zheng}}},\ }\bibfield  {title} {\bibinfo {title} {{Boundary-bulk relation in topological orders}},\ }\href {https://doi.org/10.1016/j.nuclphysb.2017.06.023} {\bibfield  {journal} {\bibinfo  {journal} {Nuclear Physics B}\ }\textbf {\bibinfo {volume} {922}},\ \bibinfo {pages} {62} (\bibinfo {year} {2017})}\BibitemShut {NoStop}%
\bibitem [{\citenamefont {{Kong}}\ and\ \citenamefont {{Zheng}}(2018)}]{kong2018emc}%
  \BibitemOpen
  \bibfield  {author} {\bibinfo {author} {\bibfnamefont {L.}~\bibnamefont {{Kong}}}\ and\ \bibinfo {author} {\bibfnamefont {H.}~\bibnamefont {{Zheng}}},\ }\bibfield  {title} {\bibinfo {title} {{Gapless edges of 2d topological orders and enriched monoidal categories}},\ }\href {https://doi.org/10.1016/j.nuclphysb.2017.12.007} {\bibfield  {journal} {\bibinfo  {journal} {Nuclear Physics B}\ }\textbf {\bibinfo {volume} {927}},\ \bibinfo {pages} {140} (\bibinfo {year} {2018})},\ \Eprint {https://arxiv.org/abs/1705.01087} {arXiv:1705.01087 [cond-mat.str-el]} \BibitemShut {NoStop}%
\bibitem [{\citenamefont {{Freed}}\ and\ \citenamefont {{Teleman}}(2018)}]{freed2018}%
  \BibitemOpen
  \bibfield  {author} {\bibinfo {author} {\bibfnamefont {D.~S.}\ \bibnamefont {{Freed}}}\ and\ \bibinfo {author} {\bibfnamefont {C.}~\bibnamefont {{Teleman}}},\ }\bibfield  {title} {\bibinfo {title} {{Topological dualities in the Ising model}},\ }\bibfield  {journal} {\bibinfo  {journal} {arXiv e-prints}\ }\href {https://doi.org/10.48550/arXiv.1806.00008} {10.48550/arXiv.1806.00008} (\bibinfo {year} {2018}),\ \Eprint {https://arxiv.org/abs/1806.00008} {arXiv:1806.00008 [math.AT]} \BibitemShut {NoStop}%
\bibitem [{\citenamefont {Kong}\ \emph {et~al.}(2020)\citenamefont {Kong}, \citenamefont {Lan}, \citenamefont {Wen}, \citenamefont {Zhang},\ and\ \citenamefont {Zheng}}]{kong2020holo}%
  \BibitemOpen
  \bibfield  {author} {\bibinfo {author} {\bibfnamefont {L.}~\bibnamefont {Kong}}, \bibinfo {author} {\bibfnamefont {T.}~\bibnamefont {Lan}}, \bibinfo {author} {\bibfnamefont {X.-G.}\ \bibnamefont {Wen}}, \bibinfo {author} {\bibfnamefont {Z.-H.}\ \bibnamefont {Zhang}},\ and\ \bibinfo {author} {\bibfnamefont {H.}~\bibnamefont {Zheng}},\ }\bibfield  {title} {\bibinfo {title} {Algebraic higher symmetry and categorical symmetry: A holographic and entanglement view of symmetry},\ }\href {https://doi.org/10.1103/PhysRevResearch.2.043086} {\bibfield  {journal} {\bibinfo  {journal} {Phys. Rev. Res.}\ }\textbf {\bibinfo {volume} {2}},\ \bibinfo {pages} {043086} (\bibinfo {year} {2020})}\BibitemShut {NoStop}%
\bibitem [{\citenamefont {{Kong}}\ \emph {et~al.}(2020)\citenamefont {{Kong}}, \citenamefont {{Lan}}, \citenamefont {{Wen}}, \citenamefont {{Zhang}},\ and\ \citenamefont {{Zheng}}}]{kong2020class}%
  \BibitemOpen
  \bibfield  {author} {\bibinfo {author} {\bibfnamefont {L.}~\bibnamefont {{Kong}}}, \bibinfo {author} {\bibfnamefont {T.}~\bibnamefont {{Lan}}}, \bibinfo {author} {\bibfnamefont {X.-G.}\ \bibnamefont {{Wen}}}, \bibinfo {author} {\bibfnamefont {Z.-H.}\ \bibnamefont {{Zhang}}},\ and\ \bibinfo {author} {\bibfnamefont {H.}~\bibnamefont {{Zheng}}},\ }\bibfield  {title} {\bibinfo {title} {{Classification of topological phases with finite internal symmetries in all dimensions}},\ }\href {https://doi.org/10.1007/JHEP09(2020)093} {\bibfield  {journal} {\bibinfo  {journal} {Journal of High Energy Physics}\ }\textbf {\bibinfo {volume} {2020}},\ \bibinfo {eid} {93} (\bibinfo {year} {2020})}\BibitemShut {NoStop}%
\bibitem [{\citenamefont {{Chatterjee}}\ \emph {et~al.}(2022)\citenamefont {{Chatterjee}}, \citenamefont {{Ji}},\ and\ \citenamefont {{Wen}}}]{chatterjee2022}%
  \BibitemOpen
  \bibfield  {author} {\bibinfo {author} {\bibfnamefont {A.}~\bibnamefont {{Chatterjee}}}, \bibinfo {author} {\bibfnamefont {W.}~\bibnamefont {{Ji}}},\ and\ \bibinfo {author} {\bibfnamefont {X.-G.}\ \bibnamefont {{Wen}}},\ }\bibfield  {title} {\bibinfo {title} {{Emergent generalized symmetry and maximal symmetry-topological-order}},\ }\bibfield  {journal} {\bibinfo  {journal} {arXiv e-prints}\ }\href {https://doi.org/10.48550/arXiv.2212.14432} {10.48550/arXiv.2212.14432} (\bibinfo {year} {2022}),\ \Eprint {https://arxiv.org/abs/2212.14432} {arXiv:2212.14432 [cond-mat.str-el]} \BibitemShut {NoStop}%
\bibitem [{\citenamefont {Ji}\ and\ \citenamefont {Wen}(2020)}]{ji2020}%
  \BibitemOpen
  \bibfield  {author} {\bibinfo {author} {\bibfnamefont {W.}~\bibnamefont {Ji}}\ and\ \bibinfo {author} {\bibfnamefont {X.-G.}\ \bibnamefont {Wen}},\ }\bibfield  {title} {\bibinfo {title} {Categorical symmetry and noninvertible anomaly in symmetry-breaking and topological phase transitions},\ }\href {https://doi.org/10.1103/PhysRevResearch.2.033417} {\bibfield  {journal} {\bibinfo  {journal} {Phys. Rev. Res.}\ }\textbf {\bibinfo {volume} {2}},\ \bibinfo {pages} {033417} (\bibinfo {year} {2020})}\BibitemShut {NoStop}%
\bibitem [{\citenamefont {{Freed}}\ \emph {et~al.}(2022)\citenamefont {{Freed}}, \citenamefont {{Moore}},\ and\ \citenamefont {{Teleman}}}]{freed2022}%
  \BibitemOpen
  \bibfield  {author} {\bibinfo {author} {\bibfnamefont {D.~S.}\ \bibnamefont {{Freed}}}, \bibinfo {author} {\bibfnamefont {G.~W.}\ \bibnamefont {{Moore}}},\ and\ \bibinfo {author} {\bibfnamefont {C.}~\bibnamefont {{Teleman}}},\ }\bibfield  {title} {\bibinfo {title} {{Topological symmetry in quantum field theory}},\ }\bibfield  {journal} {\bibinfo  {journal} {arXiv e-prints}\ }\href {https://doi.org/10.48550/arXiv.2209.07471} {10.48550/arXiv.2209.07471} (\bibinfo {year} {2022}),\ \Eprint {https://arxiv.org/abs/2209.07471} {arXiv:2209.07471 [hep-th]} \BibitemShut {NoStop}%
\bibitem [{\citenamefont {{Ji}}\ and\ \citenamefont {{Wen}}(2021)}]{ji2023}%
  \BibitemOpen
  \bibfield  {author} {\bibinfo {author} {\bibfnamefont {W.}~\bibnamefont {{Ji}}}\ and\ \bibinfo {author} {\bibfnamefont {X.-G.}\ \bibnamefont {{Wen}}},\ }\bibfield  {title} {\bibinfo {title} {{A unified view on symmetry, anomalous symmetry and non-invertible gravitational anomaly}},\ }\bibfield  {journal} {\bibinfo  {journal} {arXiv e-prints}\ }\href {https://doi.org/10.48550/arXiv.2106.02069} {10.48550/arXiv.2106.02069} (\bibinfo {year} {2021}),\ \Eprint {https://arxiv.org/abs/2106.02069} {arXiv:2106.02069 [cond-mat.str-el]} \BibitemShut {NoStop}%
\bibitem [{\citenamefont {Chatterjee}\ and\ \citenamefont {Wen}(2023{\natexlab{a}})}]{chatterjee2023shadow}%
  \BibitemOpen
  \bibfield  {author} {\bibinfo {author} {\bibfnamefont {A.}~\bibnamefont {Chatterjee}}\ and\ \bibinfo {author} {\bibfnamefont {X.-G.}\ \bibnamefont {Wen}},\ }\bibfield  {title} {\bibinfo {title} {Symmetry as a shadow of topological order and a derivation of topological holographic principle},\ }\href {https://doi.org/10.1103/PhysRevB.107.155136} {\bibfield  {journal} {\bibinfo  {journal} {Phys. Rev. B}\ }\textbf {\bibinfo {volume} {107}},\ \bibinfo {pages} {155136} (\bibinfo {year} {2023}{\natexlab{a}})}\BibitemShut {NoStop}%
\bibitem [{\citenamefont {{Apruzzi}}\ \emph {et~al.}(2023)\citenamefont {{Apruzzi}}, \citenamefont {{Bonetti}}, \citenamefont {{Garc{\'\i}a Etxebarria}}, \citenamefont {{Hosseini}},\ and\ \citenamefont {{Sch{\"a}fer-Nameki}}}]{apruzzi2023}%
  \BibitemOpen
  \bibfield  {author} {\bibinfo {author} {\bibfnamefont {F.}~\bibnamefont {{Apruzzi}}}, \bibinfo {author} {\bibfnamefont {F.}~\bibnamefont {{Bonetti}}}, \bibinfo {author} {\bibfnamefont {I.}~\bibnamefont {{Garc{\'\i}a Etxebarria}}}, \bibinfo {author} {\bibfnamefont {S.~S.}\ \bibnamefont {{Hosseini}}},\ and\ \bibinfo {author} {\bibfnamefont {S.}~\bibnamefont {{Sch{\"a}fer-Nameki}}},\ }\bibfield  {title} {\bibinfo {title} {{Symmetry TFTs from String Theory}},\ }\href {https://doi.org/10.1007/s00220-023-04737-2} {\bibfield  {journal} {\bibinfo  {journal} {Communications in Mathematical Physics}\ }\textbf {\bibinfo {volume} {402}},\ \bibinfo {pages} {895} (\bibinfo {year} {2023})},\ \Eprint {https://arxiv.org/abs/2112.02092} {arXiv:2112.02092 [hep-th]} \BibitemShut {NoStop}%
\bibitem [{\citenamefont {{Lin}}\ \emph {et~al.}(2023)\citenamefont {{Lin}}, \citenamefont {{Okada}}, \citenamefont {{Seifnashri}},\ and\ \citenamefont {{Tachikawa}}}]{lin2023}%
  \BibitemOpen
  \bibfield  {author} {\bibinfo {author} {\bibfnamefont {Y.-H.}\ \bibnamefont {{Lin}}}, \bibinfo {author} {\bibfnamefont {M.}~\bibnamefont {{Okada}}}, \bibinfo {author} {\bibfnamefont {S.}~\bibnamefont {{Seifnashri}}},\ and\ \bibinfo {author} {\bibfnamefont {Y.}~\bibnamefont {{Tachikawa}}},\ }\bibfield  {title} {\bibinfo {title} {{Asymptotic density of states in 2d CFTs with non-invertible symmetries}},\ }\href {https://doi.org/10.1007/JHEP03(2023)094} {\bibfield  {journal} {\bibinfo  {journal} {Journal of High Energy Physics}\ }\textbf {\bibinfo {volume} {2023}},\ \bibinfo {eid} {94} (\bibinfo {year} {2023})}\BibitemShut {NoStop}%
\bibitem [{\citenamefont {{Benini}}\ \emph {et~al.}(2023)\citenamefont {{Benini}}, \citenamefont {{Copetti}},\ and\ \citenamefont {{Di Pietro}}}]{benini2023}%
  \BibitemOpen
  \bibfield  {author} {\bibinfo {author} {\bibfnamefont {F.}~\bibnamefont {{Benini}}}, \bibinfo {author} {\bibfnamefont {C.}~\bibnamefont {{Copetti}}},\ and\ \bibinfo {author} {\bibfnamefont {L.}~\bibnamefont {{Di Pietro}}},\ }\bibfield  {title} {\bibinfo {title} {{Factorization and global symmetries in holography}},\ }\href {https://doi.org/10.21468/SciPostPhys.14.2.019} {\bibfield  {journal} {\bibinfo  {journal} {SciPost Physics}\ }\textbf {\bibinfo {volume} {14}},\ \bibinfo {eid} {019} (\bibinfo {year} {2023})}\BibitemShut {NoStop}%
\bibitem [{\citenamefont {{Kaidi}}\ \emph {et~al.}(2023{\natexlab{a}})\citenamefont {{Kaidi}}, \citenamefont {{Ohmori}},\ and\ \citenamefont {{Zheng}}}]{kaidi2023}%
  \BibitemOpen
  \bibfield  {author} {\bibinfo {author} {\bibfnamefont {J.}~\bibnamefont {{Kaidi}}}, \bibinfo {author} {\bibfnamefont {K.}~\bibnamefont {{Ohmori}}},\ and\ \bibinfo {author} {\bibfnamefont {Y.}~\bibnamefont {{Zheng}}},\ }\bibfield  {title} {\bibinfo {title} {{Symmetry TFTs for Non-invertible Defects}},\ }\href {https://doi.org/10.1007/s00220-023-04859-7} {\bibfield  {journal} {\bibinfo  {journal} {Communications in Mathematical Physics}\ }\textbf {\bibinfo {volume} {404}},\ \bibinfo {pages} {1021} (\bibinfo {year} {2023}{\natexlab{a}})}\BibitemShut {NoStop}%
\bibitem [{\citenamefont {{Kaidi}}\ \emph {et~al.}(2023{\natexlab{b}})\citenamefont {{Kaidi}}, \citenamefont {{Nardoni}}, \citenamefont {{Zafrir}},\ and\ \citenamefont {{Zheng}}}]{kaidi2023anomaly}%
  \BibitemOpen
  \bibfield  {author} {\bibinfo {author} {\bibfnamefont {J.}~\bibnamefont {{Kaidi}}}, \bibinfo {author} {\bibfnamefont {E.}~\bibnamefont {{Nardoni}}}, \bibinfo {author} {\bibfnamefont {G.}~\bibnamefont {{Zafrir}}},\ and\ \bibinfo {author} {\bibfnamefont {Y.}~\bibnamefont {{Zheng}}},\ }\bibfield  {title} {\bibinfo {title} {{Symmetry TFTs and anomalies of non-invertible symmetries}},\ }\href {https://doi.org/10.1007/JHEP10(2023)053} {\bibfield  {journal} {\bibinfo  {journal} {Journal of High Energy Physics}\ }\textbf {\bibinfo {volume} {2023}},\ \bibinfo {eid} {53} (\bibinfo {year} {2023}{\natexlab{b}})}\BibitemShut {NoStop}%
\bibitem [{\citenamefont {{Bhardwaj}}\ and\ \citenamefont {{Schafer-Nameki}}(2023)}]{bhardwaj2023ii}%
  \BibitemOpen
  \bibfield  {author} {\bibinfo {author} {\bibfnamefont {L.}~\bibnamefont {{Bhardwaj}}}\ and\ \bibinfo {author} {\bibfnamefont {S.}~\bibnamefont {{Schafer-Nameki}}},\ }\bibfield  {title} {\bibinfo {title} {{Generalized Charges, Part II: Non-Invertible Symmetries and the Symmetry TFT}},\ }\bibfield  {journal} {\bibinfo  {journal} {arXiv e-prints}\ }\href {https://doi.org/10.48550/arXiv.2305.17159} {10.48550/arXiv.2305.17159} (\bibinfo {year} {2023}),\ \Eprint {https://arxiv.org/abs/2305.17159} {arXiv:2305.17159 [hep-th]} \BibitemShut {NoStop}%
\bibitem [{\citenamefont {{Moradi}}\ \emph {et~al.}(2023)\citenamefont {{Moradi}}, \citenamefont {{Faroogh Moosavian}},\ and\ \citenamefont {{Tiwari}}}]{moradi2023}%
  \BibitemOpen
  \bibfield  {author} {\bibinfo {author} {\bibfnamefont {H.}~\bibnamefont {{Moradi}}}, \bibinfo {author} {\bibfnamefont {S.}~\bibnamefont {{Faroogh Moosavian}}},\ and\ \bibinfo {author} {\bibfnamefont {A.}~\bibnamefont {{Tiwari}}},\ }\bibfield  {title} {\bibinfo {title} {{Topological Holography: Towards a Unification of Landau and Beyond-Landau Physics}},\ }\href {https://doi.org/10.21468/SciPostPhysCore.6.4.066} {\bibfield  {journal} {\bibinfo  {journal} {SciPost Physics Core}\ }\textbf {\bibinfo {volume} {6}},\ \bibinfo {eid} {066} (\bibinfo {year} {2023})}\BibitemShut {NoStop}%
\bibitem [{\citenamefont {Wen}\ and\ \citenamefont {Potter}(2025)}]{Wen2023}%
  \BibitemOpen
  \bibfield  {author} {\bibinfo {author} {\bibfnamefont {R.}~\bibnamefont {Wen}}\ and\ \bibinfo {author} {\bibfnamefont {A.~C.}\ \bibnamefont {Potter}},\ }\bibfield  {title} {\bibinfo {title} {Classification of $1+1\mathrm{D}$ gapless symmetry protected phases via topological holography},\ }\href {https://doi.org/10.1103/PhysRevB.111.115161} {\bibfield  {journal} {\bibinfo  {journal} {Phys. Rev. B}\ }\textbf {\bibinfo {volume} {111}},\ \bibinfo {pages} {115161} (\bibinfo {year} {2025})}\BibitemShut {NoStop}%
\bibitem [{\citenamefont {{Huang}}\ and\ \citenamefont {{Cheng}}(2023)}]{Huang2023}%
  \BibitemOpen
  \bibfield  {author} {\bibinfo {author} {\bibfnamefont {S.-J.}\ \bibnamefont {{Huang}}}\ and\ \bibinfo {author} {\bibfnamefont {M.}~\bibnamefont {{Cheng}}},\ }\bibfield  {title} {\bibinfo {title} {{Topological holography, quantum criticality, and boundary states}},\ }\bibfield  {journal} {\bibinfo  {journal} {arXiv e-prints}\ }\href {https://doi.org/10.48550/arXiv.2310.16878} {10.48550/arXiv.2310.16878} (\bibinfo {year} {2023}),\ \Eprint {https://arxiv.org/abs/2310.16878} {arXiv:2310.16878 [cond-mat.str-el]} \BibitemShut {NoStop}%
\bibitem [{\citenamefont {Chatterjee}\ and\ \citenamefont {Wen}(2023{\natexlab{b}})}]{Chatterjee2023}%
  \BibitemOpen
  \bibfield  {author} {\bibinfo {author} {\bibfnamefont {A.}~\bibnamefont {Chatterjee}}\ and\ \bibinfo {author} {\bibfnamefont {X.-G.}\ \bibnamefont {Wen}},\ }\bibfield  {title} {\bibinfo {title} {Holographic theory for continuous phase transitions: Emergence and symmetry protection of gaplessness},\ }\href {https://doi.org/10.1103/PhysRevB.108.075105} {\bibfield  {journal} {\bibinfo  {journal} {Phys. Rev. B}\ }\textbf {\bibinfo {volume} {108}},\ \bibinfo {pages} {075105} (\bibinfo {year} {2023}{\natexlab{b}})}\BibitemShut {NoStop}%
\bibitem [{\citenamefont {{Bhardwaj}}\ \emph {et~al.}(2025{\natexlab{a}})\citenamefont {{Bhardwaj}}, \citenamefont {{Bottini}}, \citenamefont {{Pajer}},\ and\ \citenamefont {{Sch{\"a}fer-Nameki}}}]{bhardwaj2025club}%
  \BibitemOpen
  \bibfield  {author} {\bibinfo {author} {\bibfnamefont {L.}~\bibnamefont {{Bhardwaj}}}, \bibinfo {author} {\bibfnamefont {L.~E.}\ \bibnamefont {{Bottini}}}, \bibinfo {author} {\bibfnamefont {D.}~\bibnamefont {{Pajer}}},\ and\ \bibinfo {author} {\bibfnamefont {S.}~\bibnamefont {{Sch{\"a}fer-Nameki}}},\ }\bibfield  {title} {\bibinfo {title} {{The club sandwich: Gapless phases and phase transitions with non-invertible symmetries}},\ }\href {https://doi.org/10.21468/SciPostPhys.18.5.156} {\bibfield  {journal} {\bibinfo  {journal} {SciPost Physics}\ }\textbf {\bibinfo {volume} {18}},\ \bibinfo {eid} {156} (\bibinfo {year} {2025}{\natexlab{a}})}\BibitemShut {NoStop}%
\bibitem [{\citenamefont {Bhardwaj}\ \emph {et~al.}(2024)\citenamefont {Bhardwaj}, \citenamefont {Bottini}, \citenamefont {Pajer},\ and\ \citenamefont {Sch\"afer-Nameki}}]{bhardwaj2024landau}%
  \BibitemOpen
  \bibfield  {author} {\bibinfo {author} {\bibfnamefont {L.}~\bibnamefont {Bhardwaj}}, \bibinfo {author} {\bibfnamefont {L.~E.}\ \bibnamefont {Bottini}}, \bibinfo {author} {\bibfnamefont {D.}~\bibnamefont {Pajer}},\ and\ \bibinfo {author} {\bibfnamefont {S.}~\bibnamefont {Sch\"afer-Nameki}},\ }\bibfield  {title} {\bibinfo {title} {Categorical landau paradigm for gapped phases},\ }\href {https://doi.org/10.1103/PhysRevLett.133.161601} {\bibfield  {journal} {\bibinfo  {journal} {Phys. Rev. Lett.}\ }\textbf {\bibinfo {volume} {133}},\ \bibinfo {pages} {161601} (\bibinfo {year} {2024})}\BibitemShut {NoStop}%
\bibitem [{\citenamefont {{Bhardwaj}}\ \emph {et~al.}(2024{\natexlab{a}})\citenamefont {{Bhardwaj}}, \citenamefont {{Bottini}}, \citenamefont {{Schafer-Nameki}},\ and\ \citenamefont {{Tiwari}}}]{bhardwaj2024lattice}%
  \BibitemOpen
  \bibfield  {author} {\bibinfo {author} {\bibfnamefont {L.}~\bibnamefont {{Bhardwaj}}}, \bibinfo {author} {\bibfnamefont {L.~E.}\ \bibnamefont {{Bottini}}}, \bibinfo {author} {\bibfnamefont {S.}~\bibnamefont {{Schafer-Nameki}}},\ and\ \bibinfo {author} {\bibfnamefont {A.}~\bibnamefont {{Tiwari}}},\ }\bibfield  {title} {\bibinfo {title} {{Lattice Models for Phases and Transitions with Non-Invertible Symmetries}},\ }\bibfield  {journal} {\bibinfo  {journal} {arXiv e-prints}\ }\href {https://doi.org/10.48550/arXiv.2405.05964} {10.48550/arXiv.2405.05964} (\bibinfo {year} {2024}{\natexlab{a}}),\ \Eprint {https://arxiv.org/abs/2405.05964} {arXiv:2405.05964 [cond-mat.str-el]} \BibitemShut {NoStop}%
\bibitem [{\citenamefont {{Bhardwaj}}\ \emph {et~al.}(2024{\natexlab{b}})\citenamefont {{Bhardwaj}}, \citenamefont {{Pajer}}, \citenamefont {{Schafer-Nameki}},\ and\ \citenamefont {{Warman}}}]{warman2024}%
  \BibitemOpen
  \bibfield  {author} {\bibinfo {author} {\bibfnamefont {L.}~\bibnamefont {{Bhardwaj}}}, \bibinfo {author} {\bibfnamefont {D.}~\bibnamefont {{Pajer}}}, \bibinfo {author} {\bibfnamefont {S.}~\bibnamefont {{Schafer-Nameki}}},\ and\ \bibinfo {author} {\bibfnamefont {A.}~\bibnamefont {{Warman}}},\ }\bibfield  {title} {\bibinfo {title} {{Hasse Diagrams for Gapless SPT and SSB Phases with Non-Invertible Symmetries}},\ }\bibfield  {journal} {\bibinfo  {journal} {arXiv e-prints}\ }\href {https://doi.org/10.48550/arXiv.2403.00905} {10.48550/arXiv.2403.00905} (\bibinfo {year} {2024}{\natexlab{b}}),\ \Eprint {https://arxiv.org/abs/2403.00905} {arXiv:2403.00905 [cond-mat.str-el]} \BibitemShut {NoStop}%
\bibitem [{\citenamefont {{Bhardwaj}}\ \emph {et~al.}(2024{\natexlab{c}})\citenamefont {{Bhardwaj}}, \citenamefont {{Inamura}},\ and\ \citenamefont {{Tiwari}}}]{tiwari2024fermion}%
  \BibitemOpen
  \bibfield  {author} {\bibinfo {author} {\bibfnamefont {L.}~\bibnamefont {{Bhardwaj}}}, \bibinfo {author} {\bibfnamefont {K.}~\bibnamefont {{Inamura}}},\ and\ \bibinfo {author} {\bibfnamefont {A.}~\bibnamefont {{Tiwari}}},\ }\bibfield  {title} {\bibinfo {title} {{Fermionic Non-Invertible Symmetries in (1+1)d: Gapped and Gapless Phases, Transitions, and Symmetry TFTs}},\ }\bibfield  {journal} {\bibinfo  {journal} {arXiv e-prints}\ }\href {https://doi.org/10.48550/arXiv.2405.09754} {10.48550/arXiv.2405.09754} (\bibinfo {year} {2024}{\natexlab{c}}),\ \Eprint {https://arxiv.org/abs/2405.09754} {arXiv:2405.09754 [hep-th]} \BibitemShut {NoStop}%
\bibitem [{\citenamefont {{Wen}}\ \emph {et~al.}(2024)\citenamefont {{Wen}}, \citenamefont {{Ye}},\ and\ \citenamefont {{Potter}}}]{wen2024fermion}%
  \BibitemOpen
  \bibfield  {author} {\bibinfo {author} {\bibfnamefont {R.}~\bibnamefont {{Wen}}}, \bibinfo {author} {\bibfnamefont {W.}~\bibnamefont {{Ye}}},\ and\ \bibinfo {author} {\bibfnamefont {A.~C.}\ \bibnamefont {{Potter}}},\ }\bibfield  {title} {\bibinfo {title} {{Topological holography for fermions}},\ }\bibfield  {journal} {\bibinfo  {journal} {arXiv e-prints}\ }\href {https://doi.org/10.48550/arXiv.2404.19004} {10.48550/arXiv.2404.19004} (\bibinfo {year} {2024}),\ \Eprint {https://arxiv.org/abs/2404.19004} {arXiv:2404.19004 [cond-mat.str-el]} \BibitemShut {NoStop}%
\bibitem [{\citenamefont {Huang}(2025)}]{huang2025}%
  \BibitemOpen
  \bibfield  {author} {\bibinfo {author} {\bibfnamefont {S.-J.}\ \bibnamefont {Huang}},\ }\bibfield  {title} {\bibinfo {title} {Fermionic quantum criticality through the lens of topological holography},\ }\href {https://doi.org/10.1103/PhysRevB.111.155130} {\bibfield  {journal} {\bibinfo  {journal} {Phys. Rev. B}\ }\textbf {\bibinfo {volume} {111}},\ \bibinfo {pages} {155130} (\bibinfo {year} {2025})}\BibitemShut {NoStop}%
\bibitem [{\citenamefont {Kong}\ \emph {et~al.}(2020)\citenamefont {Kong}, \citenamefont {Tian},\ and\ \citenamefont {Zhou}}]{Kong_2020}%
  \BibitemOpen
  \bibfield  {author} {\bibinfo {author} {\bibfnamefont {L.}~\bibnamefont {Kong}}, \bibinfo {author} {\bibfnamefont {Y.}~\bibnamefont {Tian}},\ and\ \bibinfo {author} {\bibfnamefont {S.}~\bibnamefont {Zhou}},\ }\bibfield  {title} {\bibinfo {title} {The center of monoidal 2-categories in 3+1d dijkgraaf-witten theory},\ }\href {https://doi.org/10.1016/j.aim.2019.106928} {\bibfield  {journal} {\bibinfo  {journal} {Advances in Mathematics}\ }\textbf {\bibinfo {volume} {360}},\ \bibinfo {pages} {106928} (\bibinfo {year} {2020})}\BibitemShut {NoStop}%
\bibitem [{\citenamefont {Kong}\ and\ \citenamefont {Zheng}(2021)}]{Kong_2021}%
  \BibitemOpen
  \bibfield  {author} {\bibinfo {author} {\bibfnamefont {L.}~\bibnamefont {Kong}}\ and\ \bibinfo {author} {\bibfnamefont {H.}~\bibnamefont {Zheng}},\ }\bibfield  {title} {\bibinfo {title} {A mathematical theory of gapless edges of 2d topological orders. part ii},\ }\href {https://doi.org/10.1016/j.nuclphysb.2021.115384} {\bibfield  {journal} {\bibinfo  {journal} {Nuclear Physics B}\ }\textbf {\bibinfo {volume} {966}},\ \bibinfo {pages} {115384} (\bibinfo {year} {2021})}\BibitemShut {NoStop}%
\bibitem [{\citenamefont {{Ji}}\ and\ \citenamefont {{Chen}}(2024)}]{ji2024top}%
  \BibitemOpen
  \bibfield  {author} {\bibinfo {author} {\bibfnamefont {W.}~\bibnamefont {{Ji}}}\ and\ \bibinfo {author} {\bibfnamefont {X.}~\bibnamefont {{Chen}}},\ }\bibfield  {title} {\bibinfo {title} {{Topological defects of 2+1D systems from line excitations in 3+1D bulk}},\ }\bibfield  {journal} {\bibinfo  {journal} {arXiv e-prints}\ }\href {https://doi.org/10.48550/arXiv.2407.02488} {10.48550/arXiv.2407.02488} (\bibinfo {year} {2024}),\ \Eprint {https://arxiv.org/abs/2407.02488} {arXiv:2407.02488 [cond-mat.str-el]} \BibitemShut {NoStop}%
\bibitem [{\citenamefont {{Wen}}(2024)}]{wen2024string}%
  \BibitemOpen
  \bibfield  {author} {\bibinfo {author} {\bibfnamefont {R.}~\bibnamefont {{Wen}}},\ }\bibfield  {title} {\bibinfo {title} {{String condensation and topological holography for 2+1D gapless SPT}},\ }\bibfield  {journal} {\bibinfo  {journal} {arXiv e-prints}\ }\href {https://doi.org/10.48550/arXiv.2408.05801} {10.48550/arXiv.2408.05801} (\bibinfo {year} {2024}),\ \Eprint {https://arxiv.org/abs/2408.05801} {arXiv:2408.05801 [cond-mat.str-el]} \BibitemShut {NoStop}%
\bibitem [{\citenamefont {{Wen}}(2025)}]{wen2025top}%
  \BibitemOpen
  \bibfield  {author} {\bibinfo {author} {\bibfnamefont {R.}~\bibnamefont {{Wen}}},\ }\bibfield  {title} {\bibinfo {title} {{Topological Holography for 2+1-D Gapped and Gapless Phases with Generalized Symmetries}},\ }\bibfield  {journal} {\bibinfo  {journal} {arXiv e-prints}\ }\href {https://doi.org/10.48550/arXiv.2503.13685} {10.48550/arXiv.2503.13685} (\bibinfo {year} {2025}),\ \Eprint {https://arxiv.org/abs/2503.13685} {arXiv:2503.13685 [hep-th]} \BibitemShut {NoStop}%
\bibitem [{\citenamefont {{Antinucci}}\ \emph {et~al.}(2025)\citenamefont {{Antinucci}}, \citenamefont {{Copetti}},\ and\ \citenamefont {{Sch{\"a}fer-Nameki}}}]{antinucci2025sym}%
  \BibitemOpen
  \bibfield  {author} {\bibinfo {author} {\bibfnamefont {A.}~\bibnamefont {{Antinucci}}}, \bibinfo {author} {\bibfnamefont {C.}~\bibnamefont {{Copetti}}},\ and\ \bibinfo {author} {\bibfnamefont {S.}~\bibnamefont {{Sch{\"a}fer-Nameki}}},\ }\bibfield  {title} {\bibinfo {title} {{SymTFT for (3+1)d gapless SPTs and obstructions to confinement}},\ }\href {https://doi.org/10.21468/SciPostPhys.18.3.114} {\bibfield  {journal} {\bibinfo  {journal} {SciPost Physics}\ }\textbf {\bibinfo {volume} {18}},\ \bibinfo {eid} {114} (\bibinfo {year} {2025})}\BibitemShut {NoStop}%
\bibitem [{\citenamefont {{Putrov}}\ and\ \citenamefont {{Radhakrishnan}}(2025)}]{putrov2025braid}%
  \BibitemOpen
  \bibfield  {author} {\bibinfo {author} {\bibfnamefont {P.}~\bibnamefont {{Putrov}}}\ and\ \bibinfo {author} {\bibfnamefont {R.}~\bibnamefont {{Radhakrishnan}}},\ }\bibfield  {title} {\bibinfo {title} {{Braidings on topological operators, anomaly of higher-form symmetries and the SymTFT}},\ }\bibfield  {journal} {\bibinfo  {journal} {arXiv e-prints}\ }\href {https://doi.org/10.48550/arXiv.2503.13633} {10.48550/arXiv.2503.13633} (\bibinfo {year} {2025}),\ \Eprint {https://arxiv.org/abs/2503.13633} {arXiv:2503.13633 [hep-th]} \BibitemShut {NoStop}%
\bibitem [{\citenamefont {{Bhardwaj}}\ \emph {et~al.}(2025{\natexlab{b}})\citenamefont {{Bhardwaj}}, \citenamefont {{Gai}}, \citenamefont {{Huang}}, \citenamefont {{Inamura}}, \citenamefont {{Schafer-Nameki}}, \citenamefont {{Tiwari}},\ and\ \citenamefont {{Warman}}}]{bhardwaj2025gapless}%
  \BibitemOpen
  \bibfield  {author} {\bibinfo {author} {\bibfnamefont {L.}~\bibnamefont {{Bhardwaj}}}, \bibinfo {author} {\bibfnamefont {Y.}~\bibnamefont {{Gai}}}, \bibinfo {author} {\bibfnamefont {S.-J.}\ \bibnamefont {{Huang}}}, \bibinfo {author} {\bibfnamefont {K.}~\bibnamefont {{Inamura}}}, \bibinfo {author} {\bibfnamefont {S.}~\bibnamefont {{Schafer-Nameki}}}, \bibinfo {author} {\bibfnamefont {A.}~\bibnamefont {{Tiwari}}},\ and\ \bibinfo {author} {\bibfnamefont {A.}~\bibnamefont {{Warman}}},\ }\bibfield  {title} {\bibinfo {title} {{Gapless Phases in (2+1)d with Non-Invertible Symmetries}},\ }\bibfield  {journal} {\bibinfo  {journal} {arXiv e-prints}\ }\href {https://doi.org/10.48550/arXiv.2503.12699} {10.48550/arXiv.2503.12699} (\bibinfo {year} {2025}{\natexlab{b}}),\ \Eprint {https://arxiv.org/abs/2503.12699} {arXiv:2503.12699 [cond-mat.str-el]} \BibitemShut {NoStop}%
\bibitem [{\citenamefont {{Kong}}\ \emph {et~al.}(2022)\citenamefont {{Kong}}, \citenamefont {{Wen}},\ and\ \citenamefont {{Zheng}}}]{kong2022one}%
  \BibitemOpen
  \bibfield  {author} {\bibinfo {author} {\bibfnamefont {L.}~\bibnamefont {{Kong}}}, \bibinfo {author} {\bibfnamefont {X.-G.}\ \bibnamefont {{Wen}}},\ and\ \bibinfo {author} {\bibfnamefont {H.}~\bibnamefont {{Zheng}}},\ }\bibfield  {title} {\bibinfo {title} {{One dimensional gapped quantum phases and enriched fusion categories}},\ }\href {https://doi.org/10.1007/JHEP03(2022)022} {\bibfield  {journal} {\bibinfo  {journal} {Journal of High Energy Physics}\ }\textbf {\bibinfo {volume} {2022}},\ \bibinfo {eid} {22} (\bibinfo {year} {2022})}\BibitemShut {NoStop}%
\bibitem [{\citenamefont {Verresen}\ \emph {et~al.}(2021)\citenamefont {Verresen}, \citenamefont {Thorngren}, \citenamefont {Jones},\ and\ \citenamefont {Pollmann}}]{verresen2021}%
  \BibitemOpen
  \bibfield  {author} {\bibinfo {author} {\bibfnamefont {R.}~\bibnamefont {Verresen}}, \bibinfo {author} {\bibfnamefont {R.}~\bibnamefont {Thorngren}}, \bibinfo {author} {\bibfnamefont {N.~G.}\ \bibnamefont {Jones}},\ and\ \bibinfo {author} {\bibfnamefont {F.}~\bibnamefont {Pollmann}},\ }\bibfield  {title} {\bibinfo {title} {Gapless topological phases and symmetry-enriched quantum criticality},\ }\href {https://doi.org/10.1103/PhysRevX.11.041059} {\bibfield  {journal} {\bibinfo  {journal} {Phys. Rev. X}\ }\textbf {\bibinfo {volume} {11}},\ \bibinfo {pages} {041059} (\bibinfo {year} {2021})}\BibitemShut {NoStop}%
\bibitem [{\citenamefont {Thorngren}\ \emph {et~al.}(2021)\citenamefont {Thorngren}, \citenamefont {Vishwanath},\ and\ \citenamefont {Verresen}}]{Thorngren2021}%
  \BibitemOpen
  \bibfield  {author} {\bibinfo {author} {\bibfnamefont {R.}~\bibnamefont {Thorngren}}, \bibinfo {author} {\bibfnamefont {A.}~\bibnamefont {Vishwanath}},\ and\ \bibinfo {author} {\bibfnamefont {R.}~\bibnamefont {Verresen}},\ }\bibfield  {title} {\bibinfo {title} {Intrinsically gapless topological phases},\ }\href {https://doi.org/10.1103/PhysRevB.104.075132} {\bibfield  {journal} {\bibinfo  {journal} {Phys. Rev. B}\ }\textbf {\bibinfo {volume} {104}},\ \bibinfo {pages} {075132} (\bibinfo {year} {2021})}\BibitemShut {NoStop}%
\bibitem [{\citenamefont {{Satzinger}}\ \emph {et~al.}(2021)\citenamefont {{Satzinger}} \emph {et~al.}}]{satzinger2021}%
  \BibitemOpen
  \bibfield  {author} {\bibinfo {author} {\bibfnamefont {K.~J.}\ \bibnamefont {{Satzinger}}} \emph {et~al.},\ }\bibfield  {title} {\bibinfo {title} {{Realizing topologically ordered states on a quantum processor}},\ }\href {https://doi.org/10.1126/science.abi8378} {\bibfield  {journal} {\bibinfo  {journal} {Science}\ }\textbf {\bibinfo {volume} {374}},\ \bibinfo {pages} {1237} (\bibinfo {year} {2021})}\BibitemShut {NoStop}%
\bibitem [{\citenamefont {{Semeghini}}\ \emph {et~al.}(2021)\citenamefont {{Semeghini}}, \citenamefont {{Levine}}, \citenamefont {{Keesling}}, \citenamefont {{Ebadi}}, \citenamefont {{Wang}}, \citenamefont {{Bluvstein}}, \citenamefont {{Verresen}}, \citenamefont {{Pichler}}, \citenamefont {{Kalinowski}}, \citenamefont {{Samajdar}}, \citenamefont {{Omran}}, \citenamefont {{Sachdev}}, \citenamefont {{Vishwanath}}, \citenamefont {{Greiner}}, \citenamefont {{Vuleti{\'c}}},\ and\ \citenamefont {{Lukin}}}]{semeghini2021}%
  \BibitemOpen
  \bibfield  {author} {\bibinfo {author} {\bibfnamefont {G.}~\bibnamefont {{Semeghini}}}, \bibinfo {author} {\bibfnamefont {H.}~\bibnamefont {{Levine}}}, \bibinfo {author} {\bibfnamefont {A.}~\bibnamefont {{Keesling}}}, \bibinfo {author} {\bibfnamefont {S.}~\bibnamefont {{Ebadi}}}, \bibinfo {author} {\bibfnamefont {T.~T.}\ \bibnamefont {{Wang}}}, \bibinfo {author} {\bibfnamefont {D.}~\bibnamefont {{Bluvstein}}}, \bibinfo {author} {\bibfnamefont {R.}~\bibnamefont {{Verresen}}}, \bibinfo {author} {\bibfnamefont {H.}~\bibnamefont {{Pichler}}}, \bibinfo {author} {\bibfnamefont {M.}~\bibnamefont {{Kalinowski}}}, \bibinfo {author} {\bibfnamefont {R.}~\bibnamefont {{Samajdar}}}, \bibinfo {author} {\bibfnamefont {A.}~\bibnamefont {{Omran}}}, \bibinfo {author} {\bibfnamefont {S.}~\bibnamefont {{Sachdev}}}, \bibinfo {author} {\bibfnamefont {A.}~\bibnamefont {{Vishwanath}}}, \bibinfo {author} {\bibfnamefont {M.}~\bibnamefont {{Greiner}}}, \bibinfo {author} {\bibfnamefont {V.}~\bibnamefont {{Vuleti{\'c}}}},\ and\
  \bibinfo {author} {\bibfnamefont {M.~D.}\ \bibnamefont {{Lukin}}},\ }\bibfield  {title} {\bibinfo {title} {{Probing topological spin liquids on a programmable quantum simulator}},\ }\href {https://doi.org/10.1126/science.abi8794} {\bibfield  {journal} {\bibinfo  {journal} {Science}\ }\textbf {\bibinfo {volume} {374}},\ \bibinfo {pages} {1242} (\bibinfo {year} {2021})}\BibitemShut {NoStop}%
\bibitem [{\citenamefont {Iqbal}\ \emph {et~al.}(2024)\citenamefont {Iqbal}, \citenamefont {Tantivasadakarn}, \citenamefont {Gatterman}, \citenamefont {Gerber}, \citenamefont {Gilmore}, \citenamefont {Gresh}, \citenamefont {Hankin}, \citenamefont {Hewitt}, \citenamefont {Horst}, \citenamefont {Matheny} \emph {et~al.}}]{iqbal2023}%
  \BibitemOpen
  \bibfield  {author} {\bibinfo {author} {\bibfnamefont {M.}~\bibnamefont {Iqbal}}, \bibinfo {author} {\bibfnamefont {N.}~\bibnamefont {Tantivasadakarn}}, \bibinfo {author} {\bibfnamefont {T.~M.}\ \bibnamefont {Gatterman}}, \bibinfo {author} {\bibfnamefont {J.~A.}\ \bibnamefont {Gerber}}, \bibinfo {author} {\bibfnamefont {K.}~\bibnamefont {Gilmore}}, \bibinfo {author} {\bibfnamefont {D.}~\bibnamefont {Gresh}}, \bibinfo {author} {\bibfnamefont {A.}~\bibnamefont {Hankin}}, \bibinfo {author} {\bibfnamefont {N.}~\bibnamefont {Hewitt}}, \bibinfo {author} {\bibfnamefont {C.~V.}\ \bibnamefont {Horst}}, \bibinfo {author} {\bibfnamefont {M.}~\bibnamefont {Matheny}}, \emph {et~al.},\ }\bibfield  {title} {\bibinfo {title} {Topological order from measurements and feed-forward on a trapped ion quantum computer},\ }\href@noop {} {\bibfield  {journal} {\bibinfo  {journal} {Communications Physics}\ }\textbf {\bibinfo {volume} {7}},\ \bibinfo {pages} {205} (\bibinfo {year} {2024})}\BibitemShut {NoStop}%
\bibitem [{\citenamefont {{Xu}}\ \emph {et~al.}(2023)\citenamefont {{Xu}} \emph {et~al.}}]{xu2023}%
  \BibitemOpen
  \bibfield  {author} {\bibinfo {author} {\bibfnamefont {S.}~\bibnamefont {{Xu}}} \emph {et~al.},\ }\bibfield  {title} {\bibinfo {title} {{Digital Simulation of Projective Non-Abelian Anyons with 68 Superconducting Qubits}},\ }\href {https://doi.org/10.1088/0256-307X/40/6/060301} {\bibfield  {journal} {\bibinfo  {journal} {Chinese Physics Letters}\ }\textbf {\bibinfo {volume} {40}},\ \bibinfo {eid} {060301} (\bibinfo {year} {2023})}\BibitemShut {NoStop}%
\bibitem [{\citenamefont {{Google Quantum AI}}\ \emph {et~al.}(2023)\citenamefont {{Google Quantum AI}} \emph {et~al.}}]{andersen2023}%
  \BibitemOpen
  \bibfield  {author} {\bibinfo {author} {\bibnamefont {{Google Quantum AI}}} \emph {et~al.},\ }\bibfield  {title} {\bibinfo {title} {{Non-Abelian braiding of graph vertices in a superconducting processor}},\ }\href {https://doi.org/10.1038/s41586-023-05954-4} {\bibfield  {journal} {\bibinfo  {journal} {\nat}\ }\textbf {\bibinfo {volume} {618}},\ \bibinfo {pages} {264} (\bibinfo {year} {2023})}\BibitemShut {NoStop}%
\bibitem [{\citenamefont {{Foss-Feig}}\ \emph {et~al.}(2023)\citenamefont {{Foss-Feig}}, \citenamefont {{Tikku}}, \citenamefont {{Lu}}, \citenamefont {{Mayer}}, \citenamefont {{Iqbal}}, \citenamefont {{Gatterman}}, \citenamefont {{Gerber}}, \citenamefont {{Gilmore}}, \citenamefont {{Gresh}}, \citenamefont {{Hankin}}, \citenamefont {{Hewitt}}, \citenamefont {{Horst}}, \citenamefont {{Matheny}}, \citenamefont {{Mengle}}, \citenamefont {{Neyenhuis}}, \citenamefont {{Dreyer}}, \citenamefont {{Hayes}}, \citenamefont {{Hsieh}},\ and\ \citenamefont {{Kim}}}]{fossfeig2023}%
  \BibitemOpen
  \bibfield  {author} {\bibinfo {author} {\bibfnamefont {M.}~\bibnamefont {{Foss-Feig}}}, \bibinfo {author} {\bibfnamefont {A.}~\bibnamefont {{Tikku}}}, \bibinfo {author} {\bibfnamefont {T.-C.}\ \bibnamefont {{Lu}}}, \bibinfo {author} {\bibfnamefont {K.}~\bibnamefont {{Mayer}}}, \bibinfo {author} {\bibfnamefont {M.}~\bibnamefont {{Iqbal}}}, \bibinfo {author} {\bibfnamefont {T.~M.}\ \bibnamefont {{Gatterman}}}, \bibinfo {author} {\bibfnamefont {J.~A.}\ \bibnamefont {{Gerber}}}, \bibinfo {author} {\bibfnamefont {K.}~\bibnamefont {{Gilmore}}}, \bibinfo {author} {\bibfnamefont {D.}~\bibnamefont {{Gresh}}}, \bibinfo {author} {\bibfnamefont {A.}~\bibnamefont {{Hankin}}}, \bibinfo {author} {\bibfnamefont {N.}~\bibnamefont {{Hewitt}}}, \bibinfo {author} {\bibfnamefont {C.~V.}\ \bibnamefont {{Horst}}}, \bibinfo {author} {\bibfnamefont {M.}~\bibnamefont {{Matheny}}}, \bibinfo {author} {\bibfnamefont {T.}~\bibnamefont {{Mengle}}}, \bibinfo {author} {\bibfnamefont {B.}~\bibnamefont {{Neyenhuis}}}, \bibinfo {author}
  {\bibfnamefont {H.}~\bibnamefont {{Dreyer}}}, \bibinfo {author} {\bibfnamefont {D.}~\bibnamefont {{Hayes}}}, \bibinfo {author} {\bibfnamefont {T.~H.}\ \bibnamefont {{Hsieh}}},\ and\ \bibinfo {author} {\bibfnamefont {I.~H.}\ \bibnamefont {{Kim}}},\ }\bibfield  {title} {\bibinfo {title} {{Experimental demonstration of the advantage of adaptive quantum circuits}},\ }\bibfield  {journal} {\bibinfo  {journal} {arXiv e-prints}\ }\href {https://doi.org/10.48550/arXiv.2302.03029} {10.48550/arXiv.2302.03029} (\bibinfo {year} {2023}),\ \Eprint {https://arxiv.org/abs/2302.03029} {arXiv:2302.03029 [quant-ph]} \BibitemShut {NoStop}%
\bibitem [{\citenamefont {Goel}\ \emph {et~al.}(2024)\citenamefont {Goel}, \citenamefont {Reynolds}, \citenamefont {Girling}, \citenamefont {McCutcheon}, \citenamefont {Leedumrongwatthanakun}, \citenamefont {Srivastav}, \citenamefont {Jennings}, \citenamefont {Malik},\ and\ \citenamefont {Pachos}}]{goel2024}%
  \BibitemOpen
  \bibfield  {author} {\bibinfo {author} {\bibfnamefont {S.}~\bibnamefont {Goel}}, \bibinfo {author} {\bibfnamefont {M.}~\bibnamefont {Reynolds}}, \bibinfo {author} {\bibfnamefont {M.}~\bibnamefont {Girling}}, \bibinfo {author} {\bibfnamefont {W.}~\bibnamefont {McCutcheon}}, \bibinfo {author} {\bibfnamefont {S.}~\bibnamefont {Leedumrongwatthanakun}}, \bibinfo {author} {\bibfnamefont {V.}~\bibnamefont {Srivastav}}, \bibinfo {author} {\bibfnamefont {D.}~\bibnamefont {Jennings}}, \bibinfo {author} {\bibfnamefont {M.}~\bibnamefont {Malik}},\ and\ \bibinfo {author} {\bibfnamefont {J.~K.}\ \bibnamefont {Pachos}},\ }\bibfield  {title} {\bibinfo {title} {Unveiling the non-abelian statistics of $d({S}_{3})$ anyons using a classical photonic simulator},\ }\href {https://doi.org/10.1103/PhysRevLett.132.110601} {\bibfield  {journal} {\bibinfo  {journal} {Phys. Rev. Lett.}\ }\textbf {\bibinfo {volume} {132}},\ \bibinfo {pages} {110601} (\bibinfo {year} {2024})}\BibitemShut {NoStop}%
\bibitem [{\citenamefont {{Xu}}\ \emph {et~al.}(2024)\citenamefont {{Xu}} \emph {et~al.}}]{fib1}%
  \BibitemOpen
  \bibfield  {author} {\bibinfo {author} {\bibfnamefont {S.}~\bibnamefont {{Xu}}} \emph {et~al.},\ }\bibfield  {title} {\bibinfo {title} {{Non-Abelian braiding of Fibonacci anyons with a superconducting processor}},\ }\href {https://doi.org/10.1038/s41567-024-02529-6} {\bibfield  {journal} {\bibinfo  {journal} {Nature Physics}\ }\textbf {\bibinfo {volume} {20}},\ \bibinfo {pages} {1469} (\bibinfo {year} {2024})}\BibitemShut {NoStop}%
\bibitem [{\citenamefont {Lee}\ \emph {et~al.}(2023)\citenamefont {Lee}, \citenamefont {Jian},\ and\ \citenamefont {Xu}}]{lee2023}%
  \BibitemOpen
  \bibfield  {author} {\bibinfo {author} {\bibfnamefont {J.~Y.}\ \bibnamefont {Lee}}, \bibinfo {author} {\bibfnamefont {C.-M.}\ \bibnamefont {Jian}},\ and\ \bibinfo {author} {\bibfnamefont {C.}~\bibnamefont {Xu}},\ }\bibfield  {title} {\bibinfo {title} {Quantum criticality under decoherence or weak measurement},\ }\href {https://doi.org/10.1103/PRXQuantum.4.030317} {\bibfield  {journal} {\bibinfo  {journal} {PRX Quantum}\ }\textbf {\bibinfo {volume} {4}},\ \bibinfo {pages} {030317} (\bibinfo {year} {2023})}\BibitemShut {NoStop}%
\bibitem [{\citenamefont {Sala}\ \emph {et~al.}(2024)\citenamefont {Sala}, \citenamefont {Gopalakrishnan}, \citenamefont {Oshikawa},\ and\ \citenamefont {You}}]{sala2024}%
  \BibitemOpen
  \bibfield  {author} {\bibinfo {author} {\bibfnamefont {P.}~\bibnamefont {Sala}}, \bibinfo {author} {\bibfnamefont {S.}~\bibnamefont {Gopalakrishnan}}, \bibinfo {author} {\bibfnamefont {M.}~\bibnamefont {Oshikawa}},\ and\ \bibinfo {author} {\bibfnamefont {Y.}~\bibnamefont {You}},\ }\bibfield  {title} {\bibinfo {title} {Spontaneous strong symmetry breaking in open systems: Purification perspective},\ }\href {https://doi.org/10.1103/PhysRevB.110.155150} {\bibfield  {journal} {\bibinfo  {journal} {Phys. Rev. B}\ }\textbf {\bibinfo {volume} {110}},\ \bibinfo {pages} {155150} (\bibinfo {year} {2024})}\BibitemShut {NoStop}%
\bibitem [{\citenamefont {Lessa}\ \emph {et~al.}(2025{\natexlab{a}})\citenamefont {Lessa}, \citenamefont {Ma}, \citenamefont {Zhang}, \citenamefont {Bi}, \citenamefont {Cheng},\ and\ \citenamefont {Wang}}]{lessa2025ssb}%
  \BibitemOpen
  \bibfield  {author} {\bibinfo {author} {\bibfnamefont {L.~A.}\ \bibnamefont {Lessa}}, \bibinfo {author} {\bibfnamefont {R.}~\bibnamefont {Ma}}, \bibinfo {author} {\bibfnamefont {J.-H.}\ \bibnamefont {Zhang}}, \bibinfo {author} {\bibfnamefont {Z.}~\bibnamefont {Bi}}, \bibinfo {author} {\bibfnamefont {M.}~\bibnamefont {Cheng}},\ and\ \bibinfo {author} {\bibfnamefont {C.}~\bibnamefont {Wang}},\ }\bibfield  {title} {\bibinfo {title} {Strong-to-weak spontaneous symmetry breaking in mixed quantum states},\ }\href {https://doi.org/10.1103/PRXQuantum.6.010344} {\bibfield  {journal} {\bibinfo  {journal} {PRX Quantum}\ }\textbf {\bibinfo {volume} {6}},\ \bibinfo {pages} {010344} (\bibinfo {year} {2025}{\natexlab{a}})}\BibitemShut {NoStop}%
\bibitem [{\citenamefont {de~Groot}\ \emph {et~al.}(2022)\citenamefont {de~Groot}, \citenamefont {Turzillo},\ and\ \citenamefont {Schuch}}]{deGroot2022}%
  \BibitemOpen
  \bibfield  {author} {\bibinfo {author} {\bibfnamefont {C.}~\bibnamefont {de~Groot}}, \bibinfo {author} {\bibfnamefont {A.}~\bibnamefont {Turzillo}},\ and\ \bibinfo {author} {\bibfnamefont {N.}~\bibnamefont {Schuch}},\ }\bibfield  {title} {\bibinfo {title} {Symmetry protected topological order in open quantum systems},\ }\href {https://doi.org/10.22331/q-2022-11-10-856} {\bibfield  {journal} {\bibinfo  {journal} {Quantum}\ }\textbf {\bibinfo {volume} {6}},\ \bibinfo {pages} {856} (\bibinfo {year} {2022})}\BibitemShut {NoStop}%
\bibitem [{\citenamefont {Lee}\ \emph {et~al.}(2025)\citenamefont {Lee}, \citenamefont {You},\ and\ \citenamefont {Xu}}]{lee2023aspt}%
  \BibitemOpen
  \bibfield  {author} {\bibinfo {author} {\bibfnamefont {J.~Y.}\ \bibnamefont {Lee}}, \bibinfo {author} {\bibfnamefont {Y.-Z.}\ \bibnamefont {You}},\ and\ \bibinfo {author} {\bibfnamefont {C.}~\bibnamefont {Xu}},\ }\bibfield  {title} {\bibinfo {title} {Symmetry protected topological phases under decoherence},\ }\href {https://doi.org/10.22331/q-2025-01-23-1607} {\bibfield  {journal} {\bibinfo  {journal} {Quantum}\ }\textbf {\bibinfo {volume} {9}},\ \bibinfo {pages} {1607} (\bibinfo {year} {2025})}\BibitemShut {NoStop}%
\bibitem [{\citenamefont {{Zhang}}\ \emph {et~al.}(2022)\citenamefont {{Zhang}}, \citenamefont {{Qi}},\ and\ \citenamefont {{Bi}}}]{zhang2022strange}%
  \BibitemOpen
  \bibfield  {author} {\bibinfo {author} {\bibfnamefont {J.-H.}\ \bibnamefont {{Zhang}}}, \bibinfo {author} {\bibfnamefont {Y.}~\bibnamefont {{Qi}}},\ and\ \bibinfo {author} {\bibfnamefont {Z.}~\bibnamefont {{Bi}}},\ }\bibfield  {title} {\bibinfo {title} {{Strange Correlation Function for Average Symmetry-Protected Topological Phases}},\ }\bibfield  {journal} {\bibinfo  {journal} {arXiv e-prints}\ }\href {https://doi.org/10.48550/arXiv.2210.17485} {10.48550/arXiv.2210.17485} (\bibinfo {year} {2022}),\ \Eprint {https://arxiv.org/abs/2210.17485} {arXiv:2210.17485 [cond-mat.str-el]} \BibitemShut {NoStop}%
\bibitem [{\citenamefont {Ma}\ and\ \citenamefont {Wang}(2023)}]{Ma2023}%
  \BibitemOpen
  \bibfield  {author} {\bibinfo {author} {\bibfnamefont {R.}~\bibnamefont {Ma}}\ and\ \bibinfo {author} {\bibfnamefont {C.}~\bibnamefont {Wang}},\ }\bibfield  {title} {\bibinfo {title} {Average symmetry-protected topological phases},\ }\bibfield  {journal} {\bibinfo  {journal} {Physical Review X}\ }\textbf {\bibinfo {volume} {13}},\ \href {https://doi.org/10.1103/physrevx.13.031016} {10.1103/physrevx.13.031016} (\bibinfo {year} {2023})\BibitemShut {NoStop}%
\bibitem [{\citenamefont {Ma}\ \emph {et~al.}(2025)\citenamefont {Ma}, \citenamefont {Zhang}, \citenamefont {Bi}, \citenamefont {Cheng},\ and\ \citenamefont {Wang}}]{Ma2025}%
  \BibitemOpen
  \bibfield  {author} {\bibinfo {author} {\bibfnamefont {R.}~\bibnamefont {Ma}}, \bibinfo {author} {\bibfnamefont {J.-H.}\ \bibnamefont {Zhang}}, \bibinfo {author} {\bibfnamefont {Z.}~\bibnamefont {Bi}}, \bibinfo {author} {\bibfnamefont {M.}~\bibnamefont {Cheng}},\ and\ \bibinfo {author} {\bibfnamefont {C.}~\bibnamefont {Wang}},\ }\bibfield  {title} {\bibinfo {title} {Topological phases with average symmetries: The decohered, the disordered, and the intrinsic},\ }\href {https://doi.org/10.1103/PhysRevX.15.021062} {\bibfield  {journal} {\bibinfo  {journal} {Phys. Rev. X}\ }\textbf {\bibinfo {volume} {15}},\ \bibinfo {pages} {021062} (\bibinfo {year} {2025})}\BibitemShut {NoStop}%
\bibitem [{\citenamefont {Sohal}\ and\ \citenamefont {Prem}(2025)}]{Sohal2025}%
  \BibitemOpen
  \bibfield  {author} {\bibinfo {author} {\bibfnamefont {R.}~\bibnamefont {Sohal}}\ and\ \bibinfo {author} {\bibfnamefont {A.}~\bibnamefont {Prem}},\ }\bibfield  {title} {\bibinfo {title} {Noisy approach to intrinsically mixed-state topological order},\ }\bibfield  {journal} {\bibinfo  {journal} {PRX Quantum}\ }\textbf {\bibinfo {volume} {6}},\ \href {https://doi.org/10.1103/prxquantum.6.010313} {10.1103/prxquantum.6.010313} (\bibinfo {year} {2025})\BibitemShut {NoStop}%
\bibitem [{\citenamefont {Ellison}\ and\ \citenamefont {Cheng}(2025)}]{Ellison2025}%
  \BibitemOpen
  \bibfield  {author} {\bibinfo {author} {\bibfnamefont {T.~D.}\ \bibnamefont {Ellison}}\ and\ \bibinfo {author} {\bibfnamefont {M.}~\bibnamefont {Cheng}},\ }\bibfield  {title} {\bibinfo {title} {Toward a classification of mixed-state topological orders in two dimensions},\ }\bibfield  {journal} {\bibinfo  {journal} {PRX Quantum}\ }\textbf {\bibinfo {volume} {6}},\ \href {https://doi.org/10.1103/prxquantum.6.010315} {10.1103/prxquantum.6.010315} (\bibinfo {year} {2025})\BibitemShut {NoStop}%
\bibitem [{\citenamefont {{Hsin}}\ \emph {et~al.}(2025)\citenamefont {{Hsin}}, \citenamefont {{Kobayashi}},\ and\ \citenamefont {{Prem}}}]{hsin2025lre}%
  \BibitemOpen
  \bibfield  {author} {\bibinfo {author} {\bibfnamefont {P.-S.}\ \bibnamefont {{Hsin}}}, \bibinfo {author} {\bibfnamefont {R.}~\bibnamefont {{Kobayashi}}},\ and\ \bibinfo {author} {\bibfnamefont {A.}~\bibnamefont {{Prem}}},\ }\bibfield  {title} {\bibinfo {title} {{Higher-Form Anomalies Imply Intrinsic Long-Range Entanglement}},\ }\bibfield  {journal} {\bibinfo  {journal} {arXiv e-prints}\ }\href {https://doi.org/10.48550/arXiv.2504.10569} {10.48550/arXiv.2504.10569} (\bibinfo {year} {2025}),\ \Eprint {https://arxiv.org/abs/2504.10569} {arXiv:2504.10569 [quant-ph]} \BibitemShut {NoStop}%
\bibitem [{\citenamefont {{Sun}}\ \emph {et~al.}(2025)\citenamefont {{Sun}}, \citenamefont {{Zhang}}, \citenamefont {{Bi}},\ and\ \citenamefont {{You}}}]{sun2025holo}%
  \BibitemOpen
  \bibfield  {author} {\bibinfo {author} {\bibfnamefont {S.}~\bibnamefont {{Sun}}}, \bibinfo {author} {\bibfnamefont {J.-H.}\ \bibnamefont {{Zhang}}}, \bibinfo {author} {\bibfnamefont {Z.}~\bibnamefont {{Bi}}},\ and\ \bibinfo {author} {\bibfnamefont {Y.}~\bibnamefont {{You}}},\ }\bibfield  {title} {\bibinfo {title} {{Holographic View of Mixed-State Symmetry-Protected Topological Phases in Open Quantum Systems}},\ }\href {https://doi.org/10.1103/PRXQuantum.6.020333} {\bibfield  {journal} {\bibinfo  {journal} {PRX Quantum}\ }\textbf {\bibinfo {volume} {6}},\ \bibinfo {eid} {020333} (\bibinfo {year} {2025})}\BibitemShut {NoStop}%
\bibitem [{\citenamefont {{Heckman}}\ \emph {et~al.}(2025)\citenamefont {{Heckman}}, \citenamefont {{H{\"u}bner}},\ and\ \citenamefont {{Murdia}}}]{heckman2025holo}%
  \BibitemOpen
  \bibfield  {author} {\bibinfo {author} {\bibfnamefont {J.~J.}\ \bibnamefont {{Heckman}}}, \bibinfo {author} {\bibfnamefont {M.}~\bibnamefont {{H{\"u}bner}}},\ and\ \bibinfo {author} {\bibfnamefont {C.}~\bibnamefont {{Murdia}}},\ }\bibfield  {title} {\bibinfo {title} {{Symmetry Theories, Wigner's Function, Compactification, and Holography}},\ }\bibfield  {journal} {\bibinfo  {journal} {arXiv e-prints}\ }\href {https://doi.org/10.48550/arXiv.2505.23887} {10.48550/arXiv.2505.23887} (\bibinfo {year} {2025}),\ \Eprint {https://arxiv.org/abs/2505.23887} {arXiv:2505.23887 [hep-th]} \BibitemShut {NoStop}%
\bibitem [{\citenamefont {Ma}\ and\ \citenamefont {Turzillo}(2025)}]{ma2024}%
  \BibitemOpen
  \bibfield  {author} {\bibinfo {author} {\bibfnamefont {R.}~\bibnamefont {Ma}}\ and\ \bibinfo {author} {\bibfnamefont {A.}~\bibnamefont {Turzillo}},\ }\bibfield  {title} {\bibinfo {title} {Symmetry-protected topological phases of mixed states in the doubled space},\ }\href {https://doi.org/10.1103/PRXQuantum.6.010348} {\bibfield  {journal} {\bibinfo  {journal} {PRX Quantum}\ }\textbf {\bibinfo {volume} {6}},\ \bibinfo {pages} {010348} (\bibinfo {year} {2025})}\BibitemShut {NoStop}%
\bibitem [{\citenamefont {Zhang}\ \emph {et~al.}(2025)\citenamefont {Zhang}, \citenamefont {Xu}, \citenamefont {Zhang}, \citenamefont {Xu}, \citenamefont {Bi},\ and\ \citenamefont {Luo}}]{zhang2025}%
  \BibitemOpen
  \bibfield  {author} {\bibinfo {author} {\bibfnamefont {C.}~\bibnamefont {Zhang}}, \bibinfo {author} {\bibfnamefont {Y.}~\bibnamefont {Xu}}, \bibinfo {author} {\bibfnamefont {J.-H.}\ \bibnamefont {Zhang}}, \bibinfo {author} {\bibfnamefont {C.}~\bibnamefont {Xu}}, \bibinfo {author} {\bibfnamefont {Z.}~\bibnamefont {Bi}},\ and\ \bibinfo {author} {\bibfnamefont {Z.-X.}\ \bibnamefont {Luo}},\ }\bibfield  {title} {\bibinfo {title} {Strong-to-weak spontaneous breaking of 1-form symmetry and intrinsically mixed topological order},\ }\href {https://doi.org/10.1103/PhysRevB.111.115137} {\bibfield  {journal} {\bibinfo  {journal} {Phys. Rev. B}\ }\textbf {\bibinfo {volume} {111}},\ \bibinfo {pages} {115137} (\bibinfo {year} {2025})}\BibitemShut {NoStop}%
\bibitem [{\citenamefont {{Burnell}}(2018)}]{burnellrev}%
  \BibitemOpen
  \bibfield  {author} {\bibinfo {author} {\bibfnamefont {F.~J.}\ \bibnamefont {{Burnell}}},\ }\bibfield  {title} {\bibinfo {title} {{Anyon Condensation and Its Applications}},\ }\href {https://doi.org/10.1146/annurev-conmatphys-033117-054154} {\bibfield  {journal} {\bibinfo  {journal} {Annual Review of Condensed Matter Physics}\ }\textbf {\bibinfo {volume} {9}},\ \bibinfo {pages} {307} (\bibinfo {year} {2018})}\BibitemShut {NoStop}%
\bibitem [{\citenamefont {Davydov}(2009)}]{Davydov2009}%
  \BibitemOpen
  \bibfield  {author} {\bibinfo {author} {\bibfnamefont {A.}~\bibnamefont {Davydov}},\ }\href@noop {} {\bibinfo {title} {Modular invariants for group-theoretical modular data. i}} (\bibinfo {year} {2009}),\ \Eprint {https://arxiv.org/abs/0908.1044} {arXiv:0908.1044 [math.QA]} \BibitemShut {NoStop}%
\bibitem [{\citenamefont {Davydov}\ and\ \citenamefont {Simmons}(2016)}]{Davydov2016}%
  \BibitemOpen
  \bibfield  {author} {\bibinfo {author} {\bibfnamefont {A.}~\bibnamefont {Davydov}}\ and\ \bibinfo {author} {\bibfnamefont {D.}~\bibnamefont {Simmons}},\ }\href@noop {} {\bibinfo {title} {On lagrangian algebras in group-theoretical braided fusion categories}} (\bibinfo {year} {2016}),\ \Eprint {https://arxiv.org/abs/1603.04650} {arXiv:1603.04650 [math.QA]} \BibitemShut {NoStop}%
\bibitem [{\citenamefont {{Ostrik}}(2001)}]{ostrik2001}%
  \BibitemOpen
  \bibfield  {author} {\bibinfo {author} {\bibfnamefont {V.}~\bibnamefont {{Ostrik}}},\ }\bibfield  {title} {\bibinfo {title} {{Module categories, weak Hopf algebras and modular invariants}},\ }\href {https://doi.org/10.48550/arXiv.math/0111139} {\bibfield  {journal} {\bibinfo  {journal} {arXiv Mathematics e-prints}\ ,\ \bibinfo {eid} {math/0111139}} (\bibinfo {year} {2001})},\ \Eprint {https://arxiv.org/abs/math/0111139} {arXiv:math/0111139 [math.QA]} \BibitemShut {NoStop}%
\bibitem [{\citenamefont {Beigi}\ \emph {et~al.}(2011)\citenamefont {Beigi}, \citenamefont {Shor},\ and\ \citenamefont {Whalen}}]{Beigi2011}%
  \BibitemOpen
  \bibfield  {author} {\bibinfo {author} {\bibfnamefont {S.}~\bibnamefont {Beigi}}, \bibinfo {author} {\bibfnamefont {P.~W.}\ \bibnamefont {Shor}},\ and\ \bibinfo {author} {\bibfnamefont {D.}~\bibnamefont {Whalen}},\ }\bibfield  {title} {\bibinfo {title} {The quantum double model with boundary: Condensations and symmetries},\ }\href {https://doi.org/10.1007/s00220-011-1294-x} {\bibfield  {journal} {\bibinfo  {journal} {Communications in Mathematical Physics}\ }\textbf {\bibinfo {volume} {306}},\ \bibinfo {pages} {663} (\bibinfo {year} {2011})}\BibitemShut {NoStop}%
\bibitem [{\citenamefont {Duivenvoorden}\ \emph {et~al.}(2017)\citenamefont {Duivenvoorden}, \citenamefont {Iqbal}, \citenamefont {Haegeman}, \citenamefont {Verstraete},\ and\ \citenamefont {Schuch}}]{Duivenvoorden2017}%
  \BibitemOpen
  \bibfield  {author} {\bibinfo {author} {\bibfnamefont {K.}~\bibnamefont {Duivenvoorden}}, \bibinfo {author} {\bibfnamefont {M.}~\bibnamefont {Iqbal}}, \bibinfo {author} {\bibfnamefont {J.}~\bibnamefont {Haegeman}}, \bibinfo {author} {\bibfnamefont {F.}~\bibnamefont {Verstraete}},\ and\ \bibinfo {author} {\bibfnamefont {N.}~\bibnamefont {Schuch}},\ }\bibfield  {title} {\bibinfo {title} {Entanglement phases as holographic duals of anyon condensates},\ }\href {https://doi.org/10.1103/PhysRevB.95.235119} {\bibfield  {journal} {\bibinfo  {journal} {Phys. Rev. B}\ }\textbf {\bibinfo {volume} {95}},\ \bibinfo {pages} {235119} (\bibinfo {year} {2017})}\BibitemShut {NoStop}%
\bibitem [{\citenamefont {Drinfeld}(1986)}]{drinfeld}%
  \BibitemOpen
  \bibfield  {author} {\bibinfo {author} {\bibfnamefont {V.~G.}\ \bibnamefont {Drinfeld}},\ }\bibfield  {title} {\bibinfo {title} {{Quantum groups}},\ }\href {https://doi.org/10.1007/BF01247086} {\bibfield  {journal} {\bibinfo  {journal} {Zap. Nauchn. Semin.}\ }\textbf {\bibinfo {volume} {155}},\ \bibinfo {pages} {18} (\bibinfo {year} {1986})}\BibitemShut {NoStop}%
\bibitem [{\citenamefont {{Davydov}}(2014)}]{davydov2014}%
  \BibitemOpen
  \bibfield  {author} {\bibinfo {author} {\bibfnamefont {A.}~\bibnamefont {{Davydov}}},\ }\bibfield  {title} {\bibinfo {title} {{Bogomolov multiplier, double class-preserving automorphisms, and modular invariants for orbifolds}},\ }\href {https://doi.org/10.1063/1.4895764} {\bibfield  {journal} {\bibinfo  {journal} {Journal of Mathematical Physics}\ }\textbf {\bibinfo {volume} {55}},\ \bibinfo {eid} {092305} (\bibinfo {year} {2014})}\BibitemShut {NoStop}%
\bibitem [{\citenamefont {{Kobayashi}}\ and\ \citenamefont {{Barkeshli}}(2025)}]{kobayashi2025soft}%
  \BibitemOpen
  \bibfield  {author} {\bibinfo {author} {\bibfnamefont {R.}~\bibnamefont {{Kobayashi}}}\ and\ \bibinfo {author} {\bibfnamefont {M.}~\bibnamefont {{Barkeshli}}},\ }\bibfield  {title} {\bibinfo {title} {{Soft symmetries of topological orders}},\ }\bibfield  {journal} {\bibinfo  {journal} {arXiv e-prints}\ }\href {https://doi.org/10.48550/arXiv.2501.03314} {10.48550/arXiv.2501.03314} (\bibinfo {year} {2025}),\ \Eprint {https://arxiv.org/abs/2501.03314} {arXiv:2501.03314 [cond-mat.str-el]} \BibitemShut {NoStop}%
\bibitem [{\citenamefont {{Cong}}\ \emph {et~al.}(2016)\citenamefont {{Cong}}, \citenamefont {{Cheng}},\ and\ \citenamefont {{Wang}}}]{cong2016top}%
  \BibitemOpen
  \bibfield  {author} {\bibinfo {author} {\bibfnamefont {I.}~\bibnamefont {{Cong}}}, \bibinfo {author} {\bibfnamefont {M.}~\bibnamefont {{Cheng}}},\ and\ \bibinfo {author} {\bibfnamefont {Z.}~\bibnamefont {{Wang}}},\ }\bibfield  {title} {\bibinfo {title} {{Topological Quantum Computation with Gapped Boundaries}},\ }\bibfield  {journal} {\bibinfo  {journal} {arXiv e-prints}\ }\href {https://doi.org/10.48550/arXiv.1609.02037} {10.48550/arXiv.1609.02037} (\bibinfo {year} {2016}),\ \Eprint {https://arxiv.org/abs/1609.02037} {arXiv:1609.02037 [quant-ph]} \BibitemShut {NoStop}%
\bibitem [{\citenamefont {Schmitz}\ \emph {et~al.}(2019)\citenamefont {Schmitz}, \citenamefont {Huang},\ and\ \citenamefont {Prem}}]{schmitz2019}%
  \BibitemOpen
  \bibfield  {author} {\bibinfo {author} {\bibfnamefont {A.~T.}\ \bibnamefont {Schmitz}}, \bibinfo {author} {\bibfnamefont {S.-J.}\ \bibnamefont {Huang}},\ and\ \bibinfo {author} {\bibfnamefont {A.}~\bibnamefont {Prem}},\ }\bibfield  {title} {\bibinfo {title} {Entanglement spectra of stabilizer codes: A window into gapped quantum phases of matter},\ }\href {https://doi.org/10.1103/PhysRevB.99.205109} {\bibfield  {journal} {\bibinfo  {journal} {Phys. Rev. B}\ }\textbf {\bibinfo {volume} {99}},\ \bibinfo {pages} {205109} (\bibinfo {year} {2019})}\BibitemShut {NoStop}%
\bibitem [{\citenamefont {Schuster}\ \emph {et~al.}(2023)\citenamefont {Schuster}, \citenamefont {Tantivasadakarn}, \citenamefont {Vishwanath},\ and\ \citenamefont {Yao}}]{schuster2023holographic}%
  \BibitemOpen
  \bibfield  {author} {\bibinfo {author} {\bibfnamefont {T.}~\bibnamefont {Schuster}}, \bibinfo {author} {\bibfnamefont {N.}~\bibnamefont {Tantivasadakarn}}, \bibinfo {author} {\bibfnamefont {A.}~\bibnamefont {Vishwanath}},\ and\ \bibinfo {author} {\bibfnamefont {N.~Y.}\ \bibnamefont {Yao}},\ }\href {https://arxiv.org/abs/2312.04617} {\bibinfo {title} {A holographic view of topological stabilizer codes}} (\bibinfo {year} {2023}),\ \Eprint {https://arxiv.org/abs/2312.04617} {arXiv:2312.04617 [cond-mat.str-el]} \BibitemShut {NoStop}%
\bibitem [{\citenamefont {Liang}\ \emph {et~al.}(2024)\citenamefont {Liang}, \citenamefont {Yang}, \citenamefont {Iosue},\ and\ \citenamefont {Chen}}]{liang2024operatoralgebra}%
  \BibitemOpen
  \bibfield  {author} {\bibinfo {author} {\bibfnamefont {Z.}~\bibnamefont {Liang}}, \bibinfo {author} {\bibfnamefont {B.}~\bibnamefont {Yang}}, \bibinfo {author} {\bibfnamefont {J.~T.}\ \bibnamefont {Iosue}},\ and\ \bibinfo {author} {\bibfnamefont {Y.-A.}\ \bibnamefont {Chen}},\ }\href {https://arxiv.org/abs/2410.11942} {\bibinfo {title} {Operator algebra and algorithmic construction of boundaries and defects in (2+1)d topological pauli stabilizer codes}} (\bibinfo {year} {2024}),\ \Eprint {https://arxiv.org/abs/2410.11942} {arXiv:2410.11942 [quant-ph]} \BibitemShut {NoStop}%
\bibitem [{\citenamefont {Raussendorf}\ and\ \citenamefont {Briegel}(2001)}]{raussendorfMBQC}%
  \BibitemOpen
  \bibfield  {author} {\bibinfo {author} {\bibfnamefont {R.}~\bibnamefont {Raussendorf}}\ and\ \bibinfo {author} {\bibfnamefont {H.~J.}\ \bibnamefont {Briegel}},\ }\bibfield  {title} {\bibinfo {title} {A one-way quantum computer},\ }\href {https://doi.org/10.1103/PhysRevLett.86.5188} {\bibfield  {journal} {\bibinfo  {journal} {Phys. Rev. Lett.}\ }\textbf {\bibinfo {volume} {86}},\ \bibinfo {pages} {5188} (\bibinfo {year} {2001})}\BibitemShut {NoStop}%
\bibitem [{\citenamefont {{Son}}\ \emph {et~al.}(2012)\citenamefont {{Son}}, \citenamefont {{Amico}},\ and\ \citenamefont {{Vedral}}}]{son2012cluster}%
  \BibitemOpen
  \bibfield  {author} {\bibinfo {author} {\bibfnamefont {W.}~\bibnamefont {{Son}}}, \bibinfo {author} {\bibfnamefont {L.}~\bibnamefont {{Amico}}},\ and\ \bibinfo {author} {\bibfnamefont {V.}~\bibnamefont {{Vedral}}},\ }\bibfield  {title} {\bibinfo {title} {{Topological order in 1D Cluster state protected by symmetry}},\ }\href {https://doi.org/10.1007/s11128-011-0346-7} {\bibfield  {journal} {\bibinfo  {journal} {Quantum Information Processing}\ }\textbf {\bibinfo {volume} {11}},\ \bibinfo {pages} {1961} (\bibinfo {year} {2012})},\ \Eprint {https://arxiv.org/abs/1111.7173} {arXiv:1111.7173 [quant-ph]} \BibitemShut {NoStop}%
\bibitem [{\citenamefont {{Freed}}(2014)}]{freed2014}%
  \BibitemOpen
  \bibfield  {author} {\bibinfo {author} {\bibfnamefont {D.~S.}\ \bibnamefont {{Freed}}},\ }\bibfield  {title} {\bibinfo {title} {{Anomalies and Invertible Field Theories}},\ }\href {https://doi.org/10.48550/arXiv.1404.7224} {\bibfield  {journal} {\bibinfo  {journal} {arXiv e-prints}\ ,\ \bibinfo {eid} {arXiv:1404.7224}} (\bibinfo {year} {2014})},\ \Eprint {https://arxiv.org/abs/1404.7224} {arXiv:1404.7224 [hep-th]} \BibitemShut {NoStop}%
\bibitem [{\citenamefont {Zhang}\ and\ \citenamefont {C\'ordova}(2024)}]{zhang2024_categorical}%
  \BibitemOpen
  \bibfield  {author} {\bibinfo {author} {\bibfnamefont {C.}~\bibnamefont {Zhang}}\ and\ \bibinfo {author} {\bibfnamefont {C.}~\bibnamefont {C\'ordova}},\ }\bibfield  {title} {\bibinfo {title} {Anomalies of $(1+1)$-dimensional categorical symmetries},\ }\href {https://doi.org/10.1103/PhysRevB.110.035155} {\bibfield  {journal} {\bibinfo  {journal} {Phys. Rev. B}\ }\textbf {\bibinfo {volume} {110}},\ \bibinfo {pages} {035155} (\bibinfo {year} {2024})}\BibitemShut {NoStop}%
\bibitem [{\citenamefont {{Seifnashri}}\ and\ \citenamefont {{Shirley}}(2025)}]{wilbursahand}%
  \BibitemOpen
  \bibfield  {author} {\bibinfo {author} {\bibfnamefont {S.}~\bibnamefont {{Seifnashri}}}\ and\ \bibinfo {author} {\bibfnamefont {W.}~\bibnamefont {{Shirley}}},\ }\bibfield  {title} {\bibinfo {title} {{Disentangling anomaly-free symmetries of quantum spin chains}},\ }\bibfield  {journal} {\bibinfo  {journal} {arXiv e-prints}\ }\href {https://doi.org/10.48550/arXiv.2503.09717} {10.48550/arXiv.2503.09717} (\bibinfo {year} {2025}),\ \Eprint {https://arxiv.org/abs/2503.09717} {arXiv:2503.09717 [cond-mat.str-el]} \BibitemShut {NoStop}%
\bibitem [{\citenamefont {Levin}\ and\ \citenamefont {Gu}(2012)}]{levingu}%
  \BibitemOpen
  \bibfield  {author} {\bibinfo {author} {\bibfnamefont {M.}~\bibnamefont {Levin}}\ and\ \bibinfo {author} {\bibfnamefont {Z.-C.}\ \bibnamefont {Gu}},\ }\bibfield  {title} {\bibinfo {title} {Braiding statistics approach to symmetry-protected topological phases},\ }\href {https://doi.org/10.1103/PhysRevB.86.115109} {\bibfield  {journal} {\bibinfo  {journal} {Phys. Rev. B}\ }\textbf {\bibinfo {volume} {86}},\ \bibinfo {pages} {115109} (\bibinfo {year} {2012})}\BibitemShut {NoStop}%
\bibitem [{\citenamefont {Letchford}\ and\ \citenamefont {S{\o}rensen}(2012)}]{Letchford2012}%
  \BibitemOpen
  \bibfield  {author} {\bibinfo {author} {\bibfnamefont {A.~N.}\ \bibnamefont {Letchford}}\ and\ \bibinfo {author} {\bibfnamefont {M.~M.}\ \bibnamefont {S{\o}rensen}},\ }\bibfield  {title} {\bibinfo {title} {Binary positive semidefinite matrices and associated integer polytopes},\ }\href {https://doi.org/10.1007/s10107-010-0352-z} {\bibfield  {journal} {\bibinfo  {journal} {Mathematical Programming}\ }\textbf {\bibinfo {volume} {131}},\ \bibinfo {pages} {253} (\bibinfo {year} {2012})}\BibitemShut {NoStop}%
\bibitem [{\citenamefont {Kaidi}\ \emph {et~al.}(2022)\citenamefont {Kaidi}, \citenamefont {Komargodski}, \citenamefont {Ohmori}, \citenamefont {Seifnashri},\ and\ \citenamefont {Shao}}]{kaidi2022higher}%
  \BibitemOpen
  \bibfield  {author} {\bibinfo {author} {\bibfnamefont {J.}~\bibnamefont {Kaidi}}, \bibinfo {author} {\bibfnamefont {Z.}~\bibnamefont {Komargodski}}, \bibinfo {author} {\bibfnamefont {K.}~\bibnamefont {Ohmori}}, \bibinfo {author} {\bibfnamefont {S.}~\bibnamefont {Seifnashri}},\ and\ \bibinfo {author} {\bibfnamefont {S.-H.}\ \bibnamefont {Shao}},\ }\bibfield  {title} {\bibinfo {title} {{Higher central charges and topological boundaries in 2+1-dimensional TQFTs}},\ }\href {https://doi.org/10.21468/SciPostPhys.13.3.067} {\bibfield  {journal} {\bibinfo  {journal} {SciPost Phys.}\ }\textbf {\bibinfo {volume} {13}},\ \bibinfo {pages} {067} (\bibinfo {year} {2022})}\BibitemShut {NoStop}%
\bibitem [{\citenamefont {Lan}\ \emph {et~al.}(2015)\citenamefont {Lan}, \citenamefont {Wang},\ and\ \citenamefont {Wen}}]{lan2015}%
  \BibitemOpen
  \bibfield  {author} {\bibinfo {author} {\bibfnamefont {T.}~\bibnamefont {Lan}}, \bibinfo {author} {\bibfnamefont {J.~C.}\ \bibnamefont {Wang}},\ and\ \bibinfo {author} {\bibfnamefont {X.-G.}\ \bibnamefont {Wen}},\ }\bibfield  {title} {\bibinfo {title} {Gapped domain walls, gapped boundaries, and topological degeneracy},\ }\href {https://doi.org/10.1103/PhysRevLett.114.076402} {\bibfield  {journal} {\bibinfo  {journal} {Phys. Rev. Lett.}\ }\textbf {\bibinfo {volume} {114}},\ \bibinfo {pages} {076402} (\bibinfo {year} {2015})}\BibitemShut {NoStop}%
\bibitem [{\citenamefont {Davydov}(2010)}]{davydov2010centrealgebra}%
  \BibitemOpen
  \bibfield  {author} {\bibinfo {author} {\bibfnamefont {A.}~\bibnamefont {Davydov}},\ }\href {https://arxiv.org/abs/0908.1250} {\bibinfo {title} {Centre of an algebra}} (\bibinfo {year} {2010}),\ \Eprint {https://arxiv.org/abs/0908.1250} {arXiv:0908.1250 [math.CT]} \BibitemShut {NoStop}%
\bibitem [{\citenamefont {Xue}\ \emph {et~al.}(2024)\citenamefont {Xue}, \citenamefont {Lee},\ and\ \citenamefont {Bao}}]{xue2024}%
  \BibitemOpen
  \bibfield  {author} {\bibinfo {author} {\bibfnamefont {H.}~\bibnamefont {Xue}}, \bibinfo {author} {\bibfnamefont {J.~Y.}\ \bibnamefont {Lee}},\ and\ \bibinfo {author} {\bibfnamefont {Y.}~\bibnamefont {Bao}},\ }\href {https://arxiv.org/abs/2403.17069} {\bibinfo {title} {Tensor network formulation of symmetry protected topological phases in mixed states}} (\bibinfo {year} {2024}),\ \Eprint {https://arxiv.org/abs/2403.17069} {arXiv:2403.17069 [cond-mat.str-el]} \BibitemShut {NoStop}%
\bibitem [{\citenamefont {Guo}\ \emph {et~al.}(2025)\citenamefont {Guo}, \citenamefont {Zhang}, \citenamefont {Zhang}, \citenamefont {Yang},\ and\ \citenamefont {Bi}}]{guo2024}%
  \BibitemOpen
  \bibfield  {author} {\bibinfo {author} {\bibfnamefont {Y.}~\bibnamefont {Guo}}, \bibinfo {author} {\bibfnamefont {J.-H.}\ \bibnamefont {Zhang}}, \bibinfo {author} {\bibfnamefont {H.-R.}\ \bibnamefont {Zhang}}, \bibinfo {author} {\bibfnamefont {S.}~\bibnamefont {Yang}},\ and\ \bibinfo {author} {\bibfnamefont {Z.}~\bibnamefont {Bi}},\ }\bibfield  {title} {\bibinfo {title} {Locally purified density operators for symmetry-protected topological phases in mixed states},\ }\href {https://doi.org/10.1103/PhysRevX.15.021060} {\bibfield  {journal} {\bibinfo  {journal} {Phys. Rev. X}\ }\textbf {\bibinfo {volume} {15}},\ \bibinfo {pages} {021060} (\bibinfo {year} {2025})}\BibitemShut {NoStop}%
\bibitem [{\citenamefont {{Bu{\v{c}}a}}\ and\ \citenamefont {{Prosen}}(2012)}]{buca2012}%
  \BibitemOpen
  \bibfield  {author} {\bibinfo {author} {\bibfnamefont {B.}~\bibnamefont {{Bu{\v{c}}a}}}\ and\ \bibinfo {author} {\bibfnamefont {T.}~\bibnamefont {{Prosen}}},\ }\bibfield  {title} {\bibinfo {title} {{A note on symmetry reductions of the Lindblad equation: transport in constrained open spin chains}},\ }\href {https://doi.org/10.1088/1367-2630/14/7/073007} {\bibfield  {journal} {\bibinfo  {journal} {New Journal of Physics}\ }\textbf {\bibinfo {volume} {14}},\ \bibinfo {eid} {073007} (\bibinfo {year} {2012})}\BibitemShut {NoStop}%
\bibitem [{\citenamefont {Albert}\ and\ \citenamefont {Jiang}(2014)}]{albert2014sym}%
  \BibitemOpen
  \bibfield  {author} {\bibinfo {author} {\bibfnamefont {V.~V.}\ \bibnamefont {Albert}}\ and\ \bibinfo {author} {\bibfnamefont {L.}~\bibnamefont {Jiang}},\ }\bibfield  {title} {\bibinfo {title} {Symmetries and conserved quantities in lindblad master equations},\ }\href {https://doi.org/10.1103/PhysRevA.89.022118} {\bibfield  {journal} {\bibinfo  {journal} {Phys. Rev. A}\ }\textbf {\bibinfo {volume} {89}},\ \bibinfo {pages} {022118} (\bibinfo {year} {2014})}\BibitemShut {NoStop}%
\bibitem [{\citenamefont {Choi}(1975)}]{choi1975}%
  \BibitemOpen
  \bibfield  {author} {\bibinfo {author} {\bibfnamefont {M.-D.}\ \bibnamefont {Choi}},\ }\bibfield  {title} {\bibinfo {title} {Completely positive linear maps on complex matrices},\ }\href {https://doi.org/https://doi.org/10.1016/0024-3795(75)90075-0} {\bibfield  {journal} {\bibinfo  {journal} {Linear Algebra and its Applications}\ }\textbf {\bibinfo {volume} {10}},\ \bibinfo {pages} {285} (\bibinfo {year} {1975})}\BibitemShut {NoStop}%
\bibitem [{\citenamefont {{Jamio{\l}kowski}}(1972)}]{jamio1972}%
  \BibitemOpen
  \bibfield  {author} {\bibinfo {author} {\bibfnamefont {A.}~\bibnamefont {{Jamio{\l}kowski}}},\ }\bibfield  {title} {\bibinfo {title} {{Linear transformations which preserve trace and positive semidefiniteness of operators}},\ }\href {https://doi.org/10.1016/0034-4877(72)90011-0} {\bibfield  {journal} {\bibinfo  {journal} {Reports on Mathematical Physics}\ }\textbf {\bibinfo {volume} {3}},\ \bibinfo {pages} {275} (\bibinfo {year} {1972})}\BibitemShut {NoStop}%
\bibitem [{\citenamefont {{Feng}}\ \emph {et~al.}(2025)\citenamefont {{Feng}}, \citenamefont {{Cheng}},\ and\ \citenamefont {{Ippoliti}}}]{ippoliti2025}%
  \BibitemOpen
  \bibfield  {author} {\bibinfo {author} {\bibfnamefont {X.}~\bibnamefont {{Feng}}}, \bibinfo {author} {\bibfnamefont {Z.}~\bibnamefont {{Cheng}}},\ and\ \bibinfo {author} {\bibfnamefont {M.}~\bibnamefont {{Ippoliti}}},\ }\bibfield  {title} {\bibinfo {title} {{Hardness of observing strong-to-weak symmetry breaking}},\ }\bibfield  {journal} {\bibinfo  {journal} {arXiv e-prints}\ }\href {https://doi.org/10.48550/arXiv.2504.12233} {10.48550/arXiv.2504.12233} (\bibinfo {year} {2025}),\ \Eprint {https://arxiv.org/abs/2504.12233} {arXiv:2504.12233 [quant-ph]} \BibitemShut {NoStop}%
\bibitem [{\citenamefont {Weinstein}(2025)}]{weinstein2025}%
  \BibitemOpen
  \bibfield  {author} {\bibinfo {author} {\bibfnamefont {Z.}~\bibnamefont {Weinstein}},\ }\bibfield  {title} {\bibinfo {title} {Efficient detection of strong-to-weak spontaneous symmetry breaking via the rényi-1 correlator},\ }\bibfield  {journal} {\bibinfo  {journal} {Physical Review Letters}\ }\textbf {\bibinfo {volume} {134}},\ \href {https://doi.org/10.1103/physrevlett.134.150405} {10.1103/physrevlett.134.150405} (\bibinfo {year} {2025})\BibitemShut {NoStop}%
\bibitem [{\citenamefont {Liu}\ \emph {et~al.}(2024)\citenamefont {Liu}, \citenamefont {Chen}, \citenamefont {Zhang}, \citenamefont {Zhou},\ and\ \citenamefont {Zhang}}]{liu2024diagnosing}%
  \BibitemOpen
  \bibfield  {author} {\bibinfo {author} {\bibfnamefont {Z.}~\bibnamefont {Liu}}, \bibinfo {author} {\bibfnamefont {L.}~\bibnamefont {Chen}}, \bibinfo {author} {\bibfnamefont {Y.}~\bibnamefont {Zhang}}, \bibinfo {author} {\bibfnamefont {S.}~\bibnamefont {Zhou}},\ and\ \bibinfo {author} {\bibfnamefont {P.}~\bibnamefont {Zhang}},\ }\href {https://arxiv.org/abs/2410.09327} {\bibinfo {title} {Diagnosing strong-to-weak symmetry breaking via wightman correlators}} (\bibinfo {year} {2024}),\ \Eprint {https://arxiv.org/abs/2410.09327} {arXiv:2410.09327 [quant-ph]} \BibitemShut {NoStop}%
\bibitem [{\citenamefont {{Li}}\ and\ \citenamefont {{Mong}}(2024)}]{li2024replica}%
  \BibitemOpen
  \bibfield  {author} {\bibinfo {author} {\bibfnamefont {Z.}~\bibnamefont {{Li}}}\ and\ \bibinfo {author} {\bibfnamefont {R.~S.~K.}\ \bibnamefont {{Mong}}},\ }\bibfield  {title} {\bibinfo {title} {{Replica topological order in quantum mixed states and quantum error correction}},\ }\href {https://doi.org/10.48550/arXiv.2402.09516} {\bibfield  {journal} {\bibinfo  {journal} {arXiv e-prints}\ ,\ \bibinfo {eid} {arXiv:2402.09516}} (\bibinfo {year} {2024})},\ \Eprint {https://arxiv.org/abs/2402.09516} {arXiv:2402.09516 [quant-ph]} \BibitemShut {NoStop}%
\bibitem [{\citenamefont {{Hastings}}(2007)}]{hastings2007}%
  \BibitemOpen
  \bibfield  {author} {\bibinfo {author} {\bibfnamefont {M.~B.}\ \bibnamefont {{Hastings}}},\ }\bibfield  {title} {\bibinfo {title} {{An area law for one-dimensional quantum systems}},\ }\href {https://doi.org/10.1088/1742-5468/2007/08/P08024} {\bibfield  {journal} {\bibinfo  {journal} {Journal of Statistical Mechanics: Theory and Experiment}\ }\textbf {\bibinfo {volume} {2007}},\ \bibinfo {pages} {08024} (\bibinfo {year} {2007})}\BibitemShut {NoStop}%
\bibitem [{\citenamefont {Sang}\ \emph {et~al.}(2024)\citenamefont {Sang}, \citenamefont {Zou},\ and\ \citenamefont {Hsieh}}]{sang2023mixed}%
  \BibitemOpen
  \bibfield  {author} {\bibinfo {author} {\bibfnamefont {S.}~\bibnamefont {Sang}}, \bibinfo {author} {\bibfnamefont {Y.}~\bibnamefont {Zou}},\ and\ \bibinfo {author} {\bibfnamefont {T.~H.}\ \bibnamefont {Hsieh}},\ }\bibfield  {title} {\bibinfo {title} {Mixed-state quantum phases: Renormalization and quantum error correction},\ }\href {https://doi.org/10.1103/PhysRevX.14.031044} {\bibfield  {journal} {\bibinfo  {journal} {Phys. Rev. X}\ }\textbf {\bibinfo {volume} {14}},\ \bibinfo {pages} {031044} (\bibinfo {year} {2024})}\BibitemShut {NoStop}%
\bibitem [{\citenamefont {Sang}\ \emph {et~al.}(2025)\citenamefont {Sang}, \citenamefont {Lessa}, \citenamefont {Mong}, \citenamefont {Grover}, \citenamefont {Wang},\ and\ \citenamefont {Hsieh}}]{sang2024def}%
  \BibitemOpen
  \bibfield  {author} {\bibinfo {author} {\bibfnamefont {S.}~\bibnamefont {Sang}}, \bibinfo {author} {\bibfnamefont {L.~A.}\ \bibnamefont {Lessa}}, \bibinfo {author} {\bibfnamefont {R.~S.~K.}\ \bibnamefont {Mong}}, \bibinfo {author} {\bibfnamefont {T.}~\bibnamefont {Grover}}, \bibinfo {author} {\bibfnamefont {C.}~\bibnamefont {Wang}},\ and\ \bibinfo {author} {\bibfnamefont {T.~H.}\ \bibnamefont {Hsieh}},\ }\href {https://arxiv.org/abs/2507.02292} {\bibinfo {title} {Mixed-state phases from local reversibility}} (\bibinfo {year} {2025}),\ \Eprint {https://arxiv.org/abs/2507.02292} {arXiv:2507.02292 [quant-ph]} \BibitemShut {NoStop}%
\bibitem [{\citenamefont {Rakovszky}\ \emph {et~al.}(2024)\citenamefont {Rakovszky}, \citenamefont {Gopalakrishnan},\ and\ \citenamefont {von Keyserlingk}}]{rakovszky2023stable}%
  \BibitemOpen
  \bibfield  {author} {\bibinfo {author} {\bibfnamefont {T.}~\bibnamefont {Rakovszky}}, \bibinfo {author} {\bibfnamefont {S.}~\bibnamefont {Gopalakrishnan}},\ and\ \bibinfo {author} {\bibfnamefont {C.}~\bibnamefont {von Keyserlingk}},\ }\bibfield  {title} {\bibinfo {title} {Defining stable phases of open quantum systems},\ }\href {https://doi.org/10.1103/PhysRevX.14.041031} {\bibfield  {journal} {\bibinfo  {journal} {Phys. Rev. X}\ }\textbf {\bibinfo {volume} {14}},\ \bibinfo {pages} {041031} (\bibinfo {year} {2024})}\BibitemShut {NoStop}%
\bibitem [{\citenamefont {{Bao}}\ \emph {et~al.}(2023)\citenamefont {{Bao}}, \citenamefont {{Fan}}, \citenamefont {{Vishwanath}},\ and\ \citenamefont {{Altman}}}]{bao2023mixed}%
  \BibitemOpen
  \bibfield  {author} {\bibinfo {author} {\bibfnamefont {Y.}~\bibnamefont {{Bao}}}, \bibinfo {author} {\bibfnamefont {R.}~\bibnamefont {{Fan}}}, \bibinfo {author} {\bibfnamefont {A.}~\bibnamefont {{Vishwanath}}},\ and\ \bibinfo {author} {\bibfnamefont {E.}~\bibnamefont {{Altman}}},\ }\bibfield  {title} {\bibinfo {title} {{Mixed-state topological order and the errorfield double formulation of decoherence-induced transitions}},\ }\bibfield  {journal} {\bibinfo  {journal} {arXiv e-prints}\ }\href {https://doi.org/10.48550/arXiv.2301.05687} {10.48550/arXiv.2301.05687} (\bibinfo {year} {2023}),\ \Eprint {https://arxiv.org/abs/2301.05687} {arXiv:2301.05687 [quant-ph]} \BibitemShut {NoStop}%
\bibitem [{\citenamefont {Else}\ and\ \citenamefont {Nayak}(2014)}]{else2014}%
  \BibitemOpen
  \bibfield  {author} {\bibinfo {author} {\bibfnamefont {D.~V.}\ \bibnamefont {Else}}\ and\ \bibinfo {author} {\bibfnamefont {C.}~\bibnamefont {Nayak}},\ }\bibfield  {title} {\bibinfo {title} {Classifying symmetry-protected topological phases through the anomalous action of the symmetry on the edge},\ }\bibfield  {journal} {\bibinfo  {journal} {Physical Review B}\ }\textbf {\bibinfo {volume} {90}},\ \href {https://doi.org/10.1103/physrevb.90.235137} {10.1103/physrevb.90.235137} (\bibinfo {year} {2014})\BibitemShut {NoStop}%
\bibitem [{\citenamefont {Lam}\ \emph {et~al.}()\citenamefont {Lam}, \citenamefont {Prem}, \citenamefont {Qi},\ and\ \citenamefont {Sohal}}]{classgauge}%
  \BibitemOpen
  \bibfield  {author} {\bibinfo {author} {\bibfnamefont {H.~T.}\ \bibnamefont {Lam}}, \bibinfo {author} {\bibfnamefont {A.}~\bibnamefont {Prem}}, \bibinfo {author} {\bibfnamefont {M.}~\bibnamefont {Qi}},\ and\ \bibinfo {author} {\bibfnamefont {R.}~\bibnamefont {Sohal}},\ }\href@noop {} {\bibinfo  {journal} {To Appear}\ }\BibitemShut {NoStop}%
\bibitem [{\citenamefont {Williamson}\ \emph {et~al.}(2016)\citenamefont {Williamson}, \citenamefont {Bultinck}, \citenamefont {Mari\"en}, \citenamefont {\ifmmode \mbox{\c{S}}\else \c{S}\fi{}ahino\ifmmode~\breve{g}\else \u{g}\fi{}lu}, \citenamefont {Haegeman},\ and\ \citenamefont {Verstraete}}]{williamson2016mpo}%
  \BibitemOpen
\bibfield  {journal} {  }\bibfield  {author} {\bibinfo {author} {\bibfnamefont {D.~J.}\ \bibnamefont {Williamson}}, \bibinfo {author} {\bibfnamefont {N.}~\bibnamefont {Bultinck}}, \bibinfo {author} {\bibfnamefont {M.}~\bibnamefont {Mari\"en}}, \bibinfo {author} {\bibfnamefont {M.~B.}\ \bibnamefont {\ifmmode \mbox{\c{S}}\else \c{S}\fi{}ahino\ifmmode~\breve{g}\else \u{g}\fi{}lu}}, \bibinfo {author} {\bibfnamefont {J.}~\bibnamefont {Haegeman}},\ and\ \bibinfo {author} {\bibfnamefont {F.}~\bibnamefont {Verstraete}},\ }\bibfield  {title} {\bibinfo {title} {Matrix product operators for symmetry-protected topological phases: Gauging and edge theories},\ }\href {https://doi.org/10.1103/PhysRevB.94.205150} {\bibfield  {journal} {\bibinfo  {journal} {Phys. Rev. B}\ }\textbf {\bibinfo {volume} {94}},\ \bibinfo {pages} {205150} (\bibinfo {year} {2016})}\BibitemShut {NoStop}%
\bibitem [{\citenamefont {Prem}\ and\ \citenamefont {Williamson}(2019)}]{williamson2019gauge}%
  \BibitemOpen
  \bibfield  {author} {\bibinfo {author} {\bibfnamefont {A.}~\bibnamefont {Prem}}\ and\ \bibinfo {author} {\bibfnamefont {D.~J.}\ \bibnamefont {Williamson}},\ }\bibfield  {title} {\bibinfo {title} {{Gauging permutation symmetries as a route to non-Abelian fractons}},\ }\href {https://doi.org/10.21468/SciPostPhys.7.5.068} {\bibfield  {journal} {\bibinfo  {journal} {SciPost Phys.}\ }\textbf {\bibinfo {volume} {7}},\ \bibinfo {pages} {068} (\bibinfo {year} {2019})}\BibitemShut {NoStop}%
\bibitem [{\citenamefont {Lessa}\ \emph {et~al.}(2025{\natexlab{b}})\citenamefont {Lessa}, \citenamefont {Cheng},\ and\ \citenamefont {Wang}}]{Lessa:2024wcw}%
  \BibitemOpen
  \bibfield  {author} {\bibinfo {author} {\bibfnamefont {L.~A.}\ \bibnamefont {Lessa}}, \bibinfo {author} {\bibfnamefont {M.}~\bibnamefont {Cheng}},\ and\ \bibinfo {author} {\bibfnamefont {C.}~\bibnamefont {Wang}},\ }\bibfield  {title} {\bibinfo {title} {{Mixed-State Quantum Anomaly and Multipartite Entanglement}},\ }\href {https://doi.org/10.1103/PhysRevX.15.011069} {\bibfield  {journal} {\bibinfo  {journal} {Phys. Rev. X}\ }\textbf {\bibinfo {volume} {15}},\ \bibinfo {pages} {011069} (\bibinfo {year} {2025}{\natexlab{b}})},\ \Eprint {https://arxiv.org/abs/2401.17357} {arXiv:2401.17357 [cond-mat.str-el]} \BibitemShut {NoStop}%
\bibitem [{\citenamefont {Li}\ \emph {et~al.}(2025)\citenamefont {Li}, \citenamefont {Lee},\ and\ \citenamefont {Yoshida}}]{li2024anyon}%
  \BibitemOpen
  \bibfield  {author} {\bibinfo {author} {\bibfnamefont {Z.}~\bibnamefont {Li}}, \bibinfo {author} {\bibfnamefont {D.}~\bibnamefont {Lee}},\ and\ \bibinfo {author} {\bibfnamefont {B.}~\bibnamefont {Yoshida}},\ }\bibfield  {title} {\bibinfo {title} {How much entanglement is needed for topological codes and mixed states with anomalous symmetry?},\ }\href {https://doi.org/10.1103/pw12-kdjx} {\bibfield  {journal} {\bibinfo  {journal} {Phys. Rev. X}\ }\textbf {\bibinfo {volume} {15}},\ \bibinfo {pages} {021090} (\bibinfo {year} {2025})}\BibitemShut {NoStop}%
\bibitem [{\citenamefont {{Lessa}}\ \emph {et~al.}(2025)\citenamefont {{Lessa}}, \citenamefont {{Sang}}, \citenamefont {{Lu}}, \citenamefont {{Hsieh}},\ and\ \citenamefont {{Wang}}}]{lessa2025higher}%
  \BibitemOpen
  \bibfield  {author} {\bibinfo {author} {\bibfnamefont {L.~A.}\ \bibnamefont {{Lessa}}}, \bibinfo {author} {\bibfnamefont {S.}~\bibnamefont {{Sang}}}, \bibinfo {author} {\bibfnamefont {T.-C.}\ \bibnamefont {{Lu}}}, \bibinfo {author} {\bibfnamefont {T.~H.}\ \bibnamefont {{Hsieh}}},\ and\ \bibinfo {author} {\bibfnamefont {C.}~\bibnamefont {{Wang}}},\ }\bibfield  {title} {\bibinfo {title} {{Higher-form anomaly and long-range entanglement of mixed states}},\ }\href {https://doi.org/10.48550/arXiv.2503.12792} {\bibfield  {journal} {\bibinfo  {journal} {arXiv e-prints}\ ,\ \bibinfo {eid} {arXiv:2503.12792}} (\bibinfo {year} {2025})},\ \Eprint {https://arxiv.org/abs/2503.12792} {arXiv:2503.12792 [quant-ph]} \BibitemShut {NoStop}%
\bibitem [{\citenamefont {Wang}\ \emph {et~al.}(2025)\citenamefont {Wang}, \citenamefont {Wu},\ and\ \citenamefont {Wang}}]{Wang2025intrinsic}%
  \BibitemOpen
  \bibfield  {author} {\bibinfo {author} {\bibfnamefont {Z.}~\bibnamefont {Wang}}, \bibinfo {author} {\bibfnamefont {Z.}~\bibnamefont {Wu}},\ and\ \bibinfo {author} {\bibfnamefont {Z.}~\bibnamefont {Wang}},\ }\bibfield  {title} {\bibinfo {title} {Intrinsic mixed-state topological order},\ }\bibfield  {journal} {\bibinfo  {journal} {PRX Quantum}\ }\textbf {\bibinfo {volume} {6}},\ \href {https://doi.org/10.1103/prxquantum.6.010314} {10.1103/prxquantum.6.010314} (\bibinfo {year} {2025}),\ \Eprint {https://arxiv.org/abs/2307.13758} {arXiv:2307.13758 [quant-ph]} \BibitemShut {NoStop}%
\bibitem [{\citenamefont {Kawabata}\ \emph {et~al.}(2023)\citenamefont {Kawabata}, \citenamefont {Kulkarni}, \citenamefont {Li}, \citenamefont {Numasawa},\ and\ \citenamefont {Ryu}}]{kawabata2023sym}%
  \BibitemOpen
  \bibfield  {author} {\bibinfo {author} {\bibfnamefont {K.}~\bibnamefont {Kawabata}}, \bibinfo {author} {\bibfnamefont {A.}~\bibnamefont {Kulkarni}}, \bibinfo {author} {\bibfnamefont {J.}~\bibnamefont {Li}}, \bibinfo {author} {\bibfnamefont {T.}~\bibnamefont {Numasawa}},\ and\ \bibinfo {author} {\bibfnamefont {S.}~\bibnamefont {Ryu}},\ }\bibfield  {title} {\bibinfo {title} {Symmetry of open quantum systems: Classification of dissipative quantum chaos},\ }\href {https://doi.org/10.1103/PRXQuantum.4.030328} {\bibfield  {journal} {\bibinfo  {journal} {PRX Quantum}\ }\textbf {\bibinfo {volume} {4}},\ \bibinfo {pages} {030328} (\bibinfo {year} {2023})}\BibitemShut {NoStop}%
\bibitem [{\citenamefont {Ellison}\ \emph {et~al.}(2023)\citenamefont {Ellison}, \citenamefont {Chen}, \citenamefont {Dua}, \citenamefont {Shirley}, \citenamefont {Tantivasadakarn},\ and\ \citenamefont {Williamson}}]{Ellison2023}%
  \BibitemOpen
  \bibfield  {author} {\bibinfo {author} {\bibfnamefont {T.~D.}\ \bibnamefont {Ellison}}, \bibinfo {author} {\bibfnamefont {Y.-A.}\ \bibnamefont {Chen}}, \bibinfo {author} {\bibfnamefont {A.}~\bibnamefont {Dua}}, \bibinfo {author} {\bibfnamefont {W.}~\bibnamefont {Shirley}}, \bibinfo {author} {\bibfnamefont {N.}~\bibnamefont {Tantivasadakarn}},\ and\ \bibinfo {author} {\bibfnamefont {D.~J.}\ \bibnamefont {Williamson}},\ }\bibfield  {title} {\bibinfo {title} {Pauli topological subsystem codes from abelian anyon theories},\ }\href {https://doi.org/10.22331/q-2023-10-12-1137} {\bibfield  {journal} {\bibinfo  {journal} {Quantum}\ }\textbf {\bibinfo {volume} {7}},\ \bibinfo {pages} {1137} (\bibinfo {year} {2023})}\BibitemShut {NoStop}%
\bibitem [{\citenamefont {Landon-Cardinal}\ and\ \citenamefont {Poulin}(2013)}]{poulin2013}%
  \BibitemOpen
  \bibfield  {author} {\bibinfo {author} {\bibfnamefont {O.}~\bibnamefont {Landon-Cardinal}}\ and\ \bibinfo {author} {\bibfnamefont {D.}~\bibnamefont {Poulin}},\ }\bibfield  {title} {\bibinfo {title} {Local topological order inhibits thermal stability in 2d},\ }\href {https://doi.org/10.1103/PhysRevLett.110.090502} {\bibfield  {journal} {\bibinfo  {journal} {Phys. Rev. Lett.}\ }\textbf {\bibinfo {volume} {110}},\ \bibinfo {pages} {090502} (\bibinfo {year} {2013})}\BibitemShut {NoStop}%
\bibitem [{\citenamefont {Bombin}(2010)}]{Bombin2010}%
  \BibitemOpen
  \bibfield  {author} {\bibinfo {author} {\bibfnamefont {H.}~\bibnamefont {Bombin}},\ }\bibfield  {title} {\bibinfo {title} {Topological subsystem codes},\ }\href {https://doi.org/10.1103/PhysRevA.81.032301} {\bibfield  {journal} {\bibinfo  {journal} {Phys. Rev. A}\ }\textbf {\bibinfo {volume} {81}},\ \bibinfo {pages} {032301} (\bibinfo {year} {2010})}\BibitemShut {NoStop}%
\bibitem [{\citenamefont {Bombin}\ \emph {et~al.}(2012)\citenamefont {Bombin}, \citenamefont {Duclos-Cianci},\ and\ \citenamefont {Poulin}}]{bombin2012}%
  \BibitemOpen
  \bibfield  {author} {\bibinfo {author} {\bibfnamefont {H.}~\bibnamefont {Bombin}}, \bibinfo {author} {\bibfnamefont {G.}~\bibnamefont {Duclos-Cianci}},\ and\ \bibinfo {author} {\bibfnamefont {D.}~\bibnamefont {Poulin}},\ }\bibfield  {title} {\bibinfo {title} {Universal topological phase of two-dimensional stabilizer codes},\ }\href {https://doi.org/10.1088/1367-2630/14/7/073048} {\bibfield  {journal} {\bibinfo  {journal} {New Journal of Physics}\ }\textbf {\bibinfo {volume} {14}},\ \bibinfo {pages} {073048} (\bibinfo {year} {2012})}\BibitemShut {NoStop}%
\bibitem [{\citenamefont {Bombín}(2014)}]{Bombin2014}%
  \BibitemOpen
  \bibfield  {author} {\bibinfo {author} {\bibfnamefont {H.}~\bibnamefont {Bombín}},\ }\bibfield  {title} {\bibinfo {title} {Structure of 2d topological stabilizer codes},\ }\href {https://doi.org/10.1007/s00220-014-1893-4} {\bibfield  {journal} {\bibinfo  {journal} {Communications in Mathematical Physics}\ }\textbf {\bibinfo {volume} {327}},\ \bibinfo {pages} {387–432} (\bibinfo {year} {2014})}\BibitemShut {NoStop}%
\bibitem [{\citenamefont {Ellison}\ \emph {et~al.}(2022)\citenamefont {Ellison}, \citenamefont {Chen}, \citenamefont {Dua}, \citenamefont {Shirley}, \citenamefont {Tantivasadakarn},\ and\ \citenamefont {Williamson}}]{Ellison2022}%
  \BibitemOpen
  \bibfield  {author} {\bibinfo {author} {\bibfnamefont {T.~D.}\ \bibnamefont {Ellison}}, \bibinfo {author} {\bibfnamefont {Y.-A.}\ \bibnamefont {Chen}}, \bibinfo {author} {\bibfnamefont {A.}~\bibnamefont {Dua}}, \bibinfo {author} {\bibfnamefont {W.}~\bibnamefont {Shirley}}, \bibinfo {author} {\bibfnamefont {N.}~\bibnamefont {Tantivasadakarn}},\ and\ \bibinfo {author} {\bibfnamefont {D.~J.}\ \bibnamefont {Williamson}},\ }\bibfield  {title} {\bibinfo {title} {Pauli stabilizer models of twisted quantum doubles},\ }\bibfield  {journal} {\bibinfo  {journal} {PRX Quantum}\ }\textbf {\bibinfo {volume} {3}},\ \href {https://doi.org/10.1103/prxquantum.3.010353} {10.1103/prxquantum.3.010353} (\bibinfo {year} {2022})\BibitemShut {NoStop}%
\bibitem [{\citenamefont {Bonesteel}\ and\ \citenamefont {Yang}(2007)}]{bonesteel2007}%
  \BibitemOpen
  \bibfield  {author} {\bibinfo {author} {\bibfnamefont {N.~E.}\ \bibnamefont {Bonesteel}}\ and\ \bibinfo {author} {\bibfnamefont {K.}~\bibnamefont {Yang}},\ }\bibfield  {title} {\bibinfo {title} {Infinite-randomness fixed points for chains of non-abelian quasiparticles},\ }\href {https://doi.org/10.1103/PhysRevLett.99.140405} {\bibfield  {journal} {\bibinfo  {journal} {Phys. Rev. Lett.}\ }\textbf {\bibinfo {volume} {99}},\ \bibinfo {pages} {140405} (\bibinfo {year} {2007})}\BibitemShut {NoStop}%
\bibitem [{\citenamefont {Fidkowski}\ \emph {et~al.}(2008)\citenamefont {Fidkowski}, \citenamefont {Refael}, \citenamefont {Bonesteel},\ and\ \citenamefont {Moore}}]{fidkowski2008prb}%
  \BibitemOpen
  \bibfield  {author} {\bibinfo {author} {\bibfnamefont {L.}~\bibnamefont {Fidkowski}}, \bibinfo {author} {\bibfnamefont {G.}~\bibnamefont {Refael}}, \bibinfo {author} {\bibfnamefont {N.~E.}\ \bibnamefont {Bonesteel}},\ and\ \bibinfo {author} {\bibfnamefont {J.~E.}\ \bibnamefont {Moore}},\ }\bibfield  {title} {\bibinfo {title} {c-theorem violation for effective central charge of infinite-randomness fixed points},\ }\href {https://doi.org/10.1103/PhysRevB.78.224204} {\bibfield  {journal} {\bibinfo  {journal} {Phys. Rev. B}\ }\textbf {\bibinfo {volume} {78}},\ \bibinfo {pages} {224204} (\bibinfo {year} {2008})}\BibitemShut {NoStop}%
\bibitem [{\citenamefont {Liu}\ and\ \citenamefont {Lieu}(2024)}]{liu2024}%
  \BibitemOpen
  \bibfield  {author} {\bibinfo {author} {\bibfnamefont {Y.-J.}\ \bibnamefont {Liu}}\ and\ \bibinfo {author} {\bibfnamefont {S.}~\bibnamefont {Lieu}},\ }\bibfield  {title} {\bibinfo {title} {Dissipative phase transitions and passive error correction},\ }\href {https://doi.org/10.1103/PhysRevA.109.022422} {\bibfield  {journal} {\bibinfo  {journal} {Phys. Rev. A}\ }\textbf {\bibinfo {volume} {109}},\ \bibinfo {pages} {022422} (\bibinfo {year} {2024})}\BibitemShut {NoStop}%
\bibitem [{\citenamefont {Chirame}\ \emph {et~al.}(2025)\citenamefont {Chirame}, \citenamefont {Burnell}, \citenamefont {Gopalakrishnan},\ and\ \citenamefont {Prem}}]{chirame2024spt}%
  \BibitemOpen
  \bibfield  {author} {\bibinfo {author} {\bibfnamefont {S.}~\bibnamefont {Chirame}}, \bibinfo {author} {\bibfnamefont {F.~J.}\ \bibnamefont {Burnell}}, \bibinfo {author} {\bibfnamefont {S.}~\bibnamefont {Gopalakrishnan}},\ and\ \bibinfo {author} {\bibfnamefont {A.}~\bibnamefont {Prem}},\ }\bibfield  {title} {\bibinfo {title} {Stable symmetry-protected topological phases in systems with heralded noise},\ }\href {https://doi.org/10.1103/PhysRevLett.134.010403} {\bibfield  {journal} {\bibinfo  {journal} {Phys. Rev. Lett.}\ }\textbf {\bibinfo {volume} {134}},\ \bibinfo {pages} {010403} (\bibinfo {year} {2025})}\BibitemShut {NoStop}%
\bibitem [{\citenamefont {{Chirame}}\ \emph {et~al.}(2024)\citenamefont {{Chirame}}, \citenamefont {{Prem}}, \citenamefont {{Gopalakrishnan}},\ and\ \citenamefont {{Burnell}}}]{chirame2024to}%
  \BibitemOpen
  \bibfield  {author} {\bibinfo {author} {\bibfnamefont {S.}~\bibnamefont {{Chirame}}}, \bibinfo {author} {\bibfnamefont {A.}~\bibnamefont {{Prem}}}, \bibinfo {author} {\bibfnamefont {S.}~\bibnamefont {{Gopalakrishnan}}},\ and\ \bibinfo {author} {\bibfnamefont {F.~J.}\ \bibnamefont {{Burnell}}},\ }\bibfield  {title} {\bibinfo {title} {{Stabilizing Non-Abelian Topological Order against Heralded Noise via Local Lindbladian Dynamics}},\ }\href {https://doi.org/10.48550/arXiv.2410.21402} {\bibfield  {journal} {\bibinfo  {journal} {arXiv e-prints}\ ,\ \bibinfo {eid} {arXiv:2410.21402}} (\bibinfo {year} {2024})},\ \Eprint {https://arxiv.org/abs/2410.21402} {arXiv:2410.21402 [quant-ph]} \BibitemShut {NoStop}%
\bibitem [{\citenamefont {{Balasubramanian}}\ \emph {et~al.}(2024)\citenamefont {{Balasubramanian}}, \citenamefont {{Davydova}},\ and\ \citenamefont {{Lake}}}]{davydova2024local}%
  \BibitemOpen
  \bibfield  {author} {\bibinfo {author} {\bibfnamefont {S.}~\bibnamefont {{Balasubramanian}}}, \bibinfo {author} {\bibfnamefont {M.}~\bibnamefont {{Davydova}}},\ and\ \bibinfo {author} {\bibfnamefont {E.}~\bibnamefont {{Lake}}},\ }\bibfield  {title} {\bibinfo {title} {{A local automaton for the 2D toric code}},\ }\href {https://doi.org/10.48550/arXiv.2412.19803} {\bibfield  {journal} {\bibinfo  {journal} {arXiv e-prints}\ ,\ \bibinfo {eid} {arXiv:2412.19803}} (\bibinfo {year} {2024})},\ \Eprint {https://arxiv.org/abs/2412.19803} {arXiv:2412.19803 [quant-ph]} \BibitemShut {NoStop}%
\bibitem [{\citenamefont {Shah}\ \emph {et~al.}(2024)\citenamefont {Shah}, \citenamefont {Fechisin}, \citenamefont {Wang}, \citenamefont {Iosue}, \citenamefont {Watson}, \citenamefont {Wang}, \citenamefont {Ware}, \citenamefont {Gorshkov},\ and\ \citenamefont {Lin}}]{shah2024instability}%
  \BibitemOpen
  \bibfield  {author} {\bibinfo {author} {\bibfnamefont {J.}~\bibnamefont {Shah}}, \bibinfo {author} {\bibfnamefont {C.}~\bibnamefont {Fechisin}}, \bibinfo {author} {\bibfnamefont {Y.-X.}\ \bibnamefont {Wang}}, \bibinfo {author} {\bibfnamefont {J.~T.}\ \bibnamefont {Iosue}}, \bibinfo {author} {\bibfnamefont {J.~D.}\ \bibnamefont {Watson}}, \bibinfo {author} {\bibfnamefont {Y.-Q.}\ \bibnamefont {Wang}}, \bibinfo {author} {\bibfnamefont {B.}~\bibnamefont {Ware}}, \bibinfo {author} {\bibfnamefont {A.~V.}\ \bibnamefont {Gorshkov}},\ and\ \bibinfo {author} {\bibfnamefont {C.-J.}\ \bibnamefont {Lin}},\ }\bibfield  {title} {\bibinfo {title} {Instability of steady-state mixed-state symmetry-protected topological order to strong-to-weak spontaneous symmetry breaking},\ }\href@noop {} {\bibfield  {journal} {\bibinfo  {journal} {arXiv preprint arXiv:2410.12900}\ } (\bibinfo {year} {2024})}\BibitemShut {NoStop}%
\bibitem [{\citenamefont {{Schafer-Nameki}}\ \emph {et~al.}(2025)\citenamefont {{Schafer-Nameki}}, \citenamefont {{Tiwari}}, \citenamefont {{Warman}},\ and\ \citenamefont {{Zhang}}}]{sakuramixed}%
  \BibitemOpen
  \bibfield  {author} {\bibinfo {author} {\bibfnamefont {S.}~\bibnamefont {{Schafer-Nameki}}}, \bibinfo {author} {\bibfnamefont {A.}~\bibnamefont {{Tiwari}}}, \bibinfo {author} {\bibfnamefont {A.}~\bibnamefont {{Warman}}},\ and\ \bibinfo {author} {\bibfnamefont {C.}~\bibnamefont {{Zhang}}},\ }\bibfield  {title} {\bibinfo {title} {{SymTFT Approach for Mixed States with Non-Invertible Symmetries}},\ }\bibfield  {journal} {\bibinfo  {journal} {arXiv e-prints}\ }\href {https://doi.org/10.48550/arXiv.2507.05350} {10.48550/arXiv.2507.05350} (\bibinfo {year} {2025}),\ \Eprint {https://arxiv.org/abs/2507.05350} {arXiv:2507.05350 [quant-ph]} \BibitemShut {NoStop}%
\bibitem [{\citenamefont {{Luo}}\ \emph {et~al.}(2025)\citenamefont {{Luo}}, \citenamefont {{Wang}},\ and\ \citenamefont {{Bi}}}]{yinanmixed}%
  \BibitemOpen
  \bibfield  {author} {\bibinfo {author} {\bibfnamefont {R.}~\bibnamefont {{Luo}}}, \bibinfo {author} {\bibfnamefont {Y.-N.}\ \bibnamefont {{Wang}}},\ and\ \bibinfo {author} {\bibfnamefont {Z.}~\bibnamefont {{Bi}}},\ }\bibfield  {title} {\bibinfo {title} {{Topological Holography for Mixed-State Phases and Phase Transitions}},\ }\bibfield  {journal} {\bibinfo  {journal} {arXiv e-prints}\ }\href {https://doi.org/10.48550/arXiv.2507.06218} {10.48550/arXiv.2507.06218} (\bibinfo {year} {2025}),\ \Eprint {https://arxiv.org/abs/2507.06218} {arXiv:2507.06218 [cond-mat.str-el]} \BibitemShut {NoStop}%
\end{thebibliography}%


\end{document}